\documentclass[12pt,aps,prd,superscriptaddress,showpacs,nofootinbib,notitlepage,reprint,onecolumn]{revtex4-2}

\usepackage{graphicx}
\usepackage{amsmath,amssymb}
\usepackage{bbm}
\usepackage[dvipsnames]{xcolor}
\usepackage[colorlinks=true,allcolors=BlueViolet]{hyperref}
\usepackage[utf8]{inputenc}
\usepackage{lmodern}
\usepackage[T1]{fontenc}
\usepackage[inline]{enumitem}
\usepackage{accents}
\usepackage{slashed}
\usepackage{orcidlink}
\usepackage{soul}

\usepackage{tikz}
\usetikzlibrary{shapes.geometric}

\graphicspath{{Images/}}

\newcommand{\be}{\begin{equation}} \newcommand{\ee}{\end{equation}}
\newcommand{\ba}{\begin{array}{c}} \newcommand{\ea}{\end{array}}
\newcommand{\bea}{\begin{eqnarray}} \newcommand{\eea}{\end{eqnarray}}

\usepackage{enumitem}

\usepackage[sicmds,freestanding]{hepunits}
\DeclareSIUnit{\fm}{\femto\metre}

\newcommand{\olsi}[1]{\,\overline{\!{#1}}} 

\usepackage{physics}

\newcommand{\MyOp}[1]{\widehat{#1}}
\newcommand{\MollerOp}{\MyOp{\Omega}_{-}}

\usepackage{gentium}
\usepackage[T1]{fontenc}
\usepackage[cochineal,bigdelims,vvarbb]{newtxmath}

\newcommand{\IFIC}{Instituto de F\'{i}sica Corpuscular, Centro Mixto Universidad de Valencia-CSIC, Institutos de Investigaci\'{o}n de Paterna, Aptdo. 22085, E-46071 Valencia, Spain}
\newcommand{\DepFTUV}{Departamento de F\'isica Te\'orica}
\newcommand{\INFNCatania}{Istituto Nazionale di Fisica Nucleare, Sezione di Catania, Dipartimento di Fisica ``Ettore Majorana'', Universit\`a di Catania, Via Santa Sofia 64, I-95123 Catania, Italy}

\usepackage{lipsum}

\begin{document}

\title{Femtoscopy correlation functions and mass distributions from production experiments}

\author{M. Albaladejo\orcidlink{0000-0001-7340-9235}}
\email{Miguel.Albaladejo@ific.uv.es}
\affiliation{\IFIC}%

\author{A. Feijoo\orcidlink{0000-0002-8580-802X}}%
\email{edfeijoo@ific.uv.es}
\affiliation{Physik Department E62, Technische Universit\"at M\"unchen, 85478 Garching, Germany}%
 
\author{J. Nieves\orcidlink{0000-0002-2518-4606}}%
\email{Juan.M.Nieves@ific.uv.es}
\affiliation{\IFIC}%

\author{E. Oset\orcidlink{0000-0002-4462-7919}}%
\email{Eulogio.Oset@ific.uv.es}
\affiliation{\DepFTUV\ and \IFIC}%
 
\author{I. Vida\~na\orcidlink{0000-0001-7930-9112}}
\email{isaac.vidana@ct.infn.it}
\affiliation{\INFNCatania} 

\begin{abstract}
We discuss the relation between the Koonin--Pratt femtoscopic correlation function (CF)  and  invariant mass distributions from production experiments. We show that the equivalence is total for a zero source-size  and that a Gaussian finite-size source provides a form-factor for the virtual production of the particles.  Motivated by this remarkable relationship, we study an alternative method to the  Koonin--Pratt formula, which connects the evaluation of the CF  directly with the production mechanisms. The differences arise mostly from the $T$--matrix quadratic terms and increase with the source size. We study the case of the $D^0 D^{\ast +}$ and $D^+ D^{\ast 0}$ correlation functions of interest to unravel the dynamics of the exotic $T_ {cc}(3875)^+$, and find that these differences become quite sizable already for $1\,\fm$ sources. We  nevertheless conclude that the lack of coherence in high-multiplicity-event reactions and in the creation of the fire-ball source that emits the hadrons certainly make much more realistic the formalism based on the Koonin--Pratt equation. We finally derive an improved  Lednicky--Lyuboshits (LL) approach, which implements a Lorentz ultraviolet regulator that corrects the  pathological behavior of the LL CF in the punctual source-size limit.

\end{abstract}

\maketitle

\section{Introduction}

Given the technical limitations imposed by short-lifetime particles, there is no possibility of performing traditional scattering experiments when studying unstable hadrons. As an alternative, experimental information about their interactions is commonly drawn from reactions in which they are produced in the final state. In this context, lattice QCD is seen as a benchmark scheme to constrain and validate the effective field theories employed to  describe the relevant final state interactions in the analyzed reactions.

The femtoscopy technique, which was originally developed in an astronomical context by Hanbury-Brown and Twiss in the 1950s \cite{HanburyBrown:1954amm,HanburyBrown:1956bqd}, has been also applied since more than twenty years ago as a tool to study the possible creation and properties of the quark-gluon plasma in relativistic heavy-ion collisions. Experiments are designed to be sensitive to  correlations in momentum space for any hadron-hadron pair, and in particular to measure two-particle correlation functions (CFs). These latter observables are obtained as the  quotient of the number of pairs of particles with the same relative momentum produced in the same collision event over the reference distribution of pairs originated from mixed events~\cite{Lisa:2005dd}. In high-multiplicity events of proton-proton ($pp$), proton-nucleus ($pA$) and nucleus-nucleus ($AA$) collisions, the hadron production yields are well described by  statistical models, which makes clearer the connection between CFs and  two-hadron interactions  and scattering parameters~\cite{ALICE:2020mfd,ALICE:2022wwr,Fabbietti:2020bfg}. Theoretically, CFs can be calculated in terms of the spatial overlap between a source function, defining the probability density of finding the two hadrons of the emitted pair at a given relative distance $r$, and the square of the absolute value of the wave function of the considered hadron pair, determined from the half off-shell scattering $T-$matrix~\cite{Koonin:1977fh, Lednicky:1981su,Pratt:1986cc,Pratt:1987zz,Pratt:1990zq,Bauer:1992ffu,Morita:2014kza, Ohnishi:2016elb,Morita:2016auo, Hatsuda:2017uxk, Mihaylov:2018rva,Haidenbauer:2018jvl,Morita:2019rph, Kamiya:2019uiw,Kamiya:2021hdb,Kamiya:2022thy, Vidana:2023olz,Liu:2023uly, Albaladejo:2023pzq, Torres-Rincon:2023qll,Sarti:2023wlg,Molina:2023oeu,Molina:2023jov,Liu:2024nac,Feijoo:2024bvn}. The inverse problem of obtaining the interaction between coupled channels from the CFs of these channels was firstly studied for the interaction of the $D^0K^+, D^+K^0$ and $D^+_s\eta$~\cite{Ikeno:2023ojl}  and $D^{\ast +}D^0$ and $D^{\ast 0}D^+$~\cite{Albaladejo:2023wmv}  channels from where the $D^\ast_{s0}(2317)$ and $T_ {cc}(3875)^+$ exotic states emerge, respectively. The results in both analyses are encouraging, and show that the details of the strong interaction and the size of the source function can be simultaneously extracted with relatively good precision.

On the other hand, over the past decades experimental collaborations, such as \textsc{BaBar}, Belle, BES, \textsc{LHCb}, CMS, and ATLAS, have provided a growing number of new hadronic states, which are seen as peaks in the invariant mass distributions of the final hadrons. In particular, the spectroscopy of charmonium-like states, the so-called $XYZ$, has received an incredible boost, having the $X(3872)$ \cite{Belle:2003nnu} a prominent and pioneer role. The discovery of the $P_c$ and $P_{cs}$ baryonic pentaquark states by \textsc{LHCb} \cite{LHCb:2015yax,LHCb:2015qvk,LHCb:2019kea,LHCb:2020jpq,LHCb:2022ogu}, and more recently mesons, such as the $Z_{cs}(3985)$~\cite{BESIII:2020qkh,LHCb:2021uow}, $X(3960)$~\cite{LHCb:2022dvn,LHCb:2022aki} or  $T_{cc}(3875)^+$ \cite{LHCb:2021vvq,LHCb:2021auc}, has triggered a large activity in the hadronic community, since most of these states cannot be accommodated within simple quark models, and different theoretical interpretations of their nature have been suggested: multiquark states (tetraquarks or pentaquarks), hadroquarkonia states, hadronic molecules, cusps due to kinematic effects, or a  mixture of different configurations (see e.g. the recent reviews ~\cite{Guo:2017jvc,Brambilla:2019esw,Liu:2019zoy,Dong:2021bvy,Dong:2021rpi,Dong:2021juy,JPAC:2021rxu}).

For example, evidence for the $Z_{cs}(3985)$ state, a candidate for a charged hidden-charm tetraquark with strangeness, decaying into  $D_s^-D^{\ast 0}$ and $D_s^{\ast -}D^0$, was reported by BESIII \cite{BESIII:2020qkh} from the analysis of the $e^+e^-\to K^+D_s^-D^{\ast 0},\, K^+D_s^{\ast -}D^0$ annihilation reactions. Actually, at $e^+e^-$ center of mass energy of $4.681\,\GeV$, an excess of events was observed over the known contributions of the conventional charmed mesons near the $D_s^-D^{\ast 0}$ and $D_s^{\ast -}D^0$ mass thresholds in the $K^+$ recoil-mass spectrum. 
In turn, the narrow tetraquark-like state $T_{cc}(3875)^+$  was seen in the $D^0D^0\pi^+$ invariant mass distribution of the LHC inclusive $ pp \to X D^0D^0\pi^+$ reaction. The exotic $T_{cc}(3875)^+$ has a mass very close ($350\,\keV$ below) to the $D^0D^{\ast +}$ threshold, and it seems quite natural to interpret it as a very loosely bound isoscalar state of the pair of mesons $DD^*$, with its width due to the subsequent  $D^{\ast }\to D \pi$ decay \cite{Dong:2021bvy,Feijoo:2021ppq,Ling:2021bir,Fleming:2021wmk,Ren:2021dsi,Chen:2021cfl,Albaladejo:2021vln,Du:2021zzh,Baru:2021ldu,Santowsky:2021bhy,Deng:2021gnb,Ke:2021rxd,Agaev:2022ast,Meng:2022ozq,Abreu:2022sra,Chen:2022vpo,Albaladejo:2022sux,Dai:2023cyo,Wang:2023ovj}.

There exist experimental femtoscopy studies in the strangeness sector~\cite{ALICE:2018ysd,ALICE:2018nnl,ALICE:2019gcn,ALICE:2020wvi,ALICE:2021njx,ALICE:2021ovd,ALICE:2019buq,ALICE:2019eol,ALICE:2019hdt,ALICE:2020mfd,ALICE:2021cpv,ALICE:2022yyh,ALICE:2021cyj,ALICE:2021szj,ALICE:2023wjz,ALICE:2023eyl}, but importantly the ALICE collaboration measurement of the  $pD^{-}$, $\olsi{p}D^+$, $D^{(*)\pm}\pi^{\pm}$, and $D^{(*)\pm}K^{\pm}$ CFs in high-multiplicity $pp$ collisions at $13\,\TeV$ \cite{ALICE:2022enj,ALICE:2024bhk} paves the way to access the charm quark sector. Hence, the ALICE detector at LHC can also measure in the near future, for example, the $D^0 D^{\ast +}$ and $D^+ D^{\ast 0}$ CFs. It is therefore natural to address what is the relationship between these latter observables and the production spectrum reported by \textsc{LHCb}. 

In this work, we discuss  the  relation between femtoscopic two-particle CFs and invariant mass distributions of these particles in production experiments. We find that the equivalence is total for point-like sources (\textit{i.e.}, for a radius of the source tending to zero), whereas for extended ones we show how the source provides a form-factor for the virtual production of the particles. Motivated by this remarkable relationship, we study an alternative method to the commonly used Koonin--Pratt formula~\cite{Koonin:1977fh, Pratt:1990zq,Bauer:1992ffu} and examine the results obtained by evaluating the CF directly from the production mechanisms in standard Quantum Field Theory (QFT). The differences between  both approaches arise mostly from the $T-$matrix quadratic terms, increase with the source-size and become quite sizable already for sources of $1\,\fm$ size for the case of the $D^0 D^{\ast +}$ and $D^+ D^{\ast 0}$ CFs, which are the ones used in this work to illustrate the differences. We  nevertheless conclude that the lack of coherence in high-multiplicity-event reactions and in the creation of the fire-ball source that emits the hadrons certainly make little realistic the alternative  formalism. Thus, the analysis carried out in this work supports the traditional scheme based on the Koonin--Pratt equation.

We also critically review the Lednicky-Lyuboshits (LL) approximation~\cite{Lednicky:1981su}, showing that it is not adequate for small source-sizes since it leads to divergent CFs for point-like sources. We derive an improved LL approach, which corrects the  pathological behavior of the LL radial wave function at short distances, and hence also that of the LL CF for punctual sources. This improved model only requires the scattering length and effective range since, under certain assumptions, these observables can be used to fix the required Lorentz ultraviolet cutoff through the effective range formula. We conclude the work by pointing out some implications for experimental applications.

\section{Femtoscopy CF and spectrum from production experiments: single-channel analysis}
\label{sec:scanaly}

Experimentally, the two-particle CF is obtained as the ratio of the relative momentum distribution of pairs of particles produced in the same event and a reference distribution of pairs originated in different collisions. For simplicity, here we will focus on spinless particles for which, within certain approximations, the theoretical CF $C(\vec k\,)$ is given by (see \textit{e.g.} \cite{Bauer:1992ffu, Lisa:2005dd,Ohnishi:2016elb,Fabbietti:2020bfg}):
\begin{equation}
C(\vec k\,)=\int d^3\vec r\, S(\vec r\,)\,|\psi(\vec r; \vec k\,)|^2
\label{eq:cf}
\end{equation}
where $\psi(\vec{r};\vec{k}\,)$ is the wave function of the two-particle system. $S(\vec r\, )$ is the source function, mentioned in the Introduction, representing the distribution of the distance $r$ at which particles are emitted. 

Gaussian source profiles are typically assumed in femtoscopic
studies performed in heavy-ion collisions~\cite{Koonin:1977fh, Lisa:2005dd, Lisa:2008gf,Fabbietti:2020bfg}. However, collective expansion modes might induce
correlations between the position and momentum of the emitted particles, which might produce a decrease of the extracted Gaussian radii with increasing pair transverse momentum~\cite{Lisa:2005dd}. Indeed in Ref.~\cite{Verde:2003cx}, some evidences of non-Gaussian behavior of the source  were found in two-proton correlation functions, in the central Ar-Sc collisions, analyzed using Boltzmann-Uheling-Uhlenbeck transport theory. Realistic sources could also deviate from Gaussians by exponential tails caused by resonance decay contributions \cite{ALICE:2020ibs}.  Although the detailed non-Gaussian aspects of the correlation are important, the additional information can also cloud the main trends
in the data. In practice, Gaussian parameterizations provide the standard minimal description of experimental data, and moreover the ALICE Collaboration demonstrated in \cite{ALICE:2020ibs} that a universal Gaussian core emission of primordial baryons, particles produced during the initial collision, is present in $pp$ collisions by analyzing $p$-$p$ and $p$-$\Lambda$ CFs. The ansatz of a common Gaussian-core source was also confirmed in the meson-meson and meson-baryon sectors by studying  $\pi$-$\pi$ and $K$-$p$ pairs ~\cite{ALICE:2023sjd}.  On the other hand, the qualitative features of the findings of this work do not crucially depend on the exact details of the source. All of this has motivated us to use a spherically symmetric Gaussian distribution, normalized to unity, to model the source used to calculate the two-particle CF 

\begin{equation}
S(r) = \frac{1}{(4\pi R^2)^{3/2}} \exp\left(-\frac{r^2}{4R^2}\right),
\end{equation}
being $R$ the size of the source. The range of the source function depends on the type of reaction used, $pp$, $p-$nucleus or nucleus-nucleus collisions. In particular, it depends on the transverse mass of the system and takes values in the range of $1$--$5\,\fm$.   As a consequence, for the same produced pair of particles, the CFs are quite different for different type of reactions and therefore one can resort to different data sets to further constrain the interaction. Typically,  $R\simeq 1\,\fm$ for proton-proton collisions and $R\simeq 5\,\fm$ in the case of heavy ion collisions. In addition, $\vec k$ in Eq.~\eqref{eq:cf} is the asymptotic relative momentum of the two hadrons, \textit{i.e.}, the momentum of each particle in the center-of-mass (c.m.) frame of the pair, with the total c.m. energy given by $E=|\vec{k}|^{\, 2}/(2\mu_{ab})$ above the threshold $(m_a+m_b)$, and $\mu_{ab}=m_am_b/(m_a+m_b)$ the reduced mass of the pair. In turn, $\psi(\vec r; \vec k\,)$ is the wave function of the two particles at  $t=0$, the time at which the two hadrons have been produced and interact.

\subsection{\boldmath The wave function at  $t\to +  \infty$  }

\begin{figure}[t]
\centering
\includegraphics[height=5.5cm,keepaspectratio]{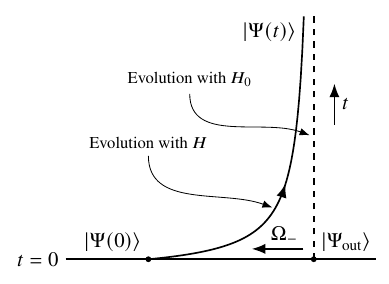}
    \caption{Schematic relation between $\ket*{ \Psi (t=0 ) }$ and $\ket*{ \Psi_{\rm out} }$.}
    \label{fig:Psiout}
\end{figure}

As it is known, in quantum mechanics (QM) the time-evolution of an eigenstate $\ket{\Psi(t=0)}$ with eigenvalue $E$ of a full conservative Hamiltonian $\MyOp{H}=\MyOp{H}_0+\MyOp{V}^{\text{QM}}$ is given by:
\begin{equation}
  \ket{ \Psi (t) } = \MyOp{U}(t)  \ket{ \Psi (t=0) } \,,
\end{equation}
where $\MyOp{U}(t)= \exponential(-it \MyOp{H}/\hbar)$ is the time evolution operator. The interaction is assumed to be described by a potential $\MyOp{V}^{\text{QM}}$ whose influence outside the interaction region becomes negligible, hence, the wave-packet will evolve freely. Therefore it is reasonable to expect that there is an eigenstate  $\ket{\Psi_{\rm out}}$ of the kinetic energy operator $\MyOp{H}_0=-\hbar^2\vec{\nabla}^{2}/(2\mu_{ab})$, with energy $E$, such that at the detector, \textit{i.e.} for $t\to +  \infty$, one has: 
\begin{equation}
  \ket{ \Psi (t\to +  \infty  ) } =   \lim_{t\to +  \infty } e^{-i\MyOp{H}_0 t/\hbar} \ket{ \Psi_{\rm out} } \label{eq:psitinf}\,,
\end{equation}
as it is schematically depicted in  Fig.~\ref{fig:Psiout}. Making use of the scattering $\MollerOp$  M\"oller operator, we have \cite{Taylor:2006,Pascual:2012} (see Eqs.~\ref{eq:a1} and \ref{eq:a3} of Appendix~\ref{app:tools}):
\begin{align}
\ket{ \Psi (t=0)} &= \MollerOp \ket{\Psi_{\rm out} } =   \left(1 -\frac{i\epsilon}{E-\MyOp{H}_0-i\epsilon} \MyOp{V}^{\text{QM}} \frac{1}{E-\MyOp{H}-i\epsilon}  \right) \ket{ \Psi_{\rm out} }   \label{eq:moller} \\
&=  \left(1 -\frac{i\epsilon}{E-\MyOp{H}_0-i\epsilon}  \MyOp{T}^{\text{QM}}(E-i\epsilon) \frac{1}{E-\MyOp{H}_0-i\epsilon}\right) \ket{ \Psi_{\rm out} } = \ket{ \Psi_{\rm out} }  + \frac{1}{E-\MyOp{H}_0-i\epsilon} \MyOp{T}^{\text{QM}}(E-i\epsilon) \ket{ \Psi_{\rm out} }\,, \nonumber
\end{align}
where the $\MyOp{T}^{\text{QM}}(z)$ operator satisfies the Lippmann-Schwinger equation (LSE):
\begin{equation}
\MyOp{T}^{\text{QM}}(z) = \MyOp{V}^{\text{QM}} + \MyOp{V}^{\text{QM}}\frac{1}{z-\MyOp{H}_0}\MyOp{T}^{\text{QM}}(z)\,,    \label{eq:lse}
\end{equation}
whose momentum-space matrix elements determine the unitary $S$-operator that gives the differential cross section~\cite{Taylor:2006,Pascual:2012}. In the derivation of Eq.~\eqref{eq:moller}, we have made use of \cite{Taylor:2006,Pascual:2012}
\begin{equation}
   \MyOp{V}^{\text{QM}} \frac{1}{z-\MyOp{H}}  =  \MyOp{T}^{\text{QM}}(z) \frac{1}{z-\MyOp{H}_0}\,. 
\end{equation}
Assuming that at the detector the wave-packet collapses  into a plane wave state of momentum $\vec{k}$ (infinite momentum resolution limit of the apparatus), one would have from Eq.~\eqref{eq:psitinf} that $\ket{\Psi_{\rm out} } = \ket*{ \vec{k}\, }$. 
On the other hand, $\psi(\vec r; \vec k\,)= N \braket{ \vec{r}\,}{ \Psi (t=0) }$, with $N=(2\pi)^{3/2}$ to keep the normalization to the unit flux.\footnote{We use for the normalization of the state with momentum $\vec p$, $\braket{ \vec p\,' }{ \vec p\,}= \delta^3(\vec p - \vec p\,')$ and therefore $\int d^3\vec p\, \op{\vec p\,}{\vec p\,} = 1$. Similarly, $\braket{ \vec r\,' }{ \vec r\,}=\delta^3(\vec r - \vec r\,')$ and then $\int d^3\vec r\, \op{\vec r\,}{ \vec r\, } = 1$. These conventions lead  in coordinate space to $\braket{ \vec r\, }{ \vec p\,} =e^{i\vec p\cdot\vec r/\hbar}/(2\pi\hbar)^{3/2}$. From now on, we will use natural units $\hbar=1=c$.} From Eq.~\eqref{eq:moller} and using $[\MyOp{T}^{\text{QM}}(z)]^\dagger=\MyOp{T}^{\text{QM}}(z^*)$ \cite{Pascual:2012}, it follows
\begin{eqnarray}
\psi^\ast(\vec r; \vec k\,)&=&  e^{-i\vec k\cdot\vec r} + N \int d^3\vec p\, 
\braket{ \vec{r}\,}{\vec p\,}^\ast 
\mel* { \vec p\, }{ \frac{1}{E-\MyOp{H}_0-i\epsilon} \MyOp{T}^{\text{QM}}(E-i\epsilon) }{\vec{k}\,}^\ast \nonumber \\
&=&  e^{-i\vec k\cdot\vec r}  + \int d^3\vec p\,e^{-i\vec p\cdot\vec r}
\mel*{ \vec k\,}{ \MyOp{T}^{\text{QM}}(E+i\epsilon)\frac{1}{E-\MyOp{H}_0+i\epsilon}
}{ \vec p\,}\nonumber \\
&=& e^{-i\vec k\cdot\vec r} +\int d^3\vec p\,\frac{e^{-i\vec p\,\cdot\vec r}}{E-\frac{\vec{p}^{\, 2}}{2\mu_{ab}}+i\epsilon}
\mel*{\vec k\,}{ \MyOp{T}^{\text{QM}}(E+i\epsilon)}{ \vec p\,}\,. \label{eq:wf1}
\end{eqnarray}
\subsection{CF of a pair of particles interacting in $S$-wave: Koonin--Pratt formula}
Now we assume here only an $S$-wave hadronic interaction, adequate in the low-momentum region. Since it only depends on the moduli of the three-momenta, we can write the $S$-wave half off-shell $T$-matrix for the transition from the initial off-shell $\ket*{ \vec p\, }$ state to the final on-shell $\ket*{ \vec k\, }$ as:
\begin{equation}
\mel*{ \vec k\,}{\MyOp{T}^{\text{QM}}(E+i\epsilon)}{\vec p\,} = T^{\text{QM}}(k\leftarrow p;E) \label{eq:thalf}\,.
\end{equation}
with $k=|\vec k|$ and $p=|\vec p|$. In principle, some ultraviolet regulator is necessary to render finite the $d^3 \vec{p}$ integration involved in Eq.~\eqref{eq:wf1}. Usually, this is introduced via an on-shell factorization, such that:
\begin{equation}\label{eq:thalf-UV}
T^{\text{QM}}(k\leftarrow p;E) = f_\text{UV}(k) f_\text{UV}(p) T^{\text{QM}}(k \leftarrow k;E)\,,
\end{equation}
where $f_\text{UV}(p)$ is, in fact, the ultraviolet regulator, satisfying that $f_\text{UV}(k) = 1$ for on-shell particles.\footnote{In principle $f_\text{UV}(k)=1$ since $ k$ is on-shell, however we keep $f_\text{UV}( k)$ explicitly to account also for the scheme of Ref.~\cite{Vidana:2023olz}, where an ultraviolet hard  cutoff $q_\text{max}$ is adopted (\textit{i.e.},  $f_\text{UV}( p)=\theta(q_\text{max}-p)$).} The normalization is such that on the mass shell:
\begin{equation}
    T^{\text{QM}}(k\leftarrow k;E) = -\frac{1}{4\pi^2\mu_{ab}}\,f_0(k), \qquad f_0(k) = \frac{1}{k\cot\delta_0(k)-ik}=\frac{1}{-1/a_0+ r_0k^2/2+ \cdots -ik}\,,
\end{equation}
with $\delta_0(k)$, $a_0$ and $r_0$ the $S$-wave phase-shift, scattering length and effective range, respectively. Now, expanding in Eq.~\eqref{eq:wf1} $e^{-i\vec p\cdot\vec r}$ in terms of spherical harmonics and keeping only the $l=0$ term one, can then perform the angular integration directly:
\begin{equation}\label{eq:wf-onshell}
    \psi^\ast(\vec r; \vec k\,) =
    e^{-i\vec k\cdot\vec r} +  \int d^3 \vec{p}\, \frac{j_0(pr)}{E-\frac{p^2}{2\mu_{ab}}+i\epsilon}T^{\text{QM}}(k\leftarrow p;E) =
    e^{-i\vec k\cdot\vec r} +  4\pi \int dp p^2\frac{j_0(pr)}{E-\frac{p^2}{2\mu_{ab}}+i\epsilon}T^{\text{QM}}(k\leftarrow p;E)
    \end{equation}
with $j_0(x)$ the Bessel spherical function of zeroth order. Hence, it is straightforward to write:
\begin{equation}
    |\psi(\vec r; \vec k\,)|^2 =  1+2\text{Re}\left( e^{i\vec k\cdot\vec r} \int d^3 \vec{p}\,\frac{j_0(pr)}{E-\frac{p^2}{2\mu_{ab}}+i\epsilon}T^{\text{QM}}(k\leftarrow p;E)\right)+\left |  \int d^3 \vec{p}\,\frac{j_0(pr)}{E-\frac{p^2}{2\mu_{ab}}+i\epsilon}T^{\text{QM}}(k\leftarrow p;E)\right|^2 \label{eq:psiq2}
\end{equation}
Inserting the above expression in Eq.~\eqref{eq:cf} we see that, when only an $S$-wave interaction is considered, the CF depends only on $k=|\vec k\,|$. We can then easily single out the ``one'' in Eq.~\eqref{eq:psiq2} to obtain the conventional Koonin--Pratt formula~\cite{Koonin:1977fh, Pratt:1990zq,Bauer:1992ffu}
\begin{eqnarray}
  C(k)&=& 1 + 4\pi \int dr r^2 S(r) \left\{\left|j_0(kr)+ \int d^3 \vec{p}\,\frac{j_0(pr)}{E-\frac{p^2}{2\mu_{ab}}+i\epsilon}T^{\text{QM}}(k\leftarrow p;E)\right|^2- j^2_0(kr)\right\}  \label{eq:KPnorel}
\end{eqnarray}
where we have added and subtracted $j^2_0(kr)$ to obtain the term unity in the CF. This is practical since for relatively large values of $k$, the CF goes to 1.

We follow next a different approach and first perform the $d^3r$ integration in Eq.~\eqref{eq:cf} using
\begin{eqnarray}
  F_R(q,q') & = & \int d^3\vec{r}\, S(r) j_0(qr)j_0(q'r)= \frac{e^{-(q^2+q^{\prime 2})R^2}\sinh(2qq'R^2)}{2qq'R^2}\simeq 1 + {\cal O}(q^2 R^2, \, q^{\,\prime 2}R^2)\,. \label{eq:defFR}
\end{eqnarray} 
In this way we can rewrite $C(k)$ as
\begin{align}
C(k) & = 1 + 2\text{Re}  \left(\int d^3 \vec{p}\,\frac{T^{\text{QM}}(k\leftarrow p;E)}{E-\frac{p^2}{2\mu_{ab}}+i\epsilon}F_R(k,p)\right) + \int d^3 \vec{p}\, d^3 \vec{p}^{\prime}\frac{T^{\text{QM}}(k\leftarrow p;E)[T^{\text{QM}}(k\leftarrow p';E)]^\ast}{\left(E-\frac{p^2}{2\mu_{ab}}+i\epsilon\right)\left(E-\frac{p^{\prime 2}}{2\mu_{ab}}-i\epsilon\right)}F_R(p,p') \nonumber \\
& = C^\text{prod}(k)+\delta C(k)\,, 
\label{eq:cknorel} 
\end{align}
with the two terms into which we have splitted $C(k)$ above, given by
\begin{eqnarray}
C^\text{prod}(k)&=&
\left|1+\int d^3 \vec{p}\,\,\frac{T^{\text{QM}}(k\leftarrow p;E)}{E-\frac{p^2}{2\mu_{ab}}+i\epsilon}\widetilde F_R(k,p)\right|^2\,,
\label{eq:ckprod} \\
\delta C(k) & = & 2\text{Re}\left(\int d^3 \vec{p}\,\frac{T^{\text{QM}}(k\leftarrow p;E)}{E-\frac{p^2}{2\mu_{ab}}+i\epsilon}\left[F_R(k,p)-\widetilde F_R(k,p)\right]\right)\nonumber \\
 & +&  \int d^3 \vec{p}\, d^3 \vec{p}^{\prime}\,\frac{T^{\text{QM}}(k\leftarrow p;E)[T^{\text{QM}}(k\leftarrow p';E)]^\ast}{\left(E-\frac{p^2}{2\mu_{ab}}+i\epsilon\right)\left(E-\frac{p^{\prime 2}}{2\mu_{ab}}-i\epsilon\right)}\left[F_R(p,p')-\widetilde F_R(k,p)\widetilde F_R(k,p')\right]\,.
\label{eq:dcknorel-aux}
\end{eqnarray}
where we have defined $\widetilde F_R(k,q)= F_R(k,q)/F_R(k,k)$. Some expansions for small size-sources of quantities related to the form-factor $F_R(q,q)'$ are given in Eqs.~\eqref{eq:a4}-\eqref{eq:ff2-deltaC} of Appendix~\ref{app:tools}.

The expression for $C(k)$ in Eqs.~\eqref{eq:KPnorel} or \eqref{eq:cknorel} is completely equivalent to the Koonin--Pratt formula~\cite{Koonin:1977fh, Pratt:1990zq,Bauer:1992ffu} for the case in which the particles interact only via $S$-wave, as shown for instance in Refs.~\cite{Vidana:2023olz,Liu:2023uly}. In addition, in the next subsection we will justify the superscript label ``prod'' that  we have given to the first term in Eq.~\eqref{eq:cknorel}.

\begin{figure*}[t]\centering
    \raisebox{-0.5\height}{\includegraphics[height=2cm]{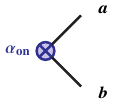}} $\qquad + \qquad$
    \raisebox{-0.5\height}{\includegraphics[height=2cm]{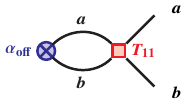}}
    \caption{Diagrams contributing to the final production of  particles $a$ and $b$. }
    \label{fig:spectrum}
\end{figure*}

\subsection{CF and invariant mass spectrum}
We will now pay attention to Eq.~\eqref{eq:cknorel}, where one can see that by neglecting $\delta C(k)$, the CF reduces to  $C^\text{prod}(k)$ (Eq.~\eqref{eq:ckprod}), which turns out to be  proportional to the invariant mass spectrum of the interacting particles depicted in Fig.~\ref{fig:spectrum}. There, we have the tree (left) and virtual propagation with re-scattering (right) diagrams contributing to the final production of the on-shell particles $a$ and $b$.  The red square stands for the half off-shell  $T^{\text{QM}}(k\leftarrow p;E)$ amplitude, while  the $\alpha_\text{on}$ and $\alpha_\text{off}$ vertices differ by the $F_R(k,p)$ form-factor that accounts for the off-shell production of $a$ and $b$ in the first step, with a regulator determined by  the size of the source\footnote{The off-shell correction $F_R(k,p)$ can be obtained by projecting a Gaussian form-factor of the type $\mathcal{F}_R(\vec k, \vec p\,)= \exp[-(\vec p-\vec k\,)^2R^2]$ into $S$-wave. This is to say, if $\vec k \cdot \vec p= kp\cos\theta$ then
\begin{equation*}
  F_R(k,p)  = \frac12 \int_{-1}^{+1} d(\cos\theta)\mathcal{F}_R(\vec k, \vec p\,)= \frac12 \int_{-1}^{+1} d(\cos\theta)e^{-(\vec p-\vec k)^2R^2}\, ,
\end{equation*}
which gives immediately the expression of Eq.~\eqref{eq:defFR} for $q,q'\equiv k,p$. This makes more natural interpreting that $F_R(k,p)$ accounts for the off-shell production of the $a$ and $b$ particles.} $R$.  The re-scattering mechanism (right diagram) requires the propagation of the virtual pair of particles and hence the correspondence of the production spectrum with the first term [$C^\text{prod}(k)$] of Eq.~\eqref{eq:cknorel} is straightforward.  This is the first important result of this work. Assuming that only the $S$-wave part of the wave function is modified by the hadronic interaction,
$C^\text{prod}(k)$ can also be written as\footnote{Note that undoing the $d^3 r$ integration in Eq.~\eqref{eq:ckprod}, we find
\begin{eqnarray}
 1+\int d^3 \vec{p}\,\,\frac{T^{\text{QM}}(k\leftarrow p;E)}{E-\frac{p^2}{2\mu_{ab}}+i\epsilon}\widetilde F_R(k,p)&=&\frac{1}{F_R(k,k)}\int d^3\vec{r}\, S(r) j_0(kr)\left[ j_0(kr)+\int d^3 \vec{p}\,\,j_0(pr)\frac{T^{\text{QM}}(k\leftarrow p;E)}{E-\frac{p^2}{2\mu_{ab}}+i\epsilon}\right] \nonumber\\
 &=&\frac{1}{F_R(k,k)}\int d^3\vec{r} \, S(r) j_0(kr)\left[ e^{-i\vec k\cdot\vec r} +\int d^3 \vec{p}\,\,e^{-i\vec p\cdot\vec r} \frac{T^{\text{QM}}(k\leftarrow p;E)}{E-\frac{p^2}{2\mu_{ab}}+i\epsilon}\right]\nonumber\\
 &=&\frac{1}{F_R(k,k)}  \int d^3\vec r\, S(r)\,j_0(kr)\,\psi^\ast(\vec r; \vec k\,)
\end{eqnarray}
}:
\begin{equation}
 C^\text{prod}(k) = \left|\frac{1}{F_R(k,k)}  \int d^3\vec r\, S(r)\,j_0(kr)\,\psi^\ast(\vec r; \vec k\,)\right|^2   \label{eq:Cprodintermswf} \ .
\end{equation}
In the production CF, the $\alpha_\text{on}$ and $\alpha_\text{off}$ vertices of  Fig.~\ref{fig:spectrum} would be represented by  $F_R(k,k)$ and $F_R(k,p)$, respectively, while the on-shell normalized   $\widetilde{F}_R(k,p) = F_R(k,p)/F_R(k,k)$ form factor  would account for the $\alpha_\text{off}/\alpha_\text{on}$ ratio. 

We look now at the additional term $\delta C(k)$. It vanishes for point-like sources ($R\to 0$) since in that limit $[F_R(k,p)-\widetilde F_R(k,p)]$ and $[F_R(p,p')-\widetilde F_R(k,p)\widetilde F_R(k,p')]$ tend to zero, as deduced from Eqs.~\eqref{eq:ff1-deltaC} and \eqref{eq:ff2-deltaC}. The previous statement is correct only when the loop integrals are  renormalized and they are finite, which in the present scheme is achieved thanks to the ultraviolet regulator  $f_\text{UV}(p)$ introduced in the half off-shell $T$--matrix in Eq.~\eqref{eq:thalf}. One can also prove that $\delta C(k)$ is zero for point-like sources using that $\lim_{R\to 0}  S(r)=\delta^3(\vec r\,)$ since then Eq.~\eqref{eq:wf-onshell} leads to:
\begin{equation}
\lim_{R\to 0}  C(k) =   \lvert \psi(\vec{r}=\vec{0}; \vec{k}\,)\rvert^2 = \left\lvert 1 +  \int d^3 \vec{p}\, \frac{T^{\text{QM}}(k\leftarrow p;E) }{E-\frac{p^2}{2\mu_{ab}}+i\epsilon}\right\rvert^2  = \lim_{R\to 0}  C^\text{prod}(k)\,.\label{eq:ckRzero}
\end{equation}
The above equation  is equivalent to the production spectrum derived from the Feynman diagrams depicted in Fig.~\ref{fig:spectrum}. We recall that the form-factor $F_R(k,p)= 1$ for a point-like source ($R= 0$). For small $R$ values, $\delta C(k)$ should not take large values at low energies, however $\delta C(k)$  can be large at low energies for extended sources as shown in Subsect.~\ref{app:k0} of the Appendix. 

\subsection{Alternative scheme to Koonin--Pratt: production CF}

From the discussion above on the connection between $C^\text{prod}(k)[=C(k)-\delta C(k)]$ and the production spectrum deduced from the Feynman diagrams depicted  in Fig.~\ref{fig:spectrum}, one might conclude that $\delta C(k)$ is an unwanted artifact stemming from the approximate Eq.~\eqref{eq:cf} to compute the CF. In the rest of this work, we study in detail the differences between Koonin--Pratt [$C(k)$] and production [$C^\text{prod}(k)$] formulae to describe the experimental CF, understood  as the ratio of the probability to find the two interacting particles  and the product of the probabilities to find each individual particle.  

\subsubsection{Coherence, incoherence, and statistical source}
\label{sec:cohe}

The above discussion, based on the Feynman diagrams of Fig.~\ref{fig:spectrum}, seems to support the use of the production CF  $[C^\text{prod}(k)]$, which implies a coherent sum of amplitudes. This is correct for exclusive processes where only the observed particles $a$ and $b$ are produced, which is obviously not the case  in  high-multiplicity event experiments.  In fact,  for  inclusive reactions, for example initiated by $pp$ collisions, the coherence is still preserved when the real and virtual production vertices, $\alpha_\text{on}$ and $\alpha_\text{off} $ in Fig.~\ref{fig:spectrum}, stand for the $pp\to X+ ab\, (\text{real}) $ and $pp\to X+ ab\, (\text{virtual}) $ processes, respectively, with the rest of the particles ($X$) in the final state being the same for both reactions. In that case, the two Feynman amplitudes of Fig.~\ref{fig:spectrum} must be coherently added since both of them contribute to the same quantum amplitude of the reaction $pp\to X + ab\, ( \text{real})$. 

However, in high-multiplicity events of $pp$, $pA$ and $AA$ collisions, where the hadron production yields are described by  statistical models which lead to the extended sources $S(\vec r\,)$ introduced in Eq.~\eqref{eq:cf}, it is not clear whether the coherent sum of amplitudes, which is assumed in $C^\text{prod}(k)$, is still appropriate.  Note that, in a sense, the Koonin--Pratt scheme implies some kind of summation of probabilities, and not of amplitudes, since the Koonin--Pratt CF is obtained from the spatial superposition between the extended source and the squared modulus of the wave-function pair $|\psi (\vec r; \vec k\,)|^2$.

\subsubsection{Relativistic effects}
In addition, Lorentz relativistic effects can be taken into account in Eq.~\eqref{eq:cknorel} using the relativistic QFT propagator and transition $T$--matrix 
\begin{eqnarray}
\int d^3 \vec{p}\,\,\frac{1}{E-\frac{p^2}{2\mu_{ab}}+i\epsilon} & \Rightarrow & G(s) = i \int\frac{d^4p}{(2\pi)^4}\frac{1}{p^2-m_a^2+i\epsilon}\frac{1}{(P_\text{c.m.}-p)^2-m_b^2+i\epsilon}\nonumber \\ 
&=&\int \frac{d^3 \vec{p}}{(2\pi)^3}\,\frac{\omega_a(p)+\omega_b(p)}{2\omega_a(p)\omega_b(p)}\frac{1}{s-\left[\omega_a(p)+\omega_b(p)\right]^2+i\epsilon} \label{eq:rel1}\\
T^{\text{QM}}(k\leftarrow p;E) & \Rightarrow& T(k\leftarrow p;s)\label{eq:rel2}
\end{eqnarray}
where $\sqrt{s}=m_a+m_b+E$, $P^\mu_\text{c.m.}=(\sqrt{s},\vec{0}\,)$, $\omega_{a,b}(p)= \sqrt{m_{a,b}^2+p^2}$ and the $p^0$ integration has been performed using Cauchy's theorem. The normalization of the QFT amplitude is determined by its relation to the  $S$-wave cross section, $T(k\leftarrow k;s)=-8\pi\sqrt{s}\ f_0(k)$ and therefore
\begin{equation}
 \Im T^{-1}(k\leftarrow k;s) =  \frac{k}{8\pi\sqrt{s}}\,, \qquad \sigma_{l=0}(s) = \frac{|T(k\leftarrow k;s)|^2}{16\pi s} = \frac{1}{16\pi s} \left | -8\pi\sqrt{s}\ f_0(k)\right|^2\,,\label{eq:secc}
\end{equation}
with the modulus of the c.m. on-shell  momentum  given by $k=\lambda^\frac12(s,m_a^2, m_b^2)/(2\sqrt{s})$, being $\lambda(x,y,z)= x^2+y^2+z^2-2xy-2yz-2xz$ the well known K\"{a}llen  (or triangle) function \cite{Kallen:1964lxa}. The half-off shell QFT amplitude $T(k\leftarrow p;s)$ should be obtained from the solution of a LSE [Eq.~\eqref{eq:lse}] using the relativistic QFT propagator $G$ and the appropriate two particle irreducible amplitude $V$,  normalized as to give  Eq.~\eqref{eq:secc}. One actually then has the Bethe-Salpeter equation (BSE) \cite{Nieves:1999bx}.

All together, the production scheme leads to an alternative evaluation of the CF in comparison to the standard Koonin--Pratt formula, \textit{i.e.},
\begin{equation} \label{eq:Cimprove}
 C^\text{prod}(s)  =    \left|1+\int \frac{d^3 \vec{p}}{(2\pi)^3}\,\frac{\omega_a(p)+\omega_b(p)}{2\omega_a(p)\omega_b(p)}\frac{T(k\leftarrow p;s)\widetilde F_R(k,p)}{s-\left[\omega_a(p)+\omega_b(p)\right]^2+i\epsilon}\right|^2\,,
\end{equation}
where the source of size $R$ enters into the definition of the form-factor $F_R(k,p)$ in Eq.~\eqref{eq:defFR}.

The discussion above illustrates the relation between experiments like ALICE, which provide femtoscopic CFs of two given particles from $pp$, $pA$ or even $AA$ collisions and other ones like BES, Belle or \textsc{LHCb}, which measure the invariant mass distribution of these two particles produced in electron-positron (BES, Belle) or proton-proton (\textsc{LHCb}) colliders.

In  Sec.~\ref{sec:res}, we will compare results for CFs calculated with $C^\text{prod}(s)$  with those obtained from the relativistic Koonin--Pratt formula derived in Refs.~\cite{Vidana:2023olz,Liu:2023uly}. The latter is obtained from Eq.~\eqref{eq:cknorel} by implementing the replacements detailed in Eqs.~\eqref{eq:rel1} and \eqref{eq:rel2}, and it reads 
\begin{align}
 C^\text{KP}(s) &=  C^\text{prod}(s)+ \delta C^\text{KP}(s)\,, \label{eq:ckprel}\\ 
 \delta C^\text{KP}(s) &= 2\text{Re}\left(\int \frac{d^3 \vec{p}}{(2\pi)^3}\frac{\omega_a(p)+\omega_b(p)}{2\omega_a(p)\omega_b(p)} \frac{T(k\leftarrow p;s)}{s-\left[\omega_a(p)+\omega_b(p)\right]^2+i\epsilon} \left[F_R(k,p)-\widetilde F_R(k,p)\right]\right) \nonumber \\
 &+ \int \frac{d^3 \vec{p}}{(2\pi)^3}\int\frac{d^3 \vec{p}^{\,\prime}}{(2\pi)^3}\left\{\frac{\omega_a(p)+\omega_b(p)}{2\omega_a(p)\omega_b(p)} \frac{T(k\leftarrow p;s)}{s-\left[\omega_a(p)+\omega_b(p)\right]^2+i\epsilon}\times\right.\label{eq:ckprel-del}\\
 &\times \left.\frac{\omega_a(p\,')+\omega_b(p\,')}{2\omega_a(p\,')\omega_b(p\,')}\frac{[T(k\leftarrow p';s)]^\ast}{s-\left[\omega_a(p\,')+\omega_b(p\,')\right]^2-i\epsilon}\right\}\left[F_R(p,p')-\widetilde F_R(k,p)\widetilde F_R(k,p')\right]\,.\nonumber
\end{align}
Alternatively, one can also obtain the same result starting from Eq.~\eqref{eq:KPnorel} and include relativistic effects, which allows to obtain $C^\text{KP}(s)$ as:
\begin{equation}
   C^\text{KP}(s) =  1 + 4\pi \int dr r^2 S(r) \left\{\left|j_0(kr)+ \int \frac{d^3 \vec{p}}{(2\pi)^3}\frac{\omega_a(p)+\omega_b(p)}{2\omega_a(p)\omega_b(p)} \frac{j_0(pr)\, T(k\leftarrow p;s)}{s-\left[\omega_a(p)+\omega_b(p)\right]^2+i\epsilon}\right|^2- j^2_0(kr)\right\}\,.  \label{eq:ckprel2}
\end{equation}

We would like to make a final comment to finish this subsection. One could take a different point of view and not consider   $\delta C(k)$ in  Eq.~\eqref{eq:cknorel} as an artifact, but instead as a necessary contribution to reproduce the measured femtoscopy CF. In that case $C(k)$ evaluated from Eq.~\eqref{eq:cknorel}, which  includes $\delta C(k)$, could be matched to the CF $C^{\prime\,\text{prod}}(k)$, inferred from the production mechanisms of Fig.~\ref{fig:spectrum}, but using a different value of the source-size $R^\prime$. This is to say, one could compute $C^{\prime\text{prod}}(k)$ from Eq.~(\ref{eq:ckprod}) using a radius of the source $R^\prime$ (with $R^\prime\neq R$), where $R$ is used in Eq.~\eqref{eq:cknorel} for $C(k)$, and tuning $R^\prime$ to reproduce $C(k)$.  Nevertheless, we do not consider this a robust theoretical scheme. This comment also applies to the expressions containing relativistic corrections [Eqs.~\eqref{eq:Cimprove} and \eqref{eq:ckprel}]. 
%
%

\section{Femtoscopy CF and spectrum from production experiments: coupled-channel analysis}
\begin{figure*}[t]\centering
    \raisebox{-0.5\height}{\includegraphics[height=2cm]{DiagramsA1.pdf}} $\qquad + \qquad$
    \raisebox{-0.5\height}{\includegraphics[height=2cm]{DiagramsA2.pdf}} $\qquad + \qquad$
    \raisebox{-0.5\height}{\includegraphics[height=2cm]{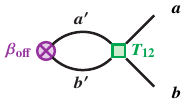}}
    \caption{Diagrams contributing to the final production of the  particles $a$ and $b$, where the coupled-channels dynamics $ab\to ab$, $ab\to a'b'$ and $a'b'\to a'b'$ is relevant.}
    \label{fig:spectrum2}
\end{figure*}
We now consider that the pair of particles $(a,b)$ strongly interact with another two particles $(a',b')$, and that the coupled-channels dynamics $ab\to ab$, $ab\to a'b'$ and $a'b'\to a'b'$ is relevant for the energy region under study. For simplicity, we will continue assuming $S$-wave scattering and that all particles are spinless, or that the spin degrees of freedom can be trivially treated. We will label the  $(a,b)$ and $(a',b')$ channels as 1 and 2, respectively, and take $(m_a,m_b)$ and $(m^\prime_a,m^\prime_b)$ for the masses with $(m^\prime_a+m^\prime_b) \geqslant (m_a+m_b)$. From the perspective of the production spectrum sketched  in Fig.~\ref{fig:spectrum2}, one would write  for the production CF of the pair $(a,b)$:
\begin{eqnarray}
 \widetilde C_1^\text{\,prod}(s) & =&   \left|1+\int \frac{d^3 \vec{p}}{(2\pi)^3}\frac{\omega_a(p)+\omega_b(p)}{2\omega_a(p)\omega_b(p)}\frac{T_{11}(k_1\leftarrow p;s)\widetilde F_R(k_1,p)}{s-\left[\omega_a(p)+\omega_b(p)\right]^2+i\epsilon} \right.\nonumber \\
 &+&\left.\sqrt{\eta_2(s)}\int \frac{d^3 \vec{p}}{(2\pi)^3}\frac{\omega^{\prime}_a(p)+\omega^{\prime}_b(p)}{2\omega^{\prime}_a(p)\omega^{\prime}_b(p)}\frac{T_{12}(k_1\leftarrow p;s)\widetilde F_R(k_1,p)}{s-\left[\omega_a^{\prime}(p)+\omega_b^{\prime}(p)\right]^2+i\epsilon}
 \right|^2 \label{eq:C1bad}
\end{eqnarray}
where $\omega^{(\prime)}_{a,b}(p)= \sqrt{m^{(\prime) 2}_{a,b}+p^2}$, and $k_1=\lambda^\frac12(s,m_a^2\ , m_b^2)/(2\sqrt{s})$  is the on-shell momentum of the pair $(a,b)$ and $\widetilde{F}_R(k_1,p)= F_R(k_1,p)/F_R(k_1,k_1)$. As compared to Fig.~\ref{fig:spectrum},  a new production mechanism is now included in Fig.~\ref{fig:spectrum2} and hence in Eq.~\eqref{eq:C1bad}. In the last Feynman diagram of this latter figure, the virtual pair $(a',b')$ is produced in the first step and it propagates and re-scatters through $T_{12}(k_1\leftarrow p;s)$ (depicted as a green square in the plot) leading  to the final on-shell particles $(a,b)$. The vertices $\alpha_\text{off}$ and  $\beta_\text{off}$ in the last two diagrams of Fig.~\ref{fig:spectrum2} are not equal owing to the different  amplitudes for producing $(a,b)$  or $(a',b')$ virtual pairs. Actually, we are assuming that $\beta_\text{off}/\alpha_\text{off} = \sqrt{\eta_2(s)}$, which would be consistent with 
\begin{equation}
    \frac{\beta_\text{on}}{\alpha_\text{on}} = \sqrt{\eta_2(s)}\, \frac{F_R(k_1,k_2)}{F_R(k_1,k_1)}
\end{equation}    
with $k_2=\lambda^\frac12(s,m^{\prime\, 2}_a, m^{\prime\,2}_b)/(2\sqrt{s})$ the on-shell momentum of the $(a',b')$ particles. The above equation would define $\eta_2(s)$. In addition, we note that the effect of the form-factor $F_R(k_1,p)$ in the two terms of Eq.~\eqref{eq:C1bad} is not the same because of the different strength that different values of $p$ could take in the $d^3 \vec{p}$ integration which accounts for the  virtual $(a,b)$ or $(a',b')$ loops.  

At this point, some remarks are in order:
\begin{itemize}[nosep,itemsep=0pt,leftmargin=0pt,itemindent=10pt,parsep=5pt]
    \item $T_{11}(k_1\leftarrow p;s)$ (red square in Fig.~\ref{fig:spectrum2}) is  the $S$-wave scattering amplitude for the transition $ab\to ab$, where initially the $(a,b) $ particles are produced in a virtual state with c.m. off-shell momentum $p$, and in the final state they are on the mass shell. $T_{11}$ contains coupled-channels effects since it involves contributions from processes of the type $ab\to a'b'\to ab$, where the pair $(a',b')$ is created in an intermediate state. Actually, the LSE (or BSE) which needs to be solved to obtain the $T$--operator is a $2\times 2$ matrix equation, in the channel space of this particular example.
    
    \item We see in Fig.~\ref{fig:spectrum2}, and therefore in Eq.~\eqref{eq:C1bad}, that the $T_{12}$ contribution of the third Feynman diagram is coherently summed to those associated to the first two ones. This is obviously only correct when both the full initial and final states are the same. Let us suppose again a $pp$ collider experiment like LHC, then the production vertices, $\alpha_\text{on}$ and $\alpha_\text{off}$ (blue) and $\beta_\text{off}$ (magenta) crossed circles in Fig.~\ref{fig:spectrum2}, stand for the $pp\to X+ ab\, (\text{real}) $, $pp\to X+ ab\, (\text{virtual}) $  and $pp\to X+ a^\prime b^\prime\, (\text{virtual})$, mechanisms  respectively. As long as the rest of the particles ($X$) in the final state are the same in all mechanisms, the three Feynman amplitudes of Fig.~\ref{fig:spectrum2} must be coherently added since they all contribute to the same quantum amplitude of the reaction $pp\to X + ab\, ( \text{real})$. Otherwise, even if the production amplitudes of the first two diagrams were still assumed to continue to be added coherently (see the discussion above in Subsect.~\ref{sec:cohe}), the square modulus of the contribution of the third diagram will have to be added to the square modulus of the sum of the amplitudes of the first two mechanisms. Such incoherent sum will give rise to a different CF of the pair $(a,b)$, namely 
\begin{eqnarray}
 C_1^\text{prod}(s) &=&    \left|1+\int \frac{d^3 \vec{p}}{(2\pi)^3}\frac{\omega_a(p)+\omega_b(p)}{2\omega_a(p)\omega_b(p)}\frac{T_{11}(k_1\leftarrow p;s)\widetilde F_R(k_1,p)}{s-\left[\omega_a(p)+\omega_b(p)\right]^2+i\epsilon}\right|^2 \nonumber \\ 
 &+& \eta_2(s)\left|\int \frac{d^3 \vec{p}}{(2\pi)^3}\frac{\omega^{\prime}_a(p) + \omega^{\prime}_b(p)}{2\omega^{\prime}_a(p)\omega^{\prime}_b(p)}\frac{T_{12}(k_1\leftarrow p;s)\widetilde F_R(k_1,p)}{s-\left[\omega_a^{\prime}(p)+\omega_b^{\prime}(p)\right]^2+i\epsilon}
 \right|^2\,. \label{eq:C1}
\end{eqnarray}
In high multiplicity collisions, this incoherent sum seems more indicated because of the large number of particles produced in each event,  which makes  reasonable to assume that the bulk of the CF comes from situations where the {\it spectator} particles ($X$) are not the same for $(a,b)$ or $(a',b')$ production. One might also argue that the interference terms, neglected in Eq.~\eqref{eq:C1}, largely cancel out when one performs the sum over all events, and for each event over all possible final configurations $X$,  to obtain the CF.  Actually, the approach of Ref.~\cite{Haidenbauer:2018jvl} is compatible with the incoherent sum assumed in Eq.~\eqref{eq:C1} (see also Refs.~\cite{Vidana:2023olz,Liu:2023uly}). In the Koonin--Pratt formalism the contribution from the second channel is added incoherently to the one of the first (observed), but with a certain weight $\eta_2$ relative to the weight $\eta_1$ = 1 for the observed channel. This overall multiplicative weight $\eta_2$ appears explicitly in the last term of Eq.~\eqref{eq:C1}, and hence one  recovers the results of Ref.~\cite{Haidenbauer:2018jvl}, but modified  because in Eq.~\eqref{eq:C1}, the $\delta C(k)$-type term discussed in the previous subsection for the single channel analysis is neglected.

\item Let us analyze now the coupled-channels CF from the wave function perspective. The ket $\ket*{\Psi(t=0)}$ will have now two components which correspond to channels 1 $(a,b)$ and 2 $(a',b')$, respectively
\begin{equation}
\begin{pmatrix}
  \Psi_1(t=0)    \\ 
   \Psi_2(t=0)   
\end{pmatrix}    
= \begin{pmatrix}
  \Psi_{\rm out;\, 1}    \\ 
   \Psi_{\rm out;\, 2}   
\end{pmatrix}  + G_0^{\text{QM}}(E-i\epsilon)\cdot T^{\text{QM}}(E-i\epsilon) \begin{pmatrix}
  \Psi_{\rm out;\, 1}    \\ 
   \Psi_{\rm out;\, 2}   
\end{pmatrix}
\end{equation}
where the resolvent of the kinetic operator $G_0^{\text{QM}}$ and  $T^{\text{QM}}$ are matrices in the coupled channel space
\begin{eqnarray}
G_0^{\text{QM}}(E-i\epsilon)&=&\begin{pmatrix}
    (E+\frac{\vec{\nabla}^{\, 2}}{2\mu_{ab}}-i\epsilon)^{-1} & 0 \\
    0 & (E+\delta m+\frac{\vec{\nabla}^{\, 2}}{2\mu^\prime_{a b}}-i\epsilon)^{-1}
\end{pmatrix}  \\
T^{\text{QM}}(E-i\epsilon) &=&
\begin{pmatrix}
T^{\text{QM}}_{11}(E-i\epsilon) & T^{\text{QM}}_{12}(E-i\epsilon) \\
T^{\text{QM}}_{21}(E-i\epsilon)& T^{\text{QM}}_{22}(E-i\epsilon)
\end{pmatrix}  
\end{eqnarray}
with $\delta m=(m_a+m_b-m_a^\prime-m_b^\prime)\leqslant 0$ and $\mu^\prime_{ab}=m_a^\prime m_b^\prime/(m_a^\prime+m_b^\prime)$. If the pair $(a,b)$ is measured in the detector, and assuming an infinite momentum resolution, we will have 
\begin{equation}
  \begin{pmatrix}
  \Psi_{\rm out;\, 1}(\vec r\,)    \\ 
   \Psi_{\rm out;\, 2} (\vec r\,) 
\end{pmatrix}  
= 
\begin{pmatrix}
    e^{i\vec k_1\cdot\vec r}/(2\pi)^\frac32 \\
    0
\end{pmatrix}\,,
\end{equation}
 which, following the steps of the derivation of  Eq.~\eqref{eq:wf1} and using again $[T^{\text{QM}}(z^*)]^\dagger=T^{\text{QM}}(z)$,  leads to
\begin{eqnarray}
\psi_1^\ast(\vec r; \vec k_1)&=&  e^{-i\vec k_1\cdot\vec r} +\int d^3 \vec{p}\,\frac{e^{-i\vec p\,\cdot\vec r}}{E-\frac{\vec{p}^{\, 2}}{2\mu_{ab}}+i\epsilon}
\mel*{ \vec k_1 }{T_{11}^{\text{QM}}(E+i\epsilon)}{\vec p} \nonumber \\
\psi_2^\ast(\vec r; \vec k_1)&=&  \int d^3 \vec{p}\,\frac{e^{-i\vec p\,\cdot\vec r}}{E+\delta m-\frac{\vec{p}^{\, 2}}{2\mu_{a^\prime b^\prime}}+i\epsilon}
\mel*{ \vec k_1 }{T_{12}^{\text{QM}}(E+i\epsilon)}{\vec p} \label{eq:wf1-couple}
\end{eqnarray}
Finally, since
\begin{equation}
  |\psi(\vec r; \vec k_1)|^2 =   |\psi_1(\vec r; \vec k_1)|^2 + |\psi_2(\vec r; \vec k_1)|^2\,, \label{eq:twowf}
\end{equation}
we see that the Koonin--Pratt  gives further support to the incoherent sum implicit in the evaluation of $C^\text{prod}_1(s)$ using Eq.~\eqref{eq:C1} as compared to $\widetilde C^\text{\,prod}_1(s)$ obtained from the coherent sum in Eq.~\eqref{eq:C1bad}. Actually, we can interpret the wave-function components $\psi_1$ and $\psi_2$ as different sources of the production vertices $\alpha_\text{on}$, $\alpha_\text{off}$ and $\beta_\text{off}$ in Fig.~\ref{fig:spectrum2}, and hence their contributions should be incoherently summed up according to Eq.~\eqref{eq:twowf}.

In the case of detecting the particles of channel 2 $(a^\prime, b^\prime)$, the contour conditions now would  be 
\begin{equation}
  \begin{pmatrix}
  \Psi_{\rm out;\, 1}(\vec r\,)    \\ 
   \Psi_{\rm out;\, 2} (\vec r\,) 
\end{pmatrix}  
= 
\begin{pmatrix}
0\\
    e^{i\vec k_2\cdot\vec r}/(2\pi)^\frac32 
\end{pmatrix}\,,
\end{equation}
with the non-relativistic c.m. on-shell momentum $k_2= \sqrt{2\mu^\prime_{ab}\left(E+\delta m\right)}$
and 
\begin{align}
\psi_1^\ast(\vec r; \vec k_2) & = \int d^3 \vec{p}\,\frac{e^{-i\vec p\,\cdot\vec r}}{E-\frac{\vec{p}^{\, 2}}{2\mu_{ab}}+i\epsilon} \mel*{ \vec k_2 }{ T_{21}^{\text{QM}}(E+i\epsilon) }{ \vec p }\,, \nonumber \\
\psi_2^\ast(\vec r; \vec k_2) & =   e^{-i\vec k_2\cdot\vec r} +\int d^3 \vec{p}\,\frac{e^{-i\vec p\,\cdot\vec r}}{E+\delta m-\frac{\vec{p}^{\, 2}}{2\mu_{a^\prime b^\prime}}+i\epsilon}
\mel*{ \vec k_2 }{ T_{22}^{\text{QM}}(E+i\epsilon) }{ \vec p }\,,
\label{eq:wf1-couple-2}
\end{align}
which would be consistent with the evaluation of the CF of the pair $(a^\prime,b^\prime)$, considering only $S$-wave interactions and incorporating relativistic corrections, as 
\begin{eqnarray}
 C^\text{prod}_2(s) &=&    \left|1+\int \frac{d^3 \vec{p}}{(2\pi)^3}\frac{\omega^{\prime}_a(p)+\omega^{\prime}_b(p)}{2\omega^{\prime}_a(p)\omega^{\prime}_b(p)}\frac{T_{22}(k_2\leftarrow p;s)\widetilde F_R(k_2,p)}{s-\left[\omega_a^{\prime}(p)+\omega_b^{\prime}(p)\right]^2+i\epsilon}\right|^2 \nonumber \\ 
 &+& \eta_1(s) \left|\int \frac{d^3 \vec{p}}{(2\pi)^3}
 \frac{\omega_a(p)+\omega_b(p)}{2\omega_a(p)\omega_b(p)}
 \frac{T_{21}(k_2\leftarrow p;s)\widetilde F_R(k_2,p)}{s-\left[\omega_a(p)+\omega_b(p)\right]^2+i\epsilon}
 \right|^2\,, \label{eq:C2}
\end{eqnarray}
with the weight $\eta_1(s)= \left(\alpha_\text{off}/\beta_\text{off}\right)^2$. We remind  here that $\widetilde F_R(k_2,p)= F_R(k_2,p)/F_R(k_2,k_2)$ and that $k_2(s)$ is the c.m. relativistic on-shell momentum of the particles $(a^\prime, b^\prime)$. 
\end{itemize}
The standard Koonin--Pratt formulae in coupled channels read~\cite{Fabbietti:2020bfg,Vidana:2023olz} 
\begin{eqnarray}
     C^\text{KP}_1 (s) &=& 1 + 4\pi \int dr r^2 S(r) \Bigg\{\, \left|\ j_0(k_1r)+ \int \frac{d^3 \vec{p}}{(2\pi)^3}\frac{\omega_a(p)+\omega_b(p)}{2\omega_a(p)\omega_b(p)} \frac{j_0(pr)\, T_{11}(k_1\leftarrow p;s)}{s-\left[\omega_a(p)+\omega_b(p)\right]^2+i\epsilon}\right|^2 \nonumber \\
     &+&\eta_2 (s) \left|\int \frac{d^3 \vec{p}}{(2\pi)^3}
 \frac{\omega_a^{\prime}(p)+\omega_b^{\prime}(p)}{2\omega_a^{\prime}(p)\omega_b^{\prime}(p)}
 \frac{j_0(pr)T_{12}(k_1\leftarrow p;s)}{s-\left[\omega_a^{\prime}(p)+\omega_b^{\prime}(p)\right]^2+i\epsilon}
 \right|^2 - j^2_0(k_1r)\Bigg\} \\
 C^\text{KP}_2 (s) &=& 1 + 4\pi \int dr r^2 S(r) \Bigg\{ \, \left|\ j_0(k_2r)+ \int \frac{d^3 \vec{p}}{(2\pi)^3}\frac{\omega_a^\prime(p)+\omega_b^\prime(p)}{2\omega_a^\prime(p)\omega_b^\prime(p)} \frac{j_0(pr)\, T_{22}(k_2\leftarrow p;s)}{s-\left[\omega_a^\prime(p)+\omega_b^\prime(p)\right]^2+i\epsilon}\right|^2 \nonumber \\
     &+&\eta_1 (s) \left|\int \frac{d^3 \vec{p}}{(2\pi)^3}
 \frac{\omega_a(p)+\omega_b(p)}{2\omega_a(p)\omega_b(p)}
 \frac{j_0(pr)T_{21}(k_2\leftarrow p;s)}{s-\left[\omega_a(p)+\omega_b(p)\right]^2+i\epsilon}
 \right|^2 - j^2_0(k_2r)\Bigg\}   \label{eq:correccion-KN-coupled-bis}
\end{eqnarray}
Therefore within the relativistic Koonin--Pratt formalism, each of the $C^\text{prod}_{i=1,2}(s)$ CFs calculated using Eqs.~\eqref{eq:C1} and \eqref{eq:C2} will receive an extra contribution $ \delta C^\text{KP}_{i=1,2}(s)$ given by
\begin{align}
     C^\text{KP}_{i=1,2}(s) & = C^\text{prod}_{i=1,2}(s) +\delta C^\text{KP}_{i=1,2}(s) \label{eq:correccion-KN-coupled}\\ 
     \delta C^\text{KP}_1(s) & = 2\text{Re}\Bigg(\int \frac{d^3 \vec{p}}{(2\pi)^3}\frac{\rho_{ab}(p)T_{11}(k_1\leftarrow p;s)}{s-(\omega_a(p)+\omega_b(p))^2+i\epsilon}\left[F_R(k_1,p)-\widetilde F_R(k_1,p)\right]\Bigg) \nonumber \\ 
     & + \int \frac{d^3 \vec{p}}{(2\pi)^3}\int\frac{d^3 \vec{p}^{\prime}}{(2\pi)^3}\, h_R(p,p';k_1)\Bigg\{ 
     \frac{\rho_{ab}(p)\rho_{ab}(p'\,)T_{11}(k_1\leftarrow p;s)[T_{11}(k_1\leftarrow p';s)]^\ast }{\left(s-(\omega_a(p)+\omega_b(p))^2+i\epsilon\right)\left(s-(\omega_a(p\,')+\omega_b(p\,'))^2-i\epsilon\right)} \nonumber\\
     & +\  \eta_2(s) \
   \frac{\rho^\prime_{ab}(p)\rho^\prime_{ab}(p'\,) T_{12}(k_1\leftarrow p;s)[T_{12}(k_1\leftarrow p';s)]^\ast }{\left(s-(\omega^\prime_a(p)+\omega^\prime_b(p))^2+i\epsilon\right)\left(s-(\omega^\prime_a(p\,')+\omega^\prime_b(p\,'))^2-i\epsilon\right)}\Bigg\} \nonumber \\
  \delta C^\text{KP}_2(s) & = 2\text{Re}\Bigg(\int \frac{d^3 \vec{p}}{(2\pi)^3}\frac{\rho^{\prime}_{ab}(p)T_{22}(k_2\leftarrow p;s)}{s-(\omega^\prime_a(p)+\omega^\prime_b(p))^2+i\epsilon}\left[F_R(k_2,p)-\widetilde F_R(k_2,p)\right]\Bigg) \nonumber \\ 
     & +  \int \frac{d^3 \vec{p}}{(2\pi)^3}\int\frac{d^3 \vec{p}^{\prime}}{(2\pi)^3}\,h_R(p,p';k_2)  \Bigg\{ 
     \frac{\rho^\prime_{ab}(p)\rho^\prime_{ab}(p'\,) T_{22}(k_2\leftarrow p;s)[T_{22}(k_2\leftarrow p';s)]^\ast }{\left(s-(\omega^\prime_a(p)+\omega^\prime_b(p))^2+i\epsilon\right)\left(s-(\omega^\prime_a(p\,')+\omega^\prime_b(p\,'))^2-i\epsilon\right)} \nonumber\\
     &+\ \eta_1(s) \
   \frac{ \rho_{ab}(p)\rho_{ab}(p'\,) T_{21}(k_2\leftarrow p;s)[T_{21}(k_2\leftarrow p';s)]^\ast }{\left(s-(\omega_a(p)+\omega_b(p))^2+i\epsilon\right)\left(s-(\omega_a(p\,')+\omega_b(p\,'))^2-i\epsilon\right)}\Bigg\} \label{eq:correccion-KN-coupled-corr} 
\end{align} 
where we have defined $\rho^{(\prime)}_{ab}(p)=[\omega^{(\prime)}_a(p)+\omega^{(\prime)}_b(p)]/[2\omega^{(\prime)}_a(p)\omega^{(\prime)}_b(p)]$ and $h_R(p,p';k) = F_R(p,p')-\widetilde F_R(k,p)\widetilde F_R(k,p')$.

As an illustration of the alternative approach to the theoretical determination of the two-particle CF, in the next section we will study $D^0 D^{\ast +}$ and $D^+ D^{\ast 0}$ CFs and discuss the effect produced by neglecting the extra $ \delta C^\text{KP}_{i=1,2}$ terms in Eq.~\eqref{eq:correccion-KN-coupled}, which will amount to use $C^\text{prod}_{i=1,2}(k)$ in accordance to the evaluation of the production Feynman diagrams of Fig.~\ref{fig:spectrum2} in QFT.

\section{\boldmath $D^0 D^{\ast +}$ and $D^+ D^{\ast 0}$ CFs}
\label{sec:res}

\begin{figure*}[t]\centering
\begin{tabular}{cc}
\includegraphics[height=5.75cm,keepaspectratio]{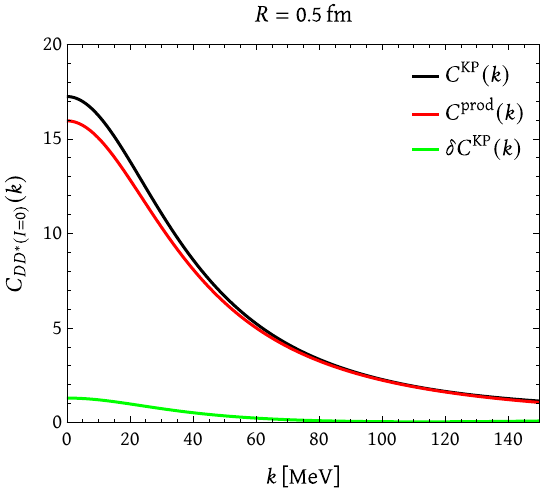} &
\includegraphics[height=5.75cm,keepaspectratio]{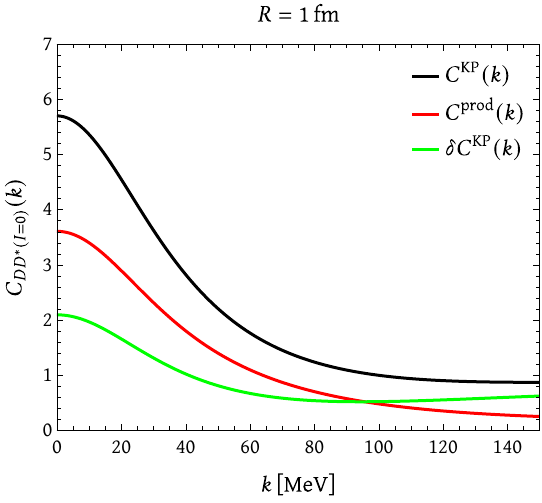} \\
\includegraphics[height=5.75cm,keepaspectratio]{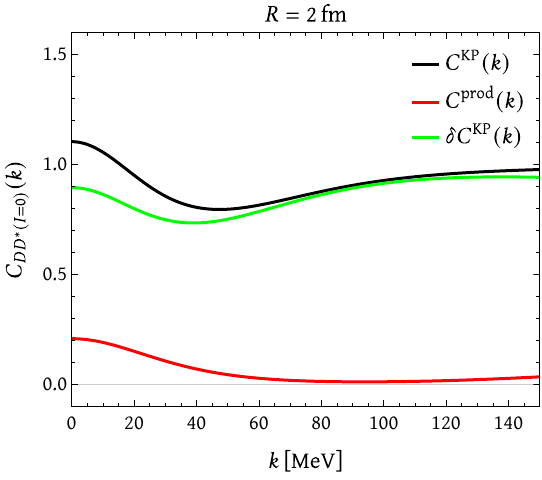} &
\includegraphics[height=5.75cm,keepaspectratio]{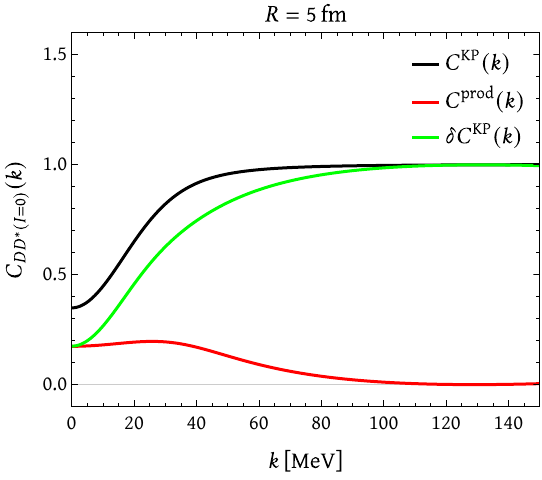}
\end{tabular}
    \caption{Predictions for the isoscalar (single channel) CF obtained from the alternative production model  [$C_{DD^{\ast}}^\text{prod}$] of  Eq.~\eqref{eq:Cimprove} (red curves)  and the relativistic Koonin--Pratt formula [$C_{DD^{*}}^\text{KP}$] of Eq.~\eqref{eq:ckprel2} (black curves) as a function of the c.m. momentum $k$. We also show the  
difference [$\delta C_{DD^{*}}^\text{KP}$] between both types of CFs (green curves)  calculated  using Eq.~\eqref{eq:ckprel-del}. For the $DD^{*}$ half-off-shell $T-$matrix, we consider a model inspired in Ref.~\cite{Feijoo:2021ppq} ($T(k\leftarrow p;E)= \theta(q_{\rm max}-p) /(V^{-1}-G(E;q_{\rm max}))$, with  $V=-437$, $q_{\rm max}=413\,\MeV$, $G(E;q_{\rm max})$ given in Eq.~(13) of that work and using isospin averaged meson masses, which gives rise to a $T_{cc}$ bound by $860\,\keV$).  In the figure we show results for four different Gaussian source-sizes.}
    \label{fig:cf-single}
\end{figure*}

Here we consider the $D^0 D^{\ast +}$ and $D^+ D^{\ast 0}$ CFs, which are of interest to unravel the dynamics of the exotic $T_ {cc}(3875)^+$. We illustrate in this system the existing differences between the Koonin--Pratt scheme and the exploratory one introduced in this work, based on the direct evaluation of the production Feynman diagrams  depicted in Fig.~\ref{fig:spectrum2}. The $T_ {cc}(3875)^+$ is a  narrow state observed in the $D^0D^0 \pi^+$ mass distribution, with a mass of $m_{\text{thr}} + \delta m_{\text{exp}}$, being $m_{\text{thr}} = 3875.09\,\MeV$ the $D^{*+} D^0$ threshold and $\delta m_{\text{exp}} = -360 \pm 40^{+4}_{-0}\,\keV$, and a width $\Gamma = 48 \pm 2^{+0}_{-14}\,\keV$ \cite{LHCb:2021auc}. The interest on the properties and nature of this tetraquark-like state is growing by the day within the hadronic community, since it obviously cannot be accommodated  within constituent quark models as a $q\bar q$ state. Among the possible interpretations,  the molecular picture \cite{Feijoo:2021ppq,Albaladejo:2021vln,Du:2021zzh,Baru:2021ldu} is quite natural based on its closeness to the $D^0D^{*+}$ and $D^+D^{*0}$ thresholds, whereas the tetraquark interpretation has been put forward \cite{Ballot:1983iv,Zouzou:1986qh}, even before its discovery. In any case, the proximity of the state to the $D^0D^{*+}$ and $D^+D^{*0}$ thresholds makes it necessary to consider the hadronic degrees of freedom for the analysis of the experimental data \cite{Dong:2021rpi}. Actually, the study carried out in Ref.~\cite{Dai:2023kwv} strongly supports the hadron composite picture of $T_ {cc}(3875)^+$.\footnote{It is argued in Ref.~\cite{Dai:2023kwv} that even starting with a genuine state of non-molecular nature, but which couples to the $D^0 D^{\ast +}$ and $D^+ D^{\ast 0}$ meson components, and if that state is responsible for a bound state appearing below the threshold, it gets dressed with a meson cloud and it becomes purely molecular in the limit case of zero binding. If one forces the non-molecular state to have the small experimental binding, the system develops $DD^*$ scattering lengths and effective range parameters in complete disagreement with present data.} Recently, the femtoscopic CFs for the $D^0D^{*+}$ and $D^+D^{*0}$ channels in heavy-ion collisions  have become of major interest, as a future source of independent experimental input. Work on that direction has been  performed for the $T_{cc}(3875)^+$ state in Refs.~\cite{Kamiya:2022thy,Vidana:2023olz,Albaladejo:2023wmv}.

For simplicity, we first neglect coupled-channel effects both on the $T$--matrix and on the CF and consider an isoscalar ($I=0$) $DD^*$ pair. In Fig.~\ref{fig:cf-single}, we compare the predictions for the  $C_{DD^{*}}$ obtained from the  production model  [$C^\text{prod}(k)$]  proposed in this work in Eq.~\eqref{eq:Cimprove} (red curves)  and the Koonin--Pratt formula [$C^\text{KP}(k)$] of Eq.~\eqref{eq:ckprel2} (black curves). We also show the difference between both of them (green curves) [$\delta C^\text{KP}(k)$] calculated  using Eq.~\eqref{eq:ckprel-del}. For the $DD^{*}$ half-off-shell $T$--matrix, we consider the model of Ref.~\cite{Feijoo:2021ppq},\footnote{In that model, the interaction between the $D^0D^{*+}$ and $D^+D^{*0}$ pairs was obtained from the exchange of  vector mesons,  within the extension of the local hidden gauge symmetry approach to the charmed sector, and using a sharp ultraviolet cutoff.  It was found that the isospin $I = 0$ combination had an attractive interaction which would bind the system, while the $I = 1$ one had a repulsive interaction. This is the reason why an isoscalar $DD^*$ pair is considered in Fig.~\ref{fig:cf-single} with charged averaged $D$ and $D^*$ masses. } which was also used in our previous work of Ref.~\cite{Vidana:2023olz}. In the figure we show results for four different Gaussian source-sizes. The differences caused by the  $\delta C^\text{KP}$ term become huge for the extended sources of $R=2$ and $5\,\fm$, but  also  for the $R=1\,\fm$ one, where we find  sizable effects in the whole range $[0,150]\,\MeV$ of c.m. momenta depicted in Fig.~\ref{fig:cf-single}.

\begin{figure*}[t]\centering
\begin{tabular}{cc}
\includegraphics[height=5.5cm,keepaspectratio]{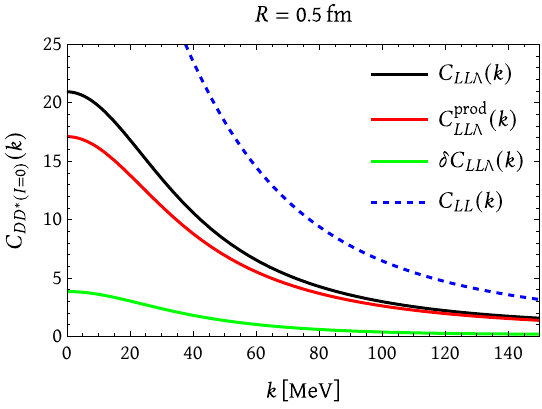} &
\includegraphics[height=5.5cm,keepaspectratio]{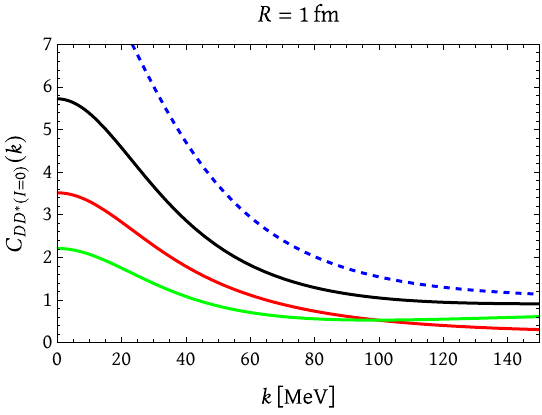} \\
\includegraphics[height=5.5cm,keepaspectratio]{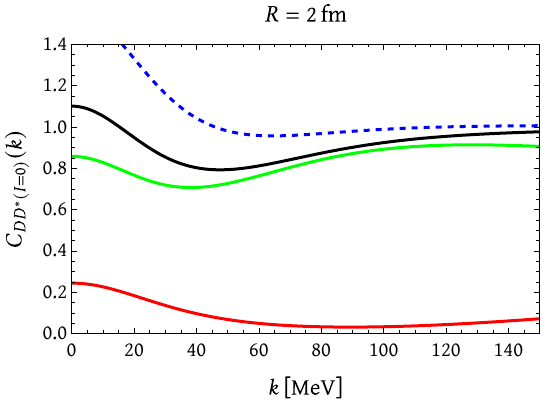} &
\includegraphics[height=5.5cm,keepaspectratio]{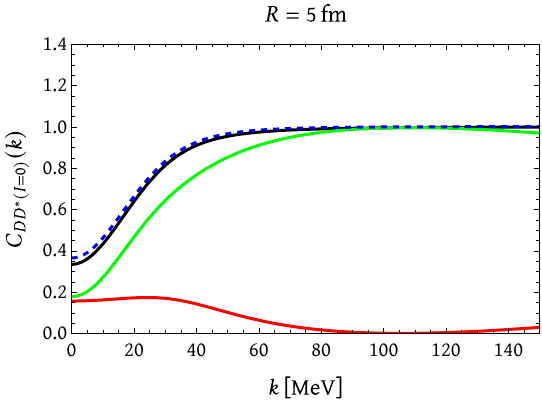}
\end{tabular}
    \caption{ Lednicky-Lyuboshits $C_{LL}$ (dashed-blue), short-distance improved $C_{LL\Lambda}$ (black) and production $C_{LL\Lambda}^\text{prod}$ (red)  CFs for $R=0.5,1,2$ and 5 fm sources as a function of the c.m. momentum $k$. The green curves shows the difference $\delta C_{LL\Lambda}$ between $C_{LL\Lambda}$ and $C_{LL\Lambda}^\text{prod}$. We have used  Eqs.~\eqref{eq:LL-KP}, \eqref{eq:LL-LKP} and \eqref{eq:LL-Lprod} and taken $a_0=5.37$ fm and $r_0=0.95$ fm from the $T_{cc}(3875)^+$ analysis carried out in Ref.~\cite{Albaladejo:2021vln} in the isospin limit for an isoscalar $DD^*$ pair and a stable $D^*$ meson.  These effective range parameters fix the $S$-wave scattering amplitude to $f_0(k)\approx (-1/a_0+r_0k^2/2-ik)^{-1}$ and the Lorentz ultraviolet cutoff $\Lambda=$570 MeV, through Eq.~\eqref{eq:MQalcance-r0}, and  also lead to a $DD^\ast$ binding energy of 860 keV.  }
    \label{fig:compTmodel}
\end{figure*}

The LL approximation is not adequate for small source-sizes and it even leads to divergent CFs for point-like sources ($R=0$), as discussed in Appendix \ref{sec:LL}. There, we also derive a short-distance improved LL-type approach ($LL\Lambda$), which implements  a Lorentz ultraviolet regulator that corrects the  pathological behaviour of the LL  CFs in the $R\to 0$ limit. We show in Fig.~\ref{fig:compTmodel} results from the $LL\Lambda$--model of Eqs.~\eqref{eq:LL-LKP}-\eqref{eq:LL-dLKP}, with an ultraviolet $\Lambda$ fixed from the scattering length and effective range through Eq.~\eqref{eq:MQalcance-r0}, for the four different sources considered in Fig.~\ref{fig:cf-single}. As we can see,  the short-distance improved LL approach is quite realistic and qualitatively describes  the general characteristics  of the production [$C_{DD^{\ast}}^\text{prod}$] and Koonin--Pratt [$ C_{DD^{*}}^\text{KP} $] CFs in Fig.~\ref{fig:cf-single}. On the other hand, for the largest sources ($R=2$ and $R=5\,\fm$), we see that $C_{LL\Lambda}^\text{prod}(k)$ for $k\sim 150\,\MeV$ has not yet reached  its asymptotic value of $1$, which however has been already reached by the short-distances improved Koonin--Pratt LL CF $[C_{LL\Lambda}]$. We observed the same behaviour in Fig.~\ref{fig:cf-single}, where the CFs were computed using the realistic model of Ref.~\cite{Feijoo:2021ppq}.  Indeed, the production CF in Fig.~\ref{fig:compTmodel}, for $R=2$ and $R=5$ fm, takes small (positive) values close to zero. For sufficiently larger c.m. momenta, $C_{LL\Lambda}^\text{prod}(k)$  should increase and  approach asymptotically to one. In Fig.~\ref{fig:compTmodel} we do not consider higher momenta  because there it is used the non-relativistic Eq.~\eqref{eq:LL-Lprod}, and moreover the scattering  amplitude used in the figure is constructed only out of the scattering length and effective range.

\begin{figure*}[t]\centering
\includegraphics[height=6.75cm,keepaspectratio]{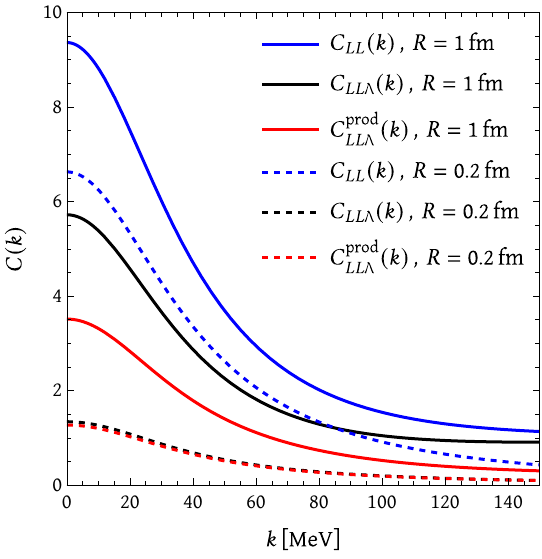}\hspace{1cm}%
\includegraphics[height=7cm,keepaspectratio]{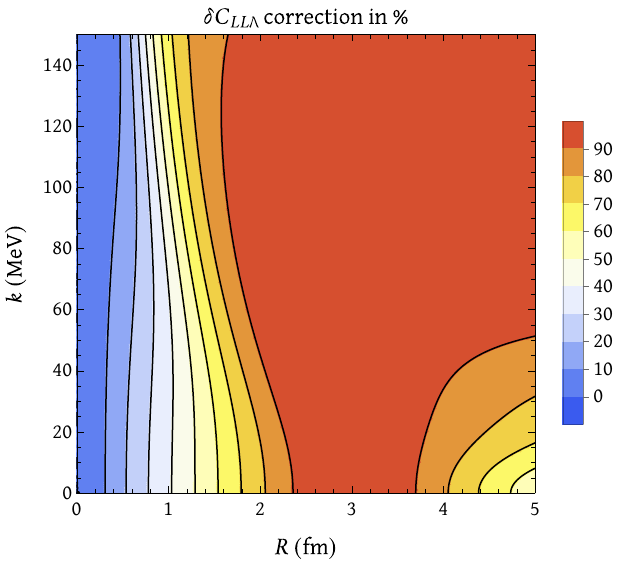}%
    \caption{Left panel: Lednicky-Lyuboshits $C_{LL}$ (blue) and short-distance improved $C_{LL\Lambda}$ (black) and production $C_{LL\Lambda}^\text{prod}$ (red)  CFs for $R=1$ fm (solid lines) and $R=0.2$ fm (dot-dashed lines) sources as a function of the c.m. momentum $k$. Results for the source size $R=0.2$ fm have been divided by a factor of fifty.  Right panel: ($R,k$) two dimensional $100\times ( C_{LL\Lambda}- C_{LL\Lambda}^\text{prod})/C_{LL\Lambda}$ distribution as function of the source size $R$ and the c.m. momentum $k$. As in Fig.~\ref{fig:compTmodel} in the top panels, we have taken $a_0=5.37$ fm and $r_0=0.95$ fm from the $T_{cc}(3875)^+$ analysis carried out in Ref.~\cite{Albaladejo:2021vln}  in the isospin limit and Lorentz ultraviolet cutoff is fixed to $\Lambda=$570 MeV.}
    \label{fig:deltaCLL2D}
\end{figure*}

In the left panel of Fig.~\ref{fig:deltaCLL2D}  we compare  $LL\Lambda-$model results for $R=0.2$ fm and $R=1$ fm sources. There, we also show  results from the common LL approximation $C_{LL}$ (blue curves), which notably deviate from those  obtained using the short-distance improved one  $C_{LL\Lambda}$ (black curves) for both sources. We also note  the large variation between the CFs computed for $R=0.2$ fm or $R=1$ fm. Moreover, we see that for the former case,  the $LL\Lambda-$ production  and Koonin--Pratt CFs are almost indistinguishable.  The ($R,k$)-two dimensional $100\times ( C_{LL\Lambda}- C_{LL\Lambda}^\text{prod})/C_{LL\Lambda}$ distribution as function of the source size $R$ and the c.m. momentum $k$ is shown in the right panel of Fig.~\ref{fig:deltaCLL2D}. These results can be used to infer how the difference of  $(C_{LL}- C_{LL}^\text{prod})$ varies as a function of  the source size  and the c.m. momentum, assuming that  the improved LL approach is realistic. For extensive  sources ($R>1$ fm), the differences between  production  and Koonin--Pratt CFs become quite large, with relative changes ranging from 40\% (low momenta) to 70\% ($k\sim 140$ MeV) already for $R=1$ fm. 

\begin{figure*}[t]\centering
\begin{tabular}{cc}
\includegraphics[height=6.5cm,keepaspectratio]{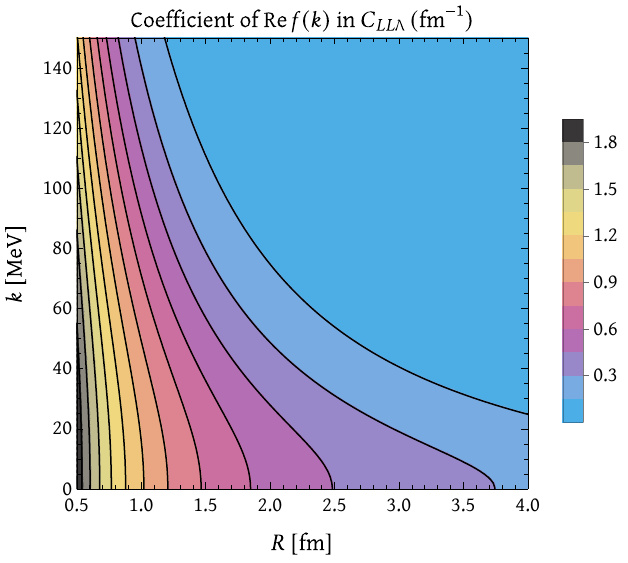} &
\includegraphics[height=6.5cm,keepaspectratio]{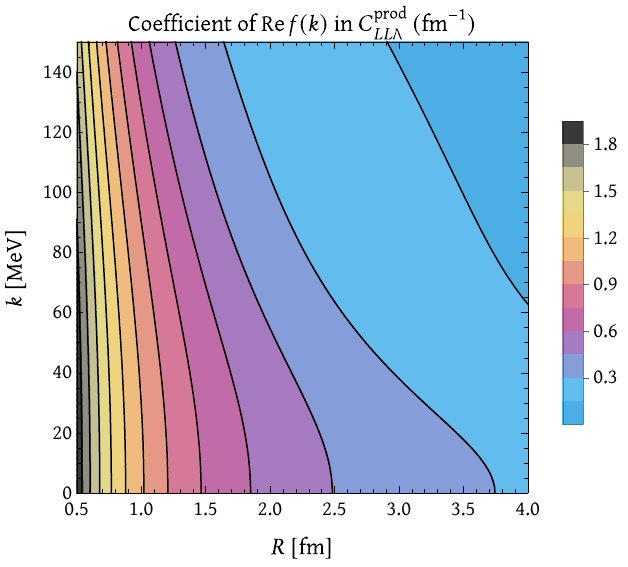}\\
\includegraphics[height=6.5cm,keepaspectratio]{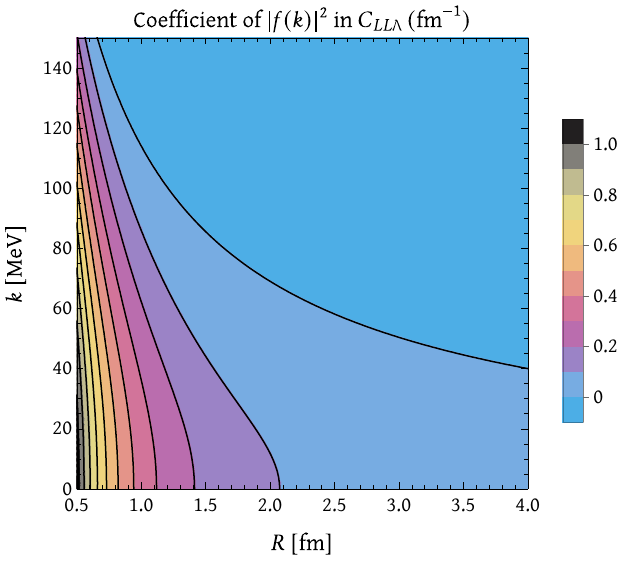} &
\includegraphics[height=6.5cm,keepaspectratio]{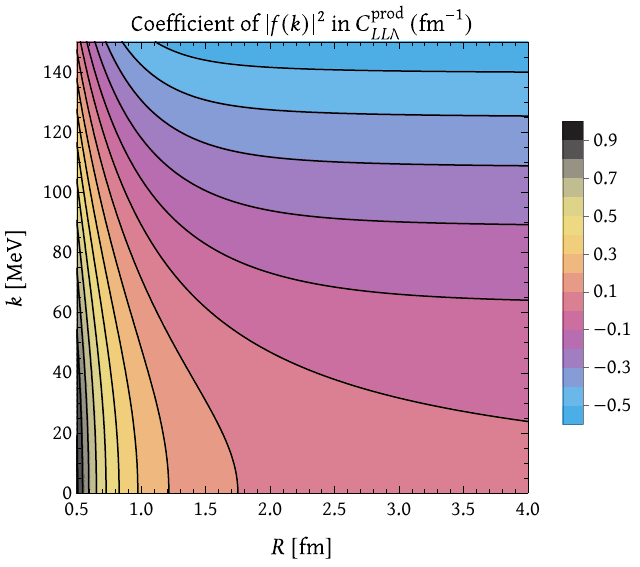}
\end{tabular}    
    \caption{Top left and right panels: ($R,k$) two dimensional  distribution of the coefficient (in fm$^{-1}$ units) of the $\text{Re}f_0(k)$ term of the LL short-distance improved $C_{LL\Lambda}$ (Eq.~\eqref{eq:LL-LKP}) and production $C_{LL\Lambda}^\text{prod}$  (Eq.~\eqref{eq:LL-Lprod}) CFs, respectively, and calculated using  $\Lambda=$570 MeV. Bottom panels: The same as in the top panels but for the coefficient (in fm$^{-2}$ units) of the  $|f_0(k)|^2$ term of the LL  short-distance improved and production CFs.}
    \label{fig:2Dcoeff}
\end{figure*}

In Fig.~\ref{fig:2Dcoeff}, we  show the ($R,k$) two-dimensional distributions of the coefficients of the $\text{Re}f_0(k)$ and $\lvert f_0(k) \rvert^2$ terms of the LL short-distance improved $C_{LL\Lambda}$ [Eq.~\eqref{eq:LL-LKP}] and production $C_{LL\Lambda}^\text{prod}$  [Eq.~\eqref{eq:LL-Lprod}] CFs. These coefficients are independent of the dynamics and they only depend on the Lorentz ultraviolet cutoff, which is fixed in the figure to $\Lambda=$570 MeV. We see large differences between the coefficient of the $|f_0(k)|^2$ term of the $C_{LL\Lambda}$ and $C_{LL\Lambda}^\text{prod}$ that increase with the source-size and c.m. momentum,  and with the latter coefficient even becoming negative in a large part of the ($R,k$)-space. We recall that, although by construction the CFs (both $C^\text{KP}$ and $C^\text{prod}$) are positive, this behaviour of the coefficient of the $|f_0(k)|^2$ term of $C_{LL\Lambda}^\text{prod}$ explains the small values (close to zero) shown for this CF in the $R=2\,\fm$ and $R=5\,\fm$ panels of Fig.~\ref{fig:compTmodel} (red curves). The differences are more moderate in the case of the coefficient of the  $\text{Re}f_0(k)$ term as long as the source size $R$ is below $2\,\fm$. 

\begin{figure*}[t]\centering
\begin{tabular}{cc}
\includegraphics[height=6cm,keepaspectratio]{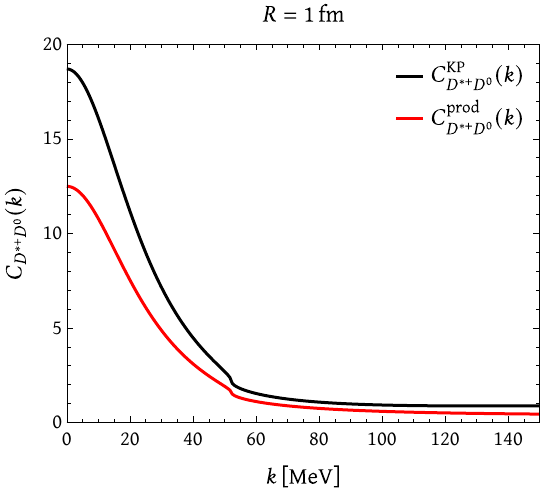} &
\includegraphics[height=6cm,keepaspectratio]{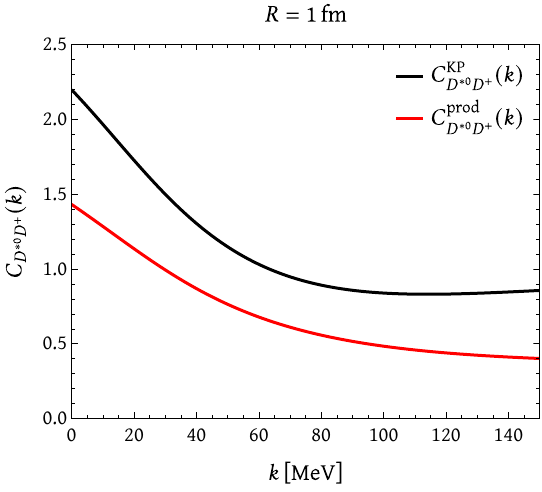} \\
\includegraphics[height=6cm,keepaspectratio]{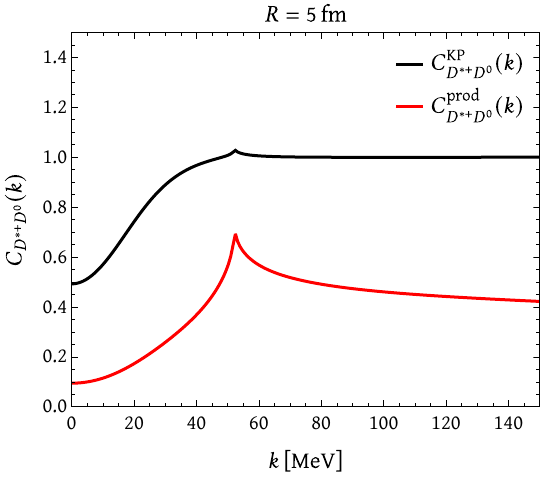} &
\includegraphics[height=6cm,keepaspectratio]{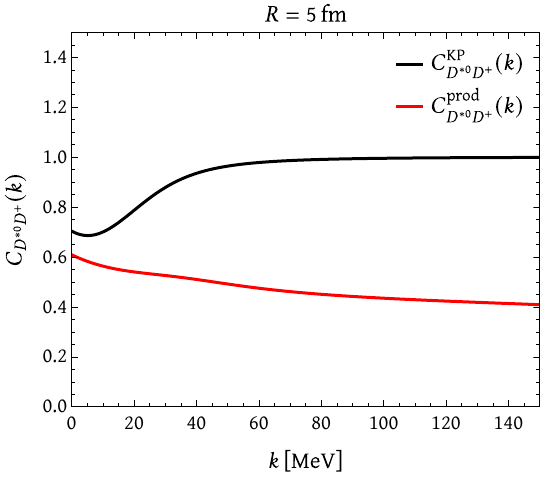}
\end{tabular}
    \caption{ $D^0 D^{\ast +}$ (left)  and $D^+ D^{\ast 0}$ (right) CFs obtained from the QFT based  $C^\text{prod}$ (Eqs.~\eqref{eq:C1} and \eqref{eq:C2}) and the relativistic $C^\text{KP}$  Koonin--Pratt (Eq.~\eqref{eq:correccion-KN-coupled}) formulae, with weights $\eta_{1,2}(s)=1$, as functions of the respective c.m. momentum. We consider 1 (top) and 5 (bottom) fm sources and a coupled-channel interaction  model inspired in Ref.~\cite{Feijoo:2021ppq}. It is the coupled-channel extension of that detailed in the caption of Fig.~\ref{fig:cf-single}, using now physical masses, the same value for $q_{\rm max}$ and a $V-$matrix constructed out of $V_{11}=V_{22}= (V_0+V_1)/2$ and $V_{12}=V_{21}=(V_0-V_1)/2$, with $V_0=-440$ and $V_1=499$, which gives a $T_{cc}$ bound by $354\,\keV$.  The black and red curves show  the $C^\text{KP}$ and the $C^\text{prod}$ results, respectively, with the differences due to the $\delta C^\text{KP}_i$ terms of Eq.~\eqref{eq:correccion-KN-coupled-corr}.  }
    \label{fig:comp.acopl}
\end{figure*}

Next in Fig.~\ref{fig:comp.acopl}, we study the full problem and consider the $D^0 D^{\ast +}$ and $D^+ D^{\ast 0}$ coupled-channel dynamics and evaluate both $C_{D^0D^{*+}}$ and $C_{D^+ D^{\ast 0}}$ CFs for two different source sizes. The  $D^+ D^{\ast 0}$ threshold is $1.4\,\MeV$ above the $D^+ D^{\ast 0}$ threshold and it is reached for a c.m. momentum $k_1\sim 52\,\MeV$. It produces a  visible cusp in the lowest ($D^0D^{\ast +}$) channel, especially for the $R=5\,\fm$ source. The Koonin--Pratt $C_{D^0D^{*+},D^+ D^{\ast 0}}^\text{KP}$ CFs  shown in the figure essentially coincide with those presented in the exploratory analysis carried out in Ref.~\cite{Vidana:2023olz}. We  see the significant differences between  production  and Koonin--Pratt results for both coupled-channels, which would certainly lead to quite different extraction of the  $D^0D^{*+}$ and $D^+ D^{\ast 0}$ scattering amplitudes in future measurements.

\section{Concluding remarks}
\subsection{Theoretical aspects}
For $S$-wave interaction   and a Gaussian source, we have shown   the existing relation between  Koonin--Pratt femtoscopic CFs [$C^\text{KP}_i(s)$] and  invariant mass distributions from production experiments. The equivalence is total for a zero source-size $R$ and for finite values of $R$, we have seen that the  Gaussian source provides a form-factor $F_R(k,p)$ for the virtual production of the particles. Motivated by this remarkable relationship, we study an alternative method to the  Koonin--Pratt formula and analyze $C^\text{prod}(s)$ in Eq.~\eqref{eq:Cimprove}, which connects the evaluation of the CF  directly with the production mechanisms depicted in Fig.~\ref{fig:spectrum}. 

The differences between  $C^\text{prod}(s)$  and $C^\text{KP}(s)$ arise mostly from the $T$--matrix quadratic terms, increase with the source-size and become quite sizable already for $R=1\,\fm$ for the case of the $D^0 D^{\ast +}$ and $D^+ D^{\ast 0}$ correlation functions of interest to unravel the dynamics of the exotic $T_ {cc}(3875)^+$.

We have  seen quite significant differences between  production  and Koonin--Pratt CFs. One might think that from a QFT perspective,  the production CF is more theoretically sound than the Koonin--Pratt one, however the presumably lack of coherence in high-multiplicity-event reactions and in the creation of the fire-ball source that emits the hadrons certainly make much more realistic a formalism based on the Koonin--Pratt equation. In the context  of coupled-channels, we have also argued that it is reasonable to sum the production probabilities instead of the production amplitudes.

We have also discussed in  Appendix \ref{sec:LL} that the  LL approximation is not adequate for small source sizes since it leads to divergent CFs for point-like sources ($R=0$), and therefore such scheme should not be used to compare Koonin--Pratt femtoscopic CFs  and invariant mass distributions from production experiments. We have derived an improved LL approach, which implements a Lorentz ultraviolet regulator that corrects the  pathological behaviour of the LL radial wave function [Eq.~\eqref{eq:wf-short}] at short distances ($r \to 0$) and hence of the corresponding LL CF in the $R\to 0$ limit. This improved model [Eqs.~\eqref{eq:LL-LKP}-\eqref{eq:LL-dLKP}] only requires the scattering length and effective range, which can be also used to approximate the scattering amplitude $f_0(k)$, since these inputs, within some approximations, might be used to fix the Lorentz ultraviolet cutoff through the effective range  formula [Eq.~\eqref{eq:MQalcance-r0}].

All results derived in this work should be easily generalized for higher angular momentum waves. In any case many of the exotic states ($X(3872), T_ {cc}(3875)^+, X(3960), Z_c(3900)\dots$), which cannot be naturally accommodated in simple constituent quark models, have  sizable/dominant $S$-wave hadron-molecular components.   

\subsection{Some implications for experimental applications}
The work that we presented here raises issues concerning the meaning of the CFs and its relationship to production measurements. To make steps forward in this problem from the experimental side it would be interesting to pay attention to the distribution of total momentum of the pairs measured. This is already available in the present measurements since the particles are detected individually and then the relative momentum is determined. Knowing the total momentum of the pair allows us to determine the invariant mass of the rest of the particles ($X$) not measured. Thus,  one has already access to the invariant mass distribution of the $X$ set of particles. The production cross section  of the measured pair depends on this mass distribution. It could be that the source function has some ties to this mass distribution, which would be different in collisions of proton with protons, with nuclei, or nucleus-nucleus collisions. This could shed some light on why the source function is so different in these cases. There is another issue worth of some attention, which is the result of the mixed event determination. This magnitude is used to normalize the correlation function dividing the probability of finding the pair in a single event by that obtained by  the mixed events measuring the relative momentum of the pair from two randomly selected different events. It is unclear whether  this distribution does not contain some energy dependence by itself, or that the ratio of the probabilities of a single event to that of the mixed event converges to unity as fast as seen in the theoretical calculations. 

There are other implications of our work concerning experimental analyses. Many correlation functions for a pair of particles are analyzed in terms of a single channel in order to determine scattering lengths and effective ranges. In some cases the consideration of coupled channels, as we have studied in the present work,  is essential. In this direction, we can call the attention to a recent work \cite{Feijoo:2024bvn}, where the $\phi p$ scattering length is determined based on the consideration of different vector-baryon channels and is found drastically smaller than the one previously obtained based on the single channel consideration of the $\phi p$ employing the LL approach \cite{ALICE:2021cpv}. These findings have repercussions in other areas of physics. Actually, based on the large scattering length obtained in the single channel analysis of \cite{ALICE:2021cpv}, the existence of a bound $\phi p$ has been claimed in \cite{Sun:2022cxf,Chizzali:2022pjd}, which would be ruled out with the result of  \cite{Feijoo:2024bvn}.

The reinforcement of the realistic character of the KP formalism to compute CFs, concluded in the present study, confers an unequivocal and compact structure that leads one to think that the femtoscopy data can provide less ambiguous constraints for the theoretical models than those coming from the weak decays, which incorporate inherent difficulties in obtaining information due to the coexistence of multiple possible mechanisms. Furthermore, compatibility studies can be performed between experimental invariant masses associated with decay processes and femtoscopic data involving the same particle pairs through theoretical bridges, as done in the recent work of Ref.~ \cite{Feijoo:2024qqg}.

\vspace{0.5cm}

\section*{Acknowledgments}
We warmly thank  M. Mikhasenko for useful discussions in the early stages of this research. This work was supported by the Spanish Ministerio de Ciencia e Innovaci\'on (MICINN) and European FEDER funds under Contracts No.\,PID2020-112777GB-I00, PID2023-147458NB-C21 and CEX2023-001292-S; by Generalitat Valenciana (GVA) under contracts PROMETEO/2020/023 and  CIPROM/2023/59. This project has received funding from the European Union Horizon 2020 research and innovation programme under the program H2020-INFRAIA-2018-1, grant agreement No.\,824093 of the STRONG-2020 project. M.\,A. is supported by GVA Grant No.\,CIDEGENT/2020/002 and MICINN Ram\'on y Cajal programme Grant No.\,RYC2022-038524-I . A.\,F. is supported by ORIGINS cluster DFG under Germany’s Excellence Strategy-EXC2094 - 390783311 and the DFG through the Grant SFB 1258 ``Neutrinos and Dark Matter in Astro and Particle Physics”.

\appendix

\section{Some mathematical details for Sect.~\ref{sec:scanaly}}
\label{app:tools}
We first consider the action of the scattering $\MollerOp$  M\"oller operator on eigenstates $\ket{\phi}$ of the kinetic energy operator $\MyOp{H}_0$ [$\MyOp{H}_0\ket{\phi}=E\ket{\phi}]$. It follows~\cite{Pascual:2012} 
\begin{equation}
    \MollerOp \ket{\phi} = \left(\lim_{t\to\
    +\infty} e^{it \MyOp{H}}e^{-it \MyOp{H}_0}\right)\ket{\phi}  = \left(\lim_{\epsilon \to 0}\, \epsilon \int_0^{+\infty} dt' e^{-t'\epsilon}e^{it' \MyOp{H}}e^{-it' \MyOp{H}_0}\right)\ket{\phi}= \lim_{\epsilon \to 0}\, \epsilon \int_0^{+\infty} dt' e^{it' (\MyOp{H}-E+i\epsilon)}\ket{\phi} = \frac{i\epsilon}{\MyOp{H}-E+i\epsilon} \ket{\phi}\label{eq:a1}
\end{equation}
The formal identity $(A+B)^{-1}= A^{-1}-A^{-1}B(A+B)^{-1}$ between the operators $A$ and  $B$ applied to the resolvent $\big(\MyOp{H}-E+i\epsilon\big)^{-1}$ leads to
\begin{equation}
\big(\MyOp{H}-E+i\epsilon\big)^{-1} = \big(\MyOp{H}_0-E+i\epsilon+ \MyOp{V}^{\rm QM}\big)^{-1} =  \frac{1}{\MyOp{H}_0-E+i\epsilon}- \frac{1}{\MyOp{H}_0-E+i\epsilon} \MyOp{V}^{\rm QM}\frac{1}{\MyOp{H}-E+i\epsilon}
\end{equation}
and thus from Eq.~\eqref{eq:a1} and using that $\MyOp{H}_0\ket{\phi}=E\ket{\phi}$, we obtain 
\begin{equation}
  \MollerOp \ket{\phi} =  \frac{i\epsilon}{\MyOp{H}-E+i\epsilon}\ket{\phi} = \left( \frac{i\epsilon}{\MyOp{H}_0-E+i\epsilon}- \frac{i\epsilon}{\MyOp{H}_0-E+i\epsilon} \MyOp{V}^{\rm QM}\frac{1}{\MyOp{H}-E+i\epsilon}\right)\ket{\phi} =  \left(1 -\frac{i\epsilon}{E-\MyOp{H}_0-i\epsilon} \MyOp{V}^{\text{QM}} \frac{1}{E-\MyOp{H}-i\epsilon}\right)  \ket{\phi}\label{eq:a3}
\end{equation}
which provides the starting point of Eq.~\eqref{eq:moller}.

Next we compile here some small-size source asymptotic behaviors of the form-factor $F_R(q,q')$ introduced in Eq.~\eqref{eq:defFR} 
\begin{eqnarray}
  F_R(q,q') & = & \int d^3\vec{r}\, S(r) j_0(qr)j_0(q'r)= \frac{e^{-(q^2+q^{\prime 2})R^2}\sinh(2qq'R^2)}{2qq'R^2}\simeq   1-(q^2+q^{\,\prime 2})R^2+ {\cal O}(\ell^4R^4)\,, \label{eq:a4}
\end{eqnarray} 
with $\ell^4=q^4, q^{\,\prime 4}, q^2q^{\,\prime 2}$. For $\widetilde F_R(k,q)= F_R(k,q)/F_R(k,k)$, one finds: 
\begin{equation}
    F_R(k,k)= (1-e^{-x^2})/x^2=1-x^2/2+{\cal O}(x^4) 
\end{equation}
with $x=2kR$. In addition, in the limit of small-size sources:  
\begin{eqnarray}
F_R(k,p)- \widetilde F_R(k,p) & = & -2k^2R^2 + \frac23 \left(k^4+3k^2p^2 \right)R^4 + {\cal O}(R^6)\label{eq:ff1-deltaC}\\
F_R(p,p')-\widetilde F_R(k,p)\widetilde F_R(k,p')&=& -2k^2R^2+\frac23\left[p^2p^{\prime 2}+k^2\left(2p^2+2p^{\prime 2}-k^2\right)\right]R^4 + {\cal O}(R^6) \label{eq:ff2-deltaC}
\end{eqnarray}

\section{Improved Lednicky-Lyuboshits  approximation using a Lorentz form-factor}
\label{sec:LL}
In the LL approximation~\cite{Lednicky:1981su}, the half off-shell $T-$matrix is approximated by the on-shell one, i.e., $T^{\text{QM}}(k\leftarrow p;E)\approx T^{\text{QM}}(k\leftarrow k;E)$. Using that 
\begin{equation}
    \int d^3 \vec{p} \frac{j_0(pr)}{E-\frac{p^2}{2\mu_{ab}}+i\epsilon} = -4\pi^2\mu_{ab}\ \frac{e^{ikr}}{r}
\end{equation}
it follows from Eq.~\eqref{eq:wf-onshell} that, within the LL  approximation, the wave function reads 
\begin{equation}
\psi^\ast_{LL}(\vec r; \vec k\,) =  e^{-i\vec k\cdot\vec r}  +f_0(k) \frac{e^{ikr}}{r} \label{eq:assy}
\end{equation}
in concordance to the asymptotic behavior ($r\to \infty$) of the $S$-wave radial wave-function~\cite{Ohnishi:2016elb,Haidenbauer:2018jvl}. Thus 
\begin{equation}
  \psi_{LL}(r; k\,) = \frac12 \int_{-1}^{+1} d(\cos{\theta}) \psi_{LL}(\vec r; \vec k\,) =\frac{1}{2ikr}\left(e^{ikr}-e^{-2i\delta_0(k)}e^{-ikr}\right)  
\end{equation}
where $\theta$ is the angle formed by $\vec r$ and $\vec k$. Using Eqs.~\eqref{eq:cf} and \eqref{eq:Cprodintermswf}, the wave-function of Eq.~\eqref{eq:assy} gives rise to
\begin{eqnarray}
C_{LL}(k)     &=&1 + \frac{2\text{Re}{f_0(k)}}{\sqrt{\pi}R} F_1(x) - \frac{\text{Im}{f_0(k)}}{R} F_2(x)
    + \frac{\left\lvert f_0(k) \right\rvert^2}{2R^2} = 1 + \frac{2\text{Re}{f_0(k)}}{\sqrt{\pi}R} F_1(x)+ \frac{\left\lvert f_0(k) \right\rvert^2}{2R^2} e^{-x^2}\label{eq:LL-KP}\\
 C_{LL}^\text{prod}(k) &=& 1 + \frac{2\text{Re}{f_0(k)}}{\sqrt{\pi}R} F_3(x) +\frac{\left\lvert f_0(k) \right\rvert^2} {2R^2}\left(-\frac{x^2}{2}+\frac{2F_3^2(x)}{\pi}\right) \label{eq:LL-dKP}
\end{eqnarray}
where $x=2kR$, $F_1(x) = \int_{0}^{x} \mathrm{d}t e^{(t^2 - x^2)}/x$, $F_3(x) = F_1(x)/F_R(k,k)= x\int_{0}^{x} \mathrm{d}t e^{(t^2-x^2) }/(1-e^{-x^2})$ and $F_2(x) = ( 1 - e^{-x^2})/x$. The asymptotic behaviors of the above functions are $F_2(x) \stackrel{x\to 0}{=} x-\frac{x^3}{2} + {\cal O}(x^5) $ and 
\begin{eqnarray}
 F_1(x) &\stackrel{x\to 0}{=}& 1-\frac{2x^2}{3} + {\cal O}(x^4)\, , \qquad F_1(x)  \stackrel{x\to  \infty}{=}\frac{1}{2x^2} + \frac{1}{4x^4} +{\cal O}(x^{-6}) \\
 F_3(x) &\stackrel{x\to 0}{=} &1-\frac{\phantom{2}x^2}{6} + {\cal O}(x^4)\,, \qquad F_3(x)\stackrel{x\to  \infty}{=} \frac{1}{2} + \frac{1}{4x^2}+ {\cal O}(x^{-4}).
\end{eqnarray}
The LL approximation does not work for point-like sources ($R=0$) since in that limit the loop integrals in Eqs.~\eqref{eq:cknorel} and \eqref{eq:dcknorel-aux} diverge. This is because these integrals are finite thanks to the ultraviolet regulator $f_\text{UV}(p)$,  included in the half off-shell $T-$matrix of Eq.~\eqref{eq:thalf}, and the  $F_R(k,p)$  and $F_R(p,p')$ form-factors which become one in the  $R\to 0$ limit. Another way to trace the origin of the divergence of the LL  CF for $R=0$ is from Eq.~\eqref{eq:ckRzero} and the non-regular behavior at $r=0$ of $\psi_{LL}(\vec r; \vec k\,)$ in Eq.~\eqref{eq:assy}.

The LL  approximation can be improved to be valid in the $R\to 0$ limit by including an ultraviolet regulator in the half off-shell $T-$matrix in Eq.~\eqref{eq:thalf}. We illustrate this for the case of a Lorentz (mono-pole) form-factor
\begin{equation}
    T^{\text{QM}}(k\leftarrow p;E)\approx T^{\text{QM}}(k\leftarrow k;E)\left(\frac{\Lambda^2+k^2}{\Lambda^2+p^2}\right) \label{eq:thalf-m}
\end{equation}
with $\Lambda$ an ultraviolet cutoff. We have now
\begin{eqnarray}
     \int d^3 \vec{p} \left(\frac{\Lambda^2+k^2}{\Lambda^2+p^2}\right)\frac{j_0(pr)}{E-\frac{p^2}{2\mu_{ab}}+i\epsilon} &=& -4\pi^2\mu_{ab}\ \left(\frac{e^{ikr}-e^{-\Lambda r}}{r}\right)
\end{eqnarray}
which leads to a regular  wave function at the origin ($r=0$)
\begin{equation}
\psi^\ast_{LL\Lambda}(\vec r; \vec k\,) =  e^{-i\vec k\cdot\vec r}  +f_0(k) \frac{e^{ikr}-e^{-\Lambda r}}{r} \label{eq:wf-short}
\end{equation}
Using in Eqs.~\eqref{eq:cf} and \eqref{eq:Cprodintermswf} the above wave-function, corrected at short distances, we now obtain
\begin{eqnarray}
C_{LL\Lambda}(k)&=&   1
    + \frac{2\text{Re} f_0(k)}{\sqrt{\pi}R} \left[F_1(x) + \frac{\sqrt{\pi}\, \text{Im}\ F_4(x_\Lambda^+ )}{x}\right]+\frac{\left\lvert f_0(k) \right\rvert^2}{2R^2} \left[e^{-x^2}
    +F_4(2x_\Lambda)-2\text{Re}\ F_4(x_\Lambda^+) \right]\label{eq:LL-LKP}\\
 C_{LL\Lambda}^\text{prod}(k) &=& 1
    + \frac{2\text{Re} f_0(k)}{\sqrt{\pi}R} \left[F_3(x) + \frac{\sqrt{\pi}\,\text{Im}\ F_4(x_\Lambda^+ )}{F_2(x)}\right] +\frac{\left\lvert f_0(k) \right\rvert^2}{2R^2}\left[\frac{2}{\pi}\left(F_3(x)+\frac{\sqrt{\pi}\,\text{Im}\ F_4(x_\Lambda^+) }{F_2(x)}\right)^2-\frac{x^2}{2}\right] \label{eq:LL-Lprod} \\
 \delta C_{LL\Lambda}(k) &=& \frac{2\text{Re} f_0(k)}{\sqrt{\pi}R} \left(\frac{F_2(x)-x}{xF_2(x)}\right)\left[x F_1(x)+ \sqrt{\pi}\,\text{Im}\ F_4(x_\Lambda^+) \right]\nonumber \\
    &&+\frac{\left\lvert f_0(k) \right\rvert^2}{2R^2} \left(e^{-x^2}
    +F_4(2x_\Lambda)-2\text{Re}\ F_4(x_\Lambda^+)  -\frac{2}{\pi}\left[F_3(x)+\frac{\sqrt{\pi}\,\text{Im}\ F_4(x_\Lambda^+) }{F_2(x)}\right]^2+\frac{x^2}{2}\right)\label{eq:LL-dLKP}
\end{eqnarray}
with $x_\Lambda= \Lambda R$, $x_\Lambda^+= x_\Lambda+ix/2$
\begin{align}
    F_4(z)&= e^{z^2}\left(1-\frac{2}{\sqrt{\pi}}\int_\Gamma e^{-t^2} dt\right)=1-\frac{2z}{\sqrt{\pi}}+z^2+ {\cal O}(z^3) \label{eq:F4}\\
    &= \frac{1}{z\sqrt{\pi}}-\frac{1}{2z^3\sqrt{\pi}}+\frac{3}{4z^5\sqrt{\pi}}+ {\cal O}(z^{-7}) \nonumber
\end{align}
where $\Gamma$ is any path in the complex plane from $t=0$ to $t=z$. It can be also obtained  from the numerical integration 
\begin{equation}
 F_4(z) =\frac{2z}{\sqrt{\pi}}\int_1^\infty dy e^{z^2(1-y^2)}  
\end{equation}
which converges for $\text{Re}(z^2)>0$ [$k<\Lambda$ in Eqs.~\eqref{eq:LL-LKP}-\eqref{eq:LL-dLKP}].

The CFs evaluated now from  Eqs.~\eqref{eq:LL-LKP} and \eqref{eq:LL-Lprod} are finite in the $R\to 0$ limit, and they should in general perform better that the original LL formulae given in Eqs.~\eqref{eq:LL-KP}-\eqref{eq:LL-dKP}. Thus, from Eqs.~\eqref{eq:ckRzero} and \eqref{eq:wf-short}, we find 
\begin{equation}
 \lim_{R\to 0 }C_{LL\Lambda}(k) =  \lim_{R\to 0 }C_{LL\Lambda}^\text{prod}(k) =  \left\lvert 1+ f_0(k)(\Lambda+ik ) \right\rvert^2 = 1 +2\Lambda \text{Re} f_0(k) + (\Lambda^2-k^2)\left\lvert f_0(k) \right\rvert^2
\end{equation}

Moreover, one might relate the Lorentz ultraviolet cutoff $\Lambda$, the scattering length and the effective range through
\begin{equation}
 r_0= \frac{3}{\Lambda}\left(1-\frac{4}{3a_0\Lambda}\right)  \label{eq:MQalcance-r0}
\end{equation}
which is obtained by using the effective range formula. It allows to calculate $r_0$  as 
\begin{equation}
r_0 = -2\int_0^\infty dr\left[u_{k=0}^2(r)-\phi_{k=0}^2(r)\right], \qquad u_{k=0}(r)= \left(1-e^{-\Lambda r}\right)-\frac{r}{a_0}
\label{eq:MQalcance}
\end{equation}
where $u_{k=0}$ is the  zero-energy reduced radial wave function and $\phi_{k=0}$ is its asymptotic form, conveniently normalized (see for instance Appendix B-2 of Ref.~\cite{Preston:1993}). The above expression gets a natural support since $\phi_{k=0}(r)=u_{k=0}(r)$ when $r$ is appreciably outside the range of hadron forces. 

In this way the only inputs needed  in Eqs.~\eqref{eq:LL-LKP} and \eqref{eq:LL-Lprod} are the scattering length and effective range, which can be also used to approximate the scattering amplitude $f_0(k)\approx (-1/a_0+r_0k^2/2-ik)^{-1}$.

However, we should point out that the effective range formula of Eq.~\eqref{eq:MQalcance} is deduced for a local and energy independent potential. The relation of Eq.~\eqref{eq:MQalcance-r0} might suffer from some corrections since the half off-shell $T-$matrix mono-pole ansatz of Eq.~\eqref{eq:thalf-m} does not derive from a local potential.   In addition,  the existence of a non-molecular state near threshold could be related to an energy dependent potential~\cite{Gamermann:2009uq}, which might induce sizable corrections to Eq.~\eqref{eq:MQalcance}, as follows from the Weinberg's compositeness relation~\cite{Weinberg:1965zz}. The additional assumption made to derive this improved LL model is the form of Eq.~\eqref{eq:thalf-m} for the half off-shell $T-$matrix.

\subsection{\boldmath The $k\to 0$ limit}
\label{app:k0}
\begin{figure*}[t]\centering
\includegraphics[height=4.50cm,keepaspectratio]{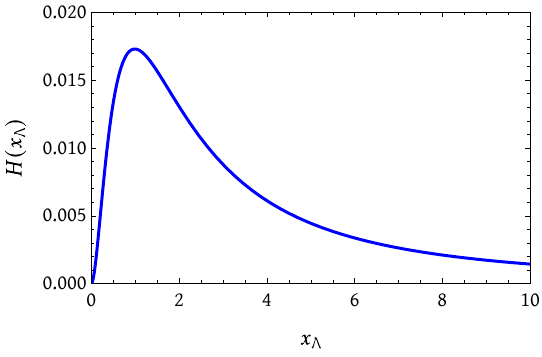}\hspace{0.75cm}%
\includegraphics[height=4.50cm,keepaspectratio]{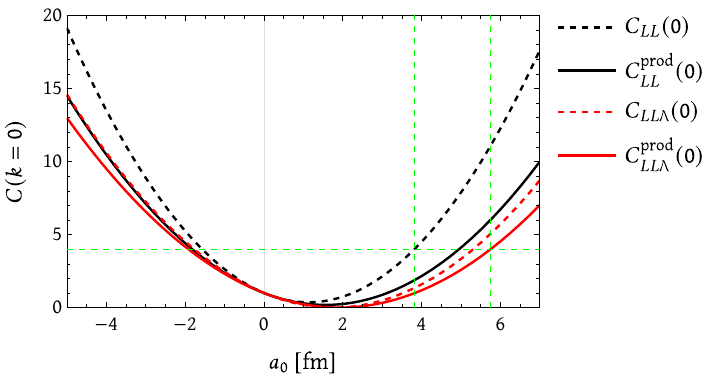}%
    \caption{ Left: Dependence of  $H(x_\Lambda)=\delta C_{LL\Lambda}(k=0)/(\Lambda a_0)^2$ [see Eq.~\eqref{eq:deltaCLLkzero}] as a function of $x_\Lambda=\Lambda R$. Right: Lednicky-Lyuboshits $C_{LL}(k=0)$ (black-dotted) and $C_{LL}^\text{prod}(k=0)$ (red-dotted), and improved $C_{LL\Lambda}(k=0)$ (black-solid) and $C_{LL\Lambda}^\text{prod}(k=0)$ (red-solid) CFs for a Gaussian source of size $R=1$ fm, as functions of the $S$-wave scattering length ($a_0$). Results for the improved $C_{LL\Lambda}(k=0)$ and $C_{LL\Lambda}^\text{prod}(k=0)$ CFs have been obtained using a Lorentz ultraviolet cutoff ($\Lambda$) of 450 MeV.   We have used  Eqs.~\eqref{eq:LL-KP}, \eqref{eq:LL-dKP}, \eqref{eq:LL-LKP} and \eqref{eq:LL-Lprod} [see also Eqs.~\eqref{eq:CLLPRODkzero}, \eqref{eq:deltaCLLkzero} and \eqref{eq:CLLPRODkzero-aux}]. The horizontal and vertical green-dashed lines guide the determination of the scattering length ($>0$) for a case in which the measured CF at threshold is 4, and  in the analysis  either the $C_{LL}$ or the $C_{LL\Lambda} ^\text{prod}$ models are used.}
    \label{fig:deltaCLLkzero}
\end{figure*}

In the strict $k\to 0$ limit, one has $[F_R(0,p)-\widetilde F_R(0,p)]=0$ and 
\begin{equation}
F_R(p,p')-\widetilde F_R(0,p)\widetilde F_R(0,p')= \frac{e^{-(p^2+p^{\prime 2})R^2}}{2pp^\prime R^2}\left(\sinh{(2pp^\prime R^2)}-2pp^\prime R^2\right)=\frac23 p^2p^{\prime 2}R^4+ {\cal O}(R^6) \label{eq:ff-deltaC-k0}
\end{equation}
which leads, within the Lorentz form-factor improved LL approximation derived above, to
\begin{eqnarray}
C^\text{prod}_{LL\Lambda}(k=0) &=& \left(1-a_0\Lambda\,F_4(x_\Lambda)\right)^2  \label{eq:CLLPRODkzero}\\
 \delta C_{LL\Lambda}(k=0) &=& (a_0\Lambda )^2\, H(x_\Lambda), \qquad H(x_\Lambda)=\frac{1-2F_4(x_\Lambda)\left[1+x_\Lambda^2 F_4(x_\Lambda)\right] + F_4(2x_\Lambda)}{2x_\Lambda^2}  \label{eq:deltaCLLkzero}
\end{eqnarray}
with $x_\Lambda=\Lambda R$. The universal function $H(x_\Lambda)$ is plotted in the left plot of Fig.~\ref{fig:deltaCLLkzero}. There we see that it has a maximum for $x_\Lambda\sim 1$ and $ \delta C_{LL\Lambda}(k=0)$ grows as the square of the scattering length. Asymptotically, we have 
\begin{eqnarray}
 C^\text{prod}_{LL\Lambda}(k=0) &\stackrel{R\to 0}{=} & \left(1-a_0\Lambda\right)^2\, , \quad  \delta C_{LL\Lambda}(k=0) \stackrel{R\to 0}{=} \left(\frac{3\pi-8}{2\pi}\right)\left( a_0 R\Lambda^2 \right)^2\left[1+ {\cal O}(\Lambda R)\right]\\
  C^\text{prod}_{LL\Lambda}(k=0) &\stackrel{\Lambda\to \infty}{=} & \left(1-\frac{a_0}{R\sqrt{\pi}}\right)^2\, , \quad \delta C_{LL\Lambda}(k=0) \stackrel{\Lambda\to \infty}{=} \left(\frac12 -\frac1\pi \right)\frac{a_0^2}{R^2}\label{eq:CLLPRODkzero-aux}
\end{eqnarray}

Note that Eq.~\eqref{eq:deltaCLLkzero} directly provides the variation of the determination of the scattering length when the Koonin--Pratt formula or the alternative production model is used to analyze the CF at zero momentum. This is illustrated in the right panel of   Fig.~\ref{fig:deltaCLLkzero}, where different LL CFs [Eqs.~\eqref{eq:CLLPRODkzero}, \eqref{eq:deltaCLLkzero} and \eqref{eq:CLLPRODkzero-aux}] at zero momentum and a source-size $R=1$ fm are shown as functions of the $S$-wave scattering length.  Results for the improved $C_{LL\Lambda}(k=0)$ and $C_{LL\Lambda}^\text{prod}(k=0)$ CFs have been obtained using a mono-pole form-factor with $\Lambda=450$ MeV. For each value of $a_0$, this cutoff can be translated into a effective range parameter $r_0$ using Eq.~\eqref{eq:MQalcance-r0}, as long as $r_0$ remains positive and the effective range formula of Eq.~\eqref{eq:MQalcance} makes sense. We see that the differences are small for $-1.5 \, \text{fm} < a_0 < 1.5 \, \text{fm}$, however for larger absolute values of the scattering length, the variation between analyses carried out using the traditional LL $C_{LL}$ (Koonin--Pratt) or the short-distance improved   
LL production   $C^\text{prod}_{LL\Lambda}$  CFs could become  large. For example, for a measured value of $C(k=0)$ of 4, one can determined $a_0=3.8$ fm or $5.8$ fm from $C_{LL}$  or $C^\text{prod}_{LL\Lambda}$ models, respectively. Exotic weakly bound states, as the $T_{cc}(3875)^+$, provide large scattering lengths which scale as $(2\mu B)^{-1/2}$, with $B(>0)$ the binding energy and $\mu$ the reduced mass of the two interacting hadrons.

\bibliographystyle{apsrev4-2}

\begin{thebibliography}{104}%
\makeatletter
\providecommand \@ifxundefined [1]{%
 \@ifx{#1\undefined}
}%
\providecommand \@ifnum [1]{%
 \ifnum #1\expandafter \@firstoftwo
 \else \expandafter \@secondoftwo
 \fi
}%
\providecommand \@ifx [1]{%
 \ifx #1\expandafter \@firstoftwo
 \else \expandafter \@secondoftwo
 \fi
}%
\providecommand \natexlab [1]{#1}%
\providecommand \enquote  [1]{``#1''}%
\providecommand \bibnamefont  [1]{#1}%
\providecommand \bibfnamefont [1]{#1}%
\providecommand \citenamefont [1]{#1}%
\providecommand \href@noop [0]{\@secondoftwo}%
\providecommand \href [0]{\begingroup \@sanitize@url \@href}%
\providecommand \@href[1]{\@@startlink{#1}\@@href}%
\providecommand \@@href[1]{\endgroup#1\@@endlink}%
\providecommand \@sanitize@url [0]{\catcode `\\12\catcode `\$12\catcode
  `\&12\catcode `\#12\catcode `\^12\catcode `\_12\catcode `\%12\relax}%
\providecommand \@@startlink[1]{}%
\providecommand \@@endlink[0]{}%
\providecommand \url  [0]{\begingroup\@sanitize@url \@url }%
\providecommand \@url [1]{\endgroup\@href {#1}{\urlprefix }}%
\providecommand \urlprefix  [0]{URL }%
\providecommand \Eprint [0]{\href }%
\providecommand \doibase [0]{https://doi.org/}%
\providecommand \selectlanguage [0]{\@gobble}%
\providecommand \bibinfo  [0]{\@secondoftwo}%
\providecommand \bibfield  [0]{\@secondoftwo}%
\providecommand \translation [1]{[#1]}%
\providecommand \BibitemOpen [0]{}%
\providecommand \bibitemStop [0]{}%
\providecommand \bibitemNoStop [0]{.\EOS\space}%
\providecommand \EOS [0]{\spacefactor3000\relax}%
\providecommand \BibitemShut  [1]{\csname bibitem#1\endcsname}%
\let\auto@bib@innerbib\@empty
\bibitem [{\citenamefont {Hanbury~Brown}\ and\ \citenamefont
  {Twiss}(1954)}]{HanburyBrown:1954amm}%
  \BibitemOpen
  \bibfield  {author} {\bibinfo {author} {\bibfnamefont {R.}~\bibnamefont
  {Hanbury~Brown}}\ and\ \bibinfo {author} {\bibfnamefont {R.~Q.}\ \bibnamefont
  {Twiss}},\ }\href {https://doi.org/10.1080/14786440708520475} {\bibfield
  {journal} {\bibinfo  {journal} {Phil. Mag. Ser. 7}\ }\textbf {\bibinfo
  {volume} {45}},\ \bibinfo {pages} {663} (\bibinfo {year} {1954})}\BibitemShut
  {NoStop}%
\bibitem [{\citenamefont {Hanbury~Brown}\ and\ \citenamefont
  {Twiss}(1956)}]{HanburyBrown:1956bqd}%
  \BibitemOpen
  \bibfield  {author} {\bibinfo {author} {\bibfnamefont {R.}~\bibnamefont
  {Hanbury~Brown}}\ and\ \bibinfo {author} {\bibfnamefont {R.~Q.}\ \bibnamefont
  {Twiss}},\ }\href {https://doi.org/10.1038/1781046a0} {\bibfield  {journal}
  {\bibinfo  {journal} {Nature}\ }\textbf {\bibinfo {volume} {178}},\ \bibinfo
  {pages} {1046} (\bibinfo {year} {1956})}\BibitemShut {NoStop}%
\bibitem [{\citenamefont {Lisa}\ \emph {et~al.}(2005)\citenamefont {Lisa},
  \citenamefont {Pratt}, \citenamefont {Soltz},\ and\ \citenamefont
  {Wiedemann}}]{Lisa:2005dd}%
  \BibitemOpen
  \bibfield  {author} {\bibinfo {author} {\bibfnamefont {M.~A.}\ \bibnamefont
  {Lisa}}, \bibinfo {author} {\bibfnamefont {S.}~\bibnamefont {Pratt}},
  \bibinfo {author} {\bibfnamefont {R.}~\bibnamefont {Soltz}},\ and\ \bibinfo
  {author} {\bibfnamefont {U.}~\bibnamefont {Wiedemann}},\ }\href
  {https://doi.org/10.1146/annurev.nucl.55.090704.151533} {\bibfield  {journal}
  {\bibinfo  {journal} {Ann. Rev. Nucl. Part. Sci.}\ }\textbf {\bibinfo
  {volume} {55}},\ \bibinfo {pages} {357} (\bibinfo {year} {2005})},\ \Eprint
  {https://arxiv.org/abs/nucl-ex/0505014} {arXiv:nucl-ex/0505014} \BibitemShut
  {NoStop}%
\bibitem [{\citenamefont {{ALICE Collab.}}(2020)}]{ALICE:2020mfd}%
  \BibitemOpen
  \bibfield  {author} {\bibinfo {author} {\bibnamefont {{ALICE Collab.}}},\
  }\href {https://doi.org/10.1038/s41586-020-3001-6} {\bibfield  {journal}
  {\bibinfo  {journal} {Nature}\ }\textbf {\bibinfo {volume} {588}},\ \bibinfo
  {pages} {232} (\bibinfo {year} {2020})},\ \bibinfo {note} {[Erratum: Nature
  590, E13 (2021)]},\ \Eprint {https://arxiv.org/abs/2005.11495}
  {arXiv:2005.11495 [nucl-ex]} \BibitemShut {NoStop}%
\bibitem [{\citenamefont {{ALICE Collab.}}(2022)}]{ALICE:2022wwr}%
  \BibitemOpen
  \bibfield  {author} {\bibinfo {author} {\bibnamefont {{ALICE Collab.}}},\
  }\href@noop {} {\  (\bibinfo {year} {2022})},\ \Eprint
  {https://arxiv.org/abs/2211.02491} {arXiv:2211.02491 [physics.ins-det]}
  \BibitemShut {NoStop}%
\bibitem [{\citenamefont {Fabbietti}\ \emph {et~al.}(2021)\citenamefont
  {Fabbietti}, \citenamefont {Mantovani~Sarti},\ and\ \citenamefont
  {Vazquez~Doce}}]{Fabbietti:2020bfg}%
  \BibitemOpen
  \bibfield  {author} {\bibinfo {author} {\bibfnamefont {L.}~\bibnamefont
  {Fabbietti}}, \bibinfo {author} {\bibfnamefont {V.}~\bibnamefont
  {Mantovani~Sarti}},\ and\ \bibinfo {author} {\bibfnamefont {O.}~\bibnamefont
  {Vazquez~Doce}},\ }\href {https://doi.org/10.1146/annurev-nucl-102419-034438}
  {\bibfield  {journal} {\bibinfo  {journal} {Ann. Rev. Nucl. Part. Sci.}\
  }\textbf {\bibinfo {volume} {71}},\ \bibinfo {pages} {377} (\bibinfo {year}
  {2021})},\ \Eprint {https://arxiv.org/abs/2012.09806} {arXiv:2012.09806
  [nucl-ex]} \BibitemShut {NoStop}%
\bibitem [{\citenamefont {Koonin}(1977)}]{Koonin:1977fh}%
  \BibitemOpen
  \bibfield  {author} {\bibinfo {author} {\bibfnamefont {S.~E.}\ \bibnamefont
  {Koonin}},\ }\href {https://doi.org/10.1016/0370-2693(77)90340-9} {\bibfield
  {journal} {\bibinfo  {journal} {Phys. Lett. B}\ }\textbf {\bibinfo {volume}
  {70}},\ \bibinfo {pages} {43} (\bibinfo {year} {1977})}\BibitemShut {NoStop}%
\bibitem [{\citenamefont {Lednicky}\ and\ \citenamefont
  {Lyuboshits}(1981)}]{Lednicky:1981su}%
  \BibitemOpen
  \bibfield  {author} {\bibinfo {author} {\bibfnamefont {R.}~\bibnamefont
  {Lednicky}}\ and\ \bibinfo {author} {\bibfnamefont {V.~L.}\ \bibnamefont
  {Lyuboshits}},\ }\href@noop {} {\bibfield  {journal} {\bibinfo  {journal}
  {Yad. Fiz.}\ }\textbf {\bibinfo {volume} {35}},\ \bibinfo {pages} {1316}
  (\bibinfo {year} {1981})}\BibitemShut {NoStop}%
\bibitem [{\citenamefont {Pratt}(1986)}]{Pratt:1986cc}%
  \BibitemOpen
  \bibfield  {author} {\bibinfo {author} {\bibfnamefont {S.}~\bibnamefont
  {Pratt}},\ }\href {https://doi.org/10.1103/PhysRevD.33.1314} {\bibfield
  {journal} {\bibinfo  {journal} {Phys. Rev. D}\ }\textbf {\bibinfo {volume}
  {33}},\ \bibinfo {pages} {1314} (\bibinfo {year} {1986})}\BibitemShut
  {NoStop}%
\bibitem [{\citenamefont {Pratt}\ and\ \citenamefont
  {Tsang}(1987)}]{Pratt:1987zz}%
  \BibitemOpen
  \bibfield  {author} {\bibinfo {author} {\bibfnamefont {S.}~\bibnamefont
  {Pratt}}\ and\ \bibinfo {author} {\bibfnamefont {M.~B.}\ \bibnamefont
  {Tsang}},\ }\href {https://doi.org/10.1103/PhysRevC.36.2390} {\bibfield
  {journal} {\bibinfo  {journal} {Phys. Rev. C}\ }\textbf {\bibinfo {volume}
  {36}},\ \bibinfo {pages} {2390} (\bibinfo {year} {1987})}\BibitemShut
  {NoStop}%
\bibitem [{\citenamefont {Pratt}\ \emph {et~al.}(1990)\citenamefont {Pratt},
  \citenamefont {Csorgo},\ and\ \citenamefont {Zimanyi}}]{Pratt:1990zq}%
  \BibitemOpen
  \bibfield  {author} {\bibinfo {author} {\bibfnamefont {S.}~\bibnamefont
  {Pratt}}, \bibinfo {author} {\bibfnamefont {T.}~\bibnamefont {Csorgo}},\ and\
  \bibinfo {author} {\bibfnamefont {J.}~\bibnamefont {Zimanyi}},\ }\href
  {https://doi.org/10.1103/PhysRevC.42.2646} {\bibfield  {journal} {\bibinfo
  {journal} {Phys. Rev. C}\ }\textbf {\bibinfo {volume} {42}},\ \bibinfo
  {pages} {2646} (\bibinfo {year} {1990})}\BibitemShut {NoStop}%
\bibitem [{\citenamefont {Bauer}\ \emph {et~al.}(1992)\citenamefont {Bauer},
  \citenamefont {Gelbke},\ and\ \citenamefont {Pratt}}]{Bauer:1992ffu}%
  \BibitemOpen
  \bibfield  {author} {\bibinfo {author} {\bibfnamefont {W.}~\bibnamefont
  {Bauer}}, \bibinfo {author} {\bibfnamefont {C.~K.}\ \bibnamefont {Gelbke}},\
  and\ \bibinfo {author} {\bibfnamefont {S.}~\bibnamefont {Pratt}},\ }\href
  {https://doi.org/10.1146/annurev.ns.42.120192.000453} {\bibfield  {journal}
  {\bibinfo  {journal} {Ann. Rev. Nucl. Part. Sci.}\ }\textbf {\bibinfo
  {volume} {42}},\ \bibinfo {pages} {77} (\bibinfo {year} {1992})}\BibitemShut
  {NoStop}%
\bibitem [{\citenamefont {Morita}\ \emph {et~al.}(2015)\citenamefont {Morita},
  \citenamefont {Furumoto},\ and\ \citenamefont {Ohnishi}}]{Morita:2014kza}%
  \BibitemOpen
  \bibfield  {author} {\bibinfo {author} {\bibfnamefont {K.}~\bibnamefont
  {Morita}}, \bibinfo {author} {\bibfnamefont {T.}~\bibnamefont {Furumoto}},\
  and\ \bibinfo {author} {\bibfnamefont {A.}~\bibnamefont {Ohnishi}},\ }\href
  {https://doi.org/10.1103/PhysRevC.91.024916} {\bibfield  {journal} {\bibinfo
  {journal} {Phys. Rev. C}\ }\textbf {\bibinfo {volume} {91}},\ \bibinfo
  {pages} {024916} (\bibinfo {year} {2015})},\ \Eprint
  {https://arxiv.org/abs/1408.6682} {arXiv:1408.6682 [nucl-th]} \BibitemShut
  {NoStop}%
\bibitem [{\citenamefont {Ohnishi}\ \emph {et~al.}(2016)\citenamefont
  {Ohnishi}, \citenamefont {Morita}, \citenamefont {Miyahara},\ and\
  \citenamefont {Hyodo}}]{Ohnishi:2016elb}%
  \BibitemOpen
  \bibfield  {author} {\bibinfo {author} {\bibfnamefont {A.}~\bibnamefont
  {Ohnishi}}, \bibinfo {author} {\bibfnamefont {K.}~\bibnamefont {Morita}},
  \bibinfo {author} {\bibfnamefont {K.}~\bibnamefont {Miyahara}},\ and\
  \bibinfo {author} {\bibfnamefont {T.}~\bibnamefont {Hyodo}},\ }\href
  {https://doi.org/10.1016/j.nuclphysa.2016.05.010} {\bibfield  {journal}
  {\bibinfo  {journal} {Nucl. Phys. A}\ }\textbf {\bibinfo {volume} {954}},\
  \bibinfo {pages} {294} (\bibinfo {year} {2016})},\ \Eprint
  {https://arxiv.org/abs/1603.05761} {arXiv:1603.05761 [nucl-th]} \BibitemShut
  {NoStop}%
\bibitem [{\citenamefont {Morita}\ \emph {et~al.}(2016)\citenamefont {Morita},
  \citenamefont {Ohnishi}, \citenamefont {Etminan},\ and\ \citenamefont
  {Hatsuda}}]{Morita:2016auo}%
  \BibitemOpen
  \bibfield  {author} {\bibinfo {author} {\bibfnamefont {K.}~\bibnamefont
  {Morita}}, \bibinfo {author} {\bibfnamefont {A.}~\bibnamefont {Ohnishi}},
  \bibinfo {author} {\bibfnamefont {F.}~\bibnamefont {Etminan}},\ and\ \bibinfo
  {author} {\bibfnamefont {T.}~\bibnamefont {Hatsuda}},\ }\href
  {https://doi.org/10.1103/PhysRevC.94.031901} {\bibfield  {journal} {\bibinfo
  {journal} {Phys. Rev. C}\ }\textbf {\bibinfo {volume} {94}},\ \bibinfo
  {pages} {031901} (\bibinfo {year} {2016})},\ \bibinfo {note} {[Erratum:
  Phys.Rev.C 100, 069902 (2019)]},\ \Eprint {https://arxiv.org/abs/1605.06765}
  {arXiv:1605.06765 [hep-ph]} \BibitemShut {NoStop}%
\bibitem [{\citenamefont {Hatsuda}\ \emph {et~al.}(2017)\citenamefont
  {Hatsuda}, \citenamefont {Morita}, \citenamefont {Ohnishi},\ and\
  \citenamefont {Sasaki}}]{Hatsuda:2017uxk}%
  \BibitemOpen
  \bibfield  {author} {\bibinfo {author} {\bibfnamefont {T.}~\bibnamefont
  {Hatsuda}}, \bibinfo {author} {\bibfnamefont {K.}~\bibnamefont {Morita}},
  \bibinfo {author} {\bibfnamefont {A.}~\bibnamefont {Ohnishi}},\ and\ \bibinfo
  {author} {\bibfnamefont {K.}~\bibnamefont {Sasaki}},\ }\href
  {https://doi.org/10.1016/j.nuclphysa.2017.04.041} {\bibfield  {journal}
  {\bibinfo  {journal} {Nucl. Phys. A}\ }\textbf {\bibinfo {volume} {967}},\
  \bibinfo {pages} {856} (\bibinfo {year} {2017})},\ \Eprint
  {https://arxiv.org/abs/1704.05225} {arXiv:1704.05225 [nucl-th]} \BibitemShut
  {NoStop}%
\bibitem [{\citenamefont {Mihaylov}\ \emph {et~al.}(2018)\citenamefont
  {Mihaylov}, \citenamefont {Mantovani~Sarti}, \citenamefont {Arnold},
  \citenamefont {Fabbietti}, \citenamefont {Hohlweger},\ and\ \citenamefont
  {Mathis}}]{Mihaylov:2018rva}%
  \BibitemOpen
  \bibfield  {author} {\bibinfo {author} {\bibfnamefont {D.~L.}\ \bibnamefont
  {Mihaylov}}, \bibinfo {author} {\bibfnamefont {V.}~\bibnamefont
  {Mantovani~Sarti}}, \bibinfo {author} {\bibfnamefont {O.~W.}\ \bibnamefont
  {Arnold}}, \bibinfo {author} {\bibfnamefont {L.}~\bibnamefont {Fabbietti}},
  \bibinfo {author} {\bibfnamefont {B.}~\bibnamefont {Hohlweger}},\ and\
  \bibinfo {author} {\bibfnamefont {A.~M.}\ \bibnamefont {Mathis}},\ }\href
  {https://doi.org/10.1140/epjc/s10052-018-5859-0} {\bibfield  {journal}
  {\bibinfo  {journal} {Eur. Phys. J. C}\ }\textbf {\bibinfo {volume} {78}},\
  \bibinfo {pages} {394} (\bibinfo {year} {2018})},\ \Eprint
  {https://arxiv.org/abs/1802.08481} {arXiv:1802.08481 [hep-ph]} \BibitemShut
  {NoStop}%
\bibitem [{\citenamefont {Haidenbauer}(2019)}]{Haidenbauer:2018jvl}%
  \BibitemOpen
  \bibfield  {author} {\bibinfo {author} {\bibfnamefont {J.}~\bibnamefont
  {Haidenbauer}},\ }\href {https://doi.org/10.1016/j.nuclphysa.2018.10.090}
  {\bibfield  {journal} {\bibinfo  {journal} {Nucl. Phys. A}\ }\textbf
  {\bibinfo {volume} {981}},\ \bibinfo {pages} {1} (\bibinfo {year} {2019})},\
  \Eprint {https://arxiv.org/abs/1808.05049} {arXiv:1808.05049 [hep-ph]}
  \BibitemShut {NoStop}%
\bibitem [{\citenamefont {Morita}\ \emph {et~al.}(2020)\citenamefont {Morita},
  \citenamefont {Gongyo}, \citenamefont {Hatsuda}, \citenamefont {Hyodo},
  \citenamefont {Kamiya},\ and\ \citenamefont {Ohnishi}}]{Morita:2019rph}%
  \BibitemOpen
  \bibfield  {author} {\bibinfo {author} {\bibfnamefont {K.}~\bibnamefont
  {Morita}}, \bibinfo {author} {\bibfnamefont {S.}~\bibnamefont {Gongyo}},
  \bibinfo {author} {\bibfnamefont {T.}~\bibnamefont {Hatsuda}}, \bibinfo
  {author} {\bibfnamefont {T.}~\bibnamefont {Hyodo}}, \bibinfo {author}
  {\bibfnamefont {Y.}~\bibnamefont {Kamiya}},\ and\ \bibinfo {author}
  {\bibfnamefont {A.}~\bibnamefont {Ohnishi}},\ }\href
  {https://doi.org/10.1103/PhysRevC.101.015201} {\bibfield  {journal} {\bibinfo
   {journal} {Phys. Rev. C}\ }\textbf {\bibinfo {volume} {101}},\ \bibinfo
  {pages} {015201} (\bibinfo {year} {2020})},\ \Eprint
  {https://arxiv.org/abs/1908.05414} {arXiv:1908.05414 [nucl-th]} \BibitemShut
  {NoStop}%
\bibitem [{\citenamefont {Kamiya}\ \emph {et~al.}(2020)\citenamefont {Kamiya},
  \citenamefont {Hyodo}, \citenamefont {Morita}, \citenamefont {Ohnishi},\ and\
  \citenamefont {Weise}}]{Kamiya:2019uiw}%
  \BibitemOpen
  \bibfield  {author} {\bibinfo {author} {\bibfnamefont {Y.}~\bibnamefont
  {Kamiya}}, \bibinfo {author} {\bibfnamefont {T.}~\bibnamefont {Hyodo}},
  \bibinfo {author} {\bibfnamefont {K.}~\bibnamefont {Morita}}, \bibinfo
  {author} {\bibfnamefont {A.}~\bibnamefont {Ohnishi}},\ and\ \bibinfo {author}
  {\bibfnamefont {W.}~\bibnamefont {Weise}},\ }\href
  {https://doi.org/10.1103/PhysRevLett.124.132501} {\bibfield  {journal}
  {\bibinfo  {journal} {Phys. Rev. Lett.}\ }\textbf {\bibinfo {volume} {124}},\
  \bibinfo {pages} {132501} (\bibinfo {year} {2020})},\ \Eprint
  {https://arxiv.org/abs/1911.01041} {arXiv:1911.01041 [nucl-th]} \BibitemShut
  {NoStop}%
\bibitem [{\citenamefont {Kamiya}\ \emph
  {et~al.}(2022{\natexlab{a}})\citenamefont {Kamiya}, \citenamefont {Sasaki},
  \citenamefont {Fukui}, \citenamefont {Hyodo}, \citenamefont {Morita},
  \citenamefont {Ogata}, \citenamefont {Ohnishi},\ and\ \citenamefont
  {Hatsuda}}]{Kamiya:2021hdb}%
  \BibitemOpen
  \bibfield  {author} {\bibinfo {author} {\bibfnamefont {Y.}~\bibnamefont
  {Kamiya}}, \bibinfo {author} {\bibfnamefont {K.}~\bibnamefont {Sasaki}},
  \bibinfo {author} {\bibfnamefont {T.}~\bibnamefont {Fukui}}, \bibinfo
  {author} {\bibfnamefont {T.}~\bibnamefont {Hyodo}}, \bibinfo {author}
  {\bibfnamefont {K.}~\bibnamefont {Morita}}, \bibinfo {author} {\bibfnamefont
  {K.}~\bibnamefont {Ogata}}, \bibinfo {author} {\bibfnamefont
  {A.}~\bibnamefont {Ohnishi}},\ and\ \bibinfo {author} {\bibfnamefont
  {T.}~\bibnamefont {Hatsuda}},\ }\href
  {https://doi.org/10.1103/PhysRevC.105.014915} {\bibfield  {journal} {\bibinfo
   {journal} {Phys. Rev. C}\ }\textbf {\bibinfo {volume} {105}},\ \bibinfo
  {pages} {014915} (\bibinfo {year} {2022}{\natexlab{a}})},\ \Eprint
  {https://arxiv.org/abs/2108.09644} {arXiv:2108.09644 [hep-ph]} \BibitemShut
  {NoStop}%
\bibitem [{\citenamefont {Kamiya}\ \emph
  {et~al.}(2022{\natexlab{b}})\citenamefont {Kamiya}, \citenamefont {Hyodo},\
  and\ \citenamefont {Ohnishi}}]{Kamiya:2022thy}%
  \BibitemOpen
  \bibfield  {author} {\bibinfo {author} {\bibfnamefont {Y.}~\bibnamefont
  {Kamiya}}, \bibinfo {author} {\bibfnamefont {T.}~\bibnamefont {Hyodo}},\ and\
  \bibinfo {author} {\bibfnamefont {A.}~\bibnamefont {Ohnishi}},\ }\href
  {https://doi.org/10.1140/epja/s10050-022-00782-y} {\bibfield  {journal}
  {\bibinfo  {journal} {Eur. Phys. J. A}\ }\textbf {\bibinfo {volume} {58}},\
  \bibinfo {pages} {131} (\bibinfo {year} {2022}{\natexlab{b}})},\ \Eprint
  {https://arxiv.org/abs/2203.13814} {arXiv:2203.13814 [hep-ph]} \BibitemShut
  {NoStop}%
\bibitem [{\citenamefont {Vida\~na}\ \emph {et~al.}(2023)\citenamefont
  {Vida\~na}, \citenamefont {Feijoo}, \citenamefont {Albaladejo}, \citenamefont
  {Nieves},\ and\ \citenamefont {Oset}}]{Vidana:2023olz}%
  \BibitemOpen
  \bibfield  {author} {\bibinfo {author} {\bibfnamefont {I.}~\bibnamefont
  {Vida\~na}}, \bibinfo {author} {\bibfnamefont {A.}~\bibnamefont {Feijoo}},
  \bibinfo {author} {\bibfnamefont {M.}~\bibnamefont {Albaladejo}}, \bibinfo
  {author} {\bibfnamefont {J.}~\bibnamefont {Nieves}},\ and\ \bibinfo {author}
  {\bibfnamefont {E.}~\bibnamefont {Oset}},\ }\href
  {https://doi.org/10.1016/j.physletb.2023.138201} {\bibfield  {journal}
  {\bibinfo  {journal} {Phys. Lett. B}\ }\textbf {\bibinfo {volume} {846}},\
  \bibinfo {pages} {138201} (\bibinfo {year} {2023})},\ \Eprint
  {https://arxiv.org/abs/2303.06079} {arXiv:2303.06079 [hep-ph]} \BibitemShut
  {NoStop}%
\bibitem [{\citenamefont {Liu}\ \emph {et~al.}(2023)\citenamefont {Liu},
  \citenamefont {Lu},\ and\ \citenamefont {Geng}}]{Liu:2023uly}%
  \BibitemOpen
  \bibfield  {author} {\bibinfo {author} {\bibfnamefont {Z.-W.}\ \bibnamefont
  {Liu}}, \bibinfo {author} {\bibfnamefont {J.-X.}\ \bibnamefont {Lu}},\ and\
  \bibinfo {author} {\bibfnamefont {L.-S.}\ \bibnamefont {Geng}},\ }\href
  {https://doi.org/10.1103/PhysRevD.107.074019} {\bibfield  {journal} {\bibinfo
   {journal} {Phys. Rev. D}\ }\textbf {\bibinfo {volume} {107}},\ \bibinfo
  {pages} {074019} (\bibinfo {year} {2023})},\ \Eprint
  {https://arxiv.org/abs/2302.01046} {arXiv:2302.01046 [hep-ph]} \BibitemShut
  {NoStop}%
\bibitem [{\citenamefont {Albaladejo}\ \emph
  {et~al.}(2023{\natexlab{a}})\citenamefont {Albaladejo}, \citenamefont
  {Nieves},\ and\ \citenamefont {Ruiz-Arriola}}]{Albaladejo:2023pzq}%
  \BibitemOpen
  \bibfield  {author} {\bibinfo {author} {\bibfnamefont {M.}~\bibnamefont
  {Albaladejo}}, \bibinfo {author} {\bibfnamefont {J.}~\bibnamefont {Nieves}},\
  and\ \bibinfo {author} {\bibfnamefont {E.}~\bibnamefont {Ruiz-Arriola}},\
  }\href {https://doi.org/10.1103/PhysRevD.108.014020} {\bibfield  {journal}
  {\bibinfo  {journal} {Phys. Rev. D}\ }\textbf {\bibinfo {volume} {108}},\
  \bibinfo {pages} {014020} (\bibinfo {year} {2023}{\natexlab{a}})},\ \Eprint
  {https://arxiv.org/abs/2304.03107} {arXiv:2304.03107 [hep-ph]} \BibitemShut
  {NoStop}%
\bibitem [{\citenamefont {Torres-Rincon}\ \emph {et~al.}(2023)\citenamefont
  {Torres-Rincon}, \citenamefont {Ramos},\ and\ \citenamefont
  {Tolos}}]{Torres-Rincon:2023qll}%
  \BibitemOpen
  \bibfield  {author} {\bibinfo {author} {\bibfnamefont {J.~M.}\ \bibnamefont
  {Torres-Rincon}}, \bibinfo {author} {\bibfnamefont {A.}~\bibnamefont
  {Ramos}},\ and\ \bibinfo {author} {\bibfnamefont {L.}~\bibnamefont {Tolos}},\
  }\href {https://doi.org/10.1103/PhysRevD.108.096008} {\bibfield  {journal}
  {\bibinfo  {journal} {Phys. Rev. D}\ }\textbf {\bibinfo {volume} {108}},\
  \bibinfo {pages} {096008} (\bibinfo {year} {2023})},\ \Eprint
  {https://arxiv.org/abs/2307.02102} {arXiv:2307.02102 [hep-ph]} \BibitemShut
  {NoStop}%
\bibitem [{\citenamefont {Sarti}\ \emph {et~al.}(2024)\citenamefont {Sarti},
  \citenamefont {Feijoo}, \citenamefont {Vida\~na}, \citenamefont {Ramos},
  \citenamefont {Giacosa}, \citenamefont {Hyodo},\ and\ \citenamefont
  {Kamiya}}]{Sarti:2023wlg}%
  \BibitemOpen
  \bibfield  {author} {\bibinfo {author} {\bibfnamefont {V.~M.}\ \bibnamefont
  {Sarti}}, \bibinfo {author} {\bibfnamefont {A.}~\bibnamefont {Feijoo}},
  \bibinfo {author} {\bibfnamefont {I.}~\bibnamefont {Vida\~na}}, \bibinfo
  {author} {\bibfnamefont {A.}~\bibnamefont {Ramos}}, \bibinfo {author}
  {\bibfnamefont {F.}~\bibnamefont {Giacosa}}, \bibinfo {author} {\bibfnamefont
  {T.}~\bibnamefont {Hyodo}},\ and\ \bibinfo {author} {\bibfnamefont
  {Y.}~\bibnamefont {Kamiya}},\ }\href
  {https://doi.org/10.1103/PhysRevD.110.L011505} {\bibfield  {journal}
  {\bibinfo  {journal} {Phys. Rev. D}\ }\textbf {\bibinfo {volume} {110}},\
  \bibinfo {pages} {L011505} (\bibinfo {year} {2024})},\ \Eprint
  {https://arxiv.org/abs/2309.08756} {arXiv:2309.08756 [hep-ph]} \BibitemShut
  {NoStop}%
\bibitem [{\citenamefont {Molina}\ \emph
  {et~al.}(2024{\natexlab{a}})\citenamefont {Molina}, \citenamefont {Liu},
  \citenamefont {Geng},\ and\ \citenamefont {Oset}}]{Molina:2023oeu}%
  \BibitemOpen
  \bibfield  {author} {\bibinfo {author} {\bibfnamefont {R.}~\bibnamefont
  {Molina}}, \bibinfo {author} {\bibfnamefont {Z.-W.}\ \bibnamefont {Liu}},
  \bibinfo {author} {\bibfnamefont {L.-S.}\ \bibnamefont {Geng}},\ and\
  \bibinfo {author} {\bibfnamefont {E.}~\bibnamefont {Oset}},\ }\href
  {https://doi.org/10.1140/epjc/s10052-024-12694-w} {\bibfield  {journal}
  {\bibinfo  {journal} {Eur. Phys. J. C}\ }\textbf {\bibinfo {volume} {84}},\
  \bibinfo {pages} {328} (\bibinfo {year} {2024}{\natexlab{a}})},\ \Eprint
  {https://arxiv.org/abs/2312.11993} {arXiv:2312.11993 [hep-ph]} \BibitemShut
  {NoStop}%
\bibitem [{\citenamefont {Molina}\ \emph
  {et~al.}(2024{\natexlab{b}})\citenamefont {Molina}, \citenamefont {Xiao},
  \citenamefont {Liang},\ and\ \citenamefont {Oset}}]{Molina:2023jov}%
  \BibitemOpen
  \bibfield  {author} {\bibinfo {author} {\bibfnamefont {R.}~\bibnamefont
  {Molina}}, \bibinfo {author} {\bibfnamefont {C.-W.}\ \bibnamefont {Xiao}},
  \bibinfo {author} {\bibfnamefont {W.-H.}\ \bibnamefont {Liang}},\ and\
  \bibinfo {author} {\bibfnamefont {E.}~\bibnamefont {Oset}},\ }\href
  {https://doi.org/10.1103/PhysRevD.109.054002} {\bibfield  {journal} {\bibinfo
   {journal} {Phys. Rev. D}\ }\textbf {\bibinfo {volume} {109}},\ \bibinfo
  {pages} {054002} (\bibinfo {year} {2024}{\natexlab{b}})},\ \Eprint
  {https://arxiv.org/abs/2310.12593} {arXiv:2310.12593 [hep-ph]} \BibitemShut
  {NoStop}%
\bibitem [{\citenamefont {Liu}\ \emph {et~al.}(2024)\citenamefont {Liu},
  \citenamefont {Lu}, \citenamefont {Liu},\ and\ \citenamefont
  {Geng}}]{Liu:2024nac}%
  \BibitemOpen
  \bibfield  {author} {\bibinfo {author} {\bibfnamefont {Z.-W.}\ \bibnamefont
  {Liu}}, \bibinfo {author} {\bibfnamefont {J.-X.}\ \bibnamefont {Lu}},
  \bibinfo {author} {\bibfnamefont {M.-Z.}\ \bibnamefont {Liu}},\ and\ \bibinfo
  {author} {\bibfnamefont {L.-S.}\ \bibnamefont {Geng}},\ }\href@noop {} {\
  (\bibinfo {year} {2024})},\ \Eprint {https://arxiv.org/abs/2404.18607}
  {arXiv:2404.18607 [hep-ph]} \BibitemShut {NoStop}%
\bibitem [{\citenamefont {Feijoo}\ \emph
  {et~al.}(2024{\natexlab{a}})\citenamefont {Feijoo}, \citenamefont
  {Korwieser},\ and\ \citenamefont {Fabbietti}}]{Feijoo:2024bvn}%
  \BibitemOpen
  \bibfield  {author} {\bibinfo {author} {\bibfnamefont {A.}~\bibnamefont
  {Feijoo}}, \bibinfo {author} {\bibfnamefont {M.}~\bibnamefont {Korwieser}},\
  and\ \bibinfo {author} {\bibfnamefont {L.}~\bibnamefont {Fabbietti}},\
  }\href@noop {} {\  (\bibinfo {year} {2024}{\natexlab{a}})},\ \Eprint
  {https://arxiv.org/abs/2407.01128} {arXiv:2407.01128 [hep-ph]} \BibitemShut
  {NoStop}%
\bibitem [{\citenamefont {Ikeno}\ \emph {et~al.}(2023)\citenamefont {Ikeno},
  \citenamefont {Toledo},\ and\ \citenamefont {Oset}}]{Ikeno:2023ojl}%
  \BibitemOpen
  \bibfield  {author} {\bibinfo {author} {\bibfnamefont {N.}~\bibnamefont
  {Ikeno}}, \bibinfo {author} {\bibfnamefont {G.}~\bibnamefont {Toledo}},\ and\
  \bibinfo {author} {\bibfnamefont {E.}~\bibnamefont {Oset}},\ }\href
  {https://doi.org/10.1016/j.physletb.2023.138281} {\bibfield  {journal}
  {\bibinfo  {journal} {Phys. Lett. B}\ }\textbf {\bibinfo {volume} {847}},\
  \bibinfo {pages} {138281} (\bibinfo {year} {2023})},\ \Eprint
  {https://arxiv.org/abs/2305.16431} {arXiv:2305.16431 [hep-ph]} \BibitemShut
  {NoStop}%
\bibitem [{\citenamefont {Albaladejo}\ \emph
  {et~al.}(2023{\natexlab{b}})\citenamefont {Albaladejo}, \citenamefont
  {Feijoo}, \citenamefont {Vida\~na}, \citenamefont {Nieves},\ and\
  \citenamefont {Oset}}]{Albaladejo:2023wmv}%
  \BibitemOpen
  \bibfield  {author} {\bibinfo {author} {\bibfnamefont {M.}~\bibnamefont
  {Albaladejo}}, \bibinfo {author} {\bibfnamefont {A.}~\bibnamefont {Feijoo}},
  \bibinfo {author} {\bibfnamefont {I.}~\bibnamefont {Vida\~na}}, \bibinfo
  {author} {\bibfnamefont {J.}~\bibnamefont {Nieves}},\ and\ \bibinfo {author}
  {\bibfnamefont {E.}~\bibnamefont {Oset}},\ }\href@noop {} {\  (\bibinfo
  {year} {2023}{\natexlab{b}})},\ \Eprint {https://arxiv.org/abs/2307.09873}
  {arXiv:2307.09873 [hep-ph]} \BibitemShut {NoStop}%
\bibitem [{\citenamefont {Choi}\ \emph {et~al.}(2003)\citenamefont {Choi} \emph
  {et~al.}}]{Belle:2003nnu}%
  \BibitemOpen
  \bibfield  {author} {\bibinfo {author} {\bibfnamefont {S.~K.}\ \bibnamefont
  {Choi}} \emph {et~al.} (\bibinfo {collaboration} {Belle}),\ }\href
  {https://doi.org/10.1103/PhysRevLett.91.262001} {\bibfield  {journal}
  {\bibinfo  {journal} {Phys. Rev. Lett.}\ }\textbf {\bibinfo {volume} {91}},\
  \bibinfo {pages} {262001} (\bibinfo {year} {2003})},\ \Eprint
  {https://arxiv.org/abs/hep-ex/0309032} {arXiv:hep-ex/0309032} \BibitemShut
  {NoStop}%
\bibitem [{\citenamefont {Aaij}\ \emph {et~al.}(2015)\citenamefont {Aaij} \emph
  {et~al.}}]{LHCb:2015yax}%
  \BibitemOpen
  \bibfield  {author} {\bibinfo {author} {\bibfnamefont {R.}~\bibnamefont
  {Aaij}} \emph {et~al.} (\bibinfo {collaboration} {LHCb}),\ }\href
  {https://doi.org/10.1103/PhysRevLett.115.072001} {\bibfield  {journal}
  {\bibinfo  {journal} {Phys. Rev. Lett.}\ }\textbf {\bibinfo {volume} {115}},\
  \bibinfo {pages} {072001} (\bibinfo {year} {2015})},\ \Eprint
  {https://arxiv.org/abs/1507.03414} {arXiv:1507.03414 [hep-ex]} \BibitemShut
  {NoStop}%
\bibitem [{\citenamefont {Aaij}\ \emph {et~al.}(2016)\citenamefont {Aaij} \emph
  {et~al.}}]{LHCb:2015qvk}%
  \BibitemOpen
  \bibfield  {author} {\bibinfo {author} {\bibfnamefont {R.}~\bibnamefont
  {Aaij}} \emph {et~al.} (\bibinfo {collaboration} {LHCb}),\ }\href
  {https://doi.org/10.1088/1674-1137/40/1/011001} {\bibfield  {journal}
  {\bibinfo  {journal} {Chin. Phys. C}\ }\textbf {\bibinfo {volume} {40}},\
  \bibinfo {pages} {011001} (\bibinfo {year} {2016})},\ \Eprint
  {https://arxiv.org/abs/1509.00292} {arXiv:1509.00292 [hep-ex]} \BibitemShut
  {NoStop}%
\bibitem [{\citenamefont {Aaij}\ \emph {et~al.}(2019)\citenamefont {Aaij} \emph
  {et~al.}}]{LHCb:2019kea}%
  \BibitemOpen
  \bibfield  {author} {\bibinfo {author} {\bibfnamefont {R.}~\bibnamefont
  {Aaij}} \emph {et~al.} (\bibinfo {collaboration} {LHCb}),\ }\href
  {https://doi.org/10.1103/PhysRevLett.122.222001} {\bibfield  {journal}
  {\bibinfo  {journal} {Phys. Rev. Lett.}\ }\textbf {\bibinfo {volume} {122}},\
  \bibinfo {pages} {222001} (\bibinfo {year} {2019})},\ \Eprint
  {https://arxiv.org/abs/1904.03947} {arXiv:1904.03947 [hep-ex]} \BibitemShut
  {NoStop}%
\bibitem [{\citenamefont {Aaij}\ \emph
  {et~al.}(2021{\natexlab{a}})\citenamefont {Aaij} \emph
  {et~al.}}]{LHCb:2020jpq}%
  \BibitemOpen
  \bibfield  {author} {\bibinfo {author} {\bibfnamefont {R.}~\bibnamefont
  {Aaij}} \emph {et~al.} (\bibinfo {collaboration} {LHCb}),\ }\href
  {https://doi.org/10.1016/j.scib.2021.02.030} {\bibfield  {journal} {\bibinfo
  {journal} {Sci. Bull.}\ }\textbf {\bibinfo {volume} {66}},\ \bibinfo {pages}
  {1278} (\bibinfo {year} {2021}{\natexlab{a}})},\ \Eprint
  {https://arxiv.org/abs/2012.10380} {arXiv:2012.10380 [hep-ex]} \BibitemShut
  {NoStop}%
\bibitem [{\citenamefont {Aaij}\ \emph
  {et~al.}(2023{\natexlab{a}})\citenamefont {Aaij} \emph
  {et~al.}}]{LHCb:2022ogu}%
  \BibitemOpen
  \bibfield  {author} {\bibinfo {author} {\bibfnamefont {R.}~\bibnamefont
  {Aaij}} \emph {et~al.} (\bibinfo {collaboration} {LHCb}),\ }\href
  {https://doi.org/10.1103/PhysRevLett.131.031901} {\bibfield  {journal}
  {\bibinfo  {journal} {Phys. Rev. Lett.}\ }\textbf {\bibinfo {volume} {131}},\
  \bibinfo {pages} {031901} (\bibinfo {year} {2023}{\natexlab{a}})},\ \Eprint
  {https://arxiv.org/abs/2210.10346} {arXiv:2210.10346 [hep-ex]} \BibitemShut
  {NoStop}%
\bibitem [{\citenamefont {Ablikim}\ \emph {et~al.}(2021)\citenamefont {Ablikim}
  \emph {et~al.}}]{BESIII:2020qkh}%
  \BibitemOpen
  \bibfield  {author} {\bibinfo {author} {\bibfnamefont {M.}~\bibnamefont
  {Ablikim}} \emph {et~al.} (\bibinfo {collaboration} {BESIII}),\ }\href
  {https://doi.org/10.1103/PhysRevLett.126.102001} {\bibfield  {journal}
  {\bibinfo  {journal} {Phys. Rev. Lett.}\ }\textbf {\bibinfo {volume} {126}},\
  \bibinfo {pages} {102001} (\bibinfo {year} {2021})},\ \Eprint
  {https://arxiv.org/abs/2011.07855} {arXiv:2011.07855 [hep-ex]} \BibitemShut
  {NoStop}%
\bibitem [{\citenamefont {Aaij}\ \emph
  {et~al.}(2021{\natexlab{b}})\citenamefont {Aaij} \emph
  {et~al.}}]{LHCb:2021uow}%
  \BibitemOpen
  \bibfield  {author} {\bibinfo {author} {\bibfnamefont {R.}~\bibnamefont
  {Aaij}} \emph {et~al.} (\bibinfo {collaboration} {LHCb}),\ }\href
  {https://doi.org/10.1103/PhysRevLett.127.082001} {\bibfield  {journal}
  {\bibinfo  {journal} {Phys. Rev. Lett.}\ }\textbf {\bibinfo {volume} {127}},\
  \bibinfo {pages} {082001} (\bibinfo {year} {2021}{\natexlab{b}})},\ \Eprint
  {https://arxiv.org/abs/2103.01803} {arXiv:2103.01803 [hep-ex]} \BibitemShut
  {NoStop}%
\bibitem [{\citenamefont {Aaij}\ \emph
  {et~al.}(2023{\natexlab{b}})\citenamefont {Aaij} \emph
  {et~al.}}]{LHCb:2022dvn}%
  \BibitemOpen
  \bibfield  {author} {\bibinfo {author} {\bibfnamefont {R.}~\bibnamefont
  {Aaij}} \emph {et~al.} (\bibinfo {collaboration} {LHCb}),\ }\href
  {https://doi.org/10.1103/PhysRevD.108.034012} {\bibfield  {journal} {\bibinfo
   {journal} {Phys. Rev. D}\ }\textbf {\bibinfo {volume} {108}},\ \bibinfo
  {pages} {034012} (\bibinfo {year} {2023}{\natexlab{b}})},\ \Eprint
  {https://arxiv.org/abs/2211.05034} {arXiv:2211.05034 [hep-ex]} \BibitemShut
  {NoStop}%
\bibitem [{\citenamefont {Aaij}\ \emph
  {et~al.}(2023{\natexlab{c}})\citenamefont {Aaij} \emph
  {et~al.}}]{LHCb:2022aki}%
  \BibitemOpen
  \bibfield  {author} {\bibinfo {author} {\bibfnamefont {R.}~\bibnamefont
  {Aaij}} \emph {et~al.} (\bibinfo {collaboration} {LHCb}),\ }\href
  {https://doi.org/10.1103/PhysRevLett.131.071901} {\bibfield  {journal}
  {\bibinfo  {journal} {Phys. Rev. Lett.}\ }\textbf {\bibinfo {volume} {131}},\
  \bibinfo {pages} {071901} (\bibinfo {year} {2023}{\natexlab{c}})},\ \Eprint
  {https://arxiv.org/abs/2210.15153} {arXiv:2210.15153 [hep-ex]} \BibitemShut
  {NoStop}%
\bibitem [{\citenamefont {Aaij}\ \emph
  {et~al.}(2022{\natexlab{a}})\citenamefont {Aaij} \emph
  {et~al.}}]{LHCb:2021vvq}%
  \BibitemOpen
  \bibfield  {author} {\bibinfo {author} {\bibfnamefont {R.}~\bibnamefont
  {Aaij}} \emph {et~al.} (\bibinfo {collaboration} {LHCb}),\ }\href
  {https://doi.org/10.1038/s41567-022-01614-y} {\bibfield  {journal} {\bibinfo
  {journal} {Nature Phys.}\ }\textbf {\bibinfo {volume} {18}},\ \bibinfo
  {pages} {751} (\bibinfo {year} {2022}{\natexlab{a}})},\ \Eprint
  {https://arxiv.org/abs/2109.01038} {arXiv:2109.01038 [hep-ex]} \BibitemShut
  {NoStop}%
\bibitem [{\citenamefont {Aaij}\ \emph
  {et~al.}(2022{\natexlab{b}})\citenamefont {Aaij} \emph
  {et~al.}}]{LHCb:2021auc}%
  \BibitemOpen
  \bibfield  {author} {\bibinfo {author} {\bibfnamefont {R.}~\bibnamefont
  {Aaij}} \emph {et~al.} (\bibinfo {collaboration} {LHCb}),\ }\href
  {https://doi.org/10.1038/s41467-022-30206-w} {\bibfield  {journal} {\bibinfo
  {journal} {Nature Commun.}\ }\textbf {\bibinfo {volume} {13}},\ \bibinfo
  {pages} {3351} (\bibinfo {year} {2022}{\natexlab{b}})},\ \Eprint
  {https://arxiv.org/abs/2109.01056} {arXiv:2109.01056 [hep-ex]} \BibitemShut
  {NoStop}%
\bibitem [{\citenamefont {Guo}\ \emph {et~al.}(2018)\citenamefont {Guo},
  \citenamefont {Hanhart}, \citenamefont {Mei\ss{}ner}, \citenamefont {Wang},
  \citenamefont {Zhao},\ and\ \citenamefont {Zou}}]{Guo:2017jvc}%
  \BibitemOpen
  \bibfield  {author} {\bibinfo {author} {\bibfnamefont {F.-K.}\ \bibnamefont
  {Guo}}, \bibinfo {author} {\bibfnamefont {C.}~\bibnamefont {Hanhart}},
  \bibinfo {author} {\bibfnamefont {U.-G.}\ \bibnamefont {Mei\ss{}ner}},
  \bibinfo {author} {\bibfnamefont {Q.}~\bibnamefont {Wang}}, \bibinfo {author}
  {\bibfnamefont {Q.}~\bibnamefont {Zhao}},\ and\ \bibinfo {author}
  {\bibfnamefont {B.-S.}\ \bibnamefont {Zou}},\ }\href
  {https://doi.org/10.1103/RevModPhys.90.015004} {\bibfield  {journal}
  {\bibinfo  {journal} {Rev. Mod. Phys.}\ }\textbf {\bibinfo {volume} {90}},\
  \bibinfo {pages} {015004} (\bibinfo {year} {2018})},\ \Eprint
  {https://arxiv.org/abs/1705.00141} {arXiv:1705.00141 [hep-ph]} \BibitemShut
  {NoStop}%
\bibitem [{\citenamefont {Brambilla}\ \emph {et~al.}(2020)\citenamefont
  {Brambilla}, \citenamefont {Eidelman}, \citenamefont {Hanhart}, \citenamefont
  {Nefediev}, \citenamefont {Shen}, \citenamefont {Thomas}, \citenamefont
  {Vairo},\ and\ \citenamefont {Yuan}}]{Brambilla:2019esw}%
  \BibitemOpen
  \bibfield  {author} {\bibinfo {author} {\bibfnamefont {N.}~\bibnamefont
  {Brambilla}}, \bibinfo {author} {\bibfnamefont {S.}~\bibnamefont {Eidelman}},
  \bibinfo {author} {\bibfnamefont {C.}~\bibnamefont {Hanhart}}, \bibinfo
  {author} {\bibfnamefont {A.}~\bibnamefont {Nefediev}}, \bibinfo {author}
  {\bibfnamefont {C.-P.}\ \bibnamefont {Shen}}, \bibinfo {author}
  {\bibfnamefont {C.~E.}\ \bibnamefont {Thomas}}, \bibinfo {author}
  {\bibfnamefont {A.}~\bibnamefont {Vairo}},\ and\ \bibinfo {author}
  {\bibfnamefont {C.-Z.}\ \bibnamefont {Yuan}},\ }\href
  {https://doi.org/10.1016/j.physrep.2020.05.001} {\bibfield  {journal}
  {\bibinfo  {journal} {Phys. Rept.}\ }\textbf {\bibinfo {volume} {873}},\
  \bibinfo {pages} {1} (\bibinfo {year} {2020})},\ \Eprint
  {https://arxiv.org/abs/1907.07583} {arXiv:1907.07583 [hep-ex]} \BibitemShut
  {NoStop}%
\bibitem [{\citenamefont {Liu}\ \emph {et~al.}(2019)\citenamefont {Liu},
  \citenamefont {Chen}, \citenamefont {Chen}, \citenamefont {Liu},\ and\
  \citenamefont {Zhu}}]{Liu:2019zoy}%
  \BibitemOpen
  \bibfield  {author} {\bibinfo {author} {\bibfnamefont {Y.-R.}\ \bibnamefont
  {Liu}}, \bibinfo {author} {\bibfnamefont {H.-X.}\ \bibnamefont {Chen}},
  \bibinfo {author} {\bibfnamefont {W.}~\bibnamefont {Chen}}, \bibinfo {author}
  {\bibfnamefont {X.}~\bibnamefont {Liu}},\ and\ \bibinfo {author}
  {\bibfnamefont {S.-L.}\ \bibnamefont {Zhu}},\ }\href
  {https://doi.org/10.1016/j.ppnp.2019.04.003} {\bibfield  {journal} {\bibinfo
  {journal} {Prog. Part. Nucl. Phys.}\ }\textbf {\bibinfo {volume} {107}},\
  \bibinfo {pages} {237} (\bibinfo {year} {2019})},\ \Eprint
  {https://arxiv.org/abs/1903.11976} {arXiv:1903.11976 [hep-ph]} \BibitemShut
  {NoStop}%
\bibitem [{\citenamefont {Dong}\ \emph
  {et~al.}(2021{\natexlab{a}})\citenamefont {Dong}, \citenamefont {Guo},\ and\
  \citenamefont {Zou}}]{Dong:2021bvy}%
  \BibitemOpen
  \bibfield  {author} {\bibinfo {author} {\bibfnamefont {X.-K.}\ \bibnamefont
  {Dong}}, \bibinfo {author} {\bibfnamefont {F.-K.}\ \bibnamefont {Guo}},\ and\
  \bibinfo {author} {\bibfnamefont {B.-S.}\ \bibnamefont {Zou}},\ }\href
  {https://doi.org/10.1088/1572-9494/ac27a2} {\bibfield  {journal} {\bibinfo
  {journal} {Commun. Theor. Phys.}\ }\textbf {\bibinfo {volume} {73}},\
  \bibinfo {pages} {125201} (\bibinfo {year} {2021}{\natexlab{a}})},\ \Eprint
  {https://arxiv.org/abs/2108.02673} {arXiv:2108.02673 [hep-ph]} \BibitemShut
  {NoStop}%
\bibitem [{\citenamefont {Dong}\ \emph
  {et~al.}(2021{\natexlab{b}})\citenamefont {Dong}, \citenamefont {Guo},\ and\
  \citenamefont {Zou}}]{Dong:2021rpi}%
  \BibitemOpen
  \bibfield  {author} {\bibinfo {author} {\bibfnamefont {X.-K.}\ \bibnamefont
  {Dong}}, \bibinfo {author} {\bibfnamefont {F.-K.}\ \bibnamefont {Guo}},\ and\
  \bibinfo {author} {\bibfnamefont {B.~S.}\ \bibnamefont {Zou}},\ }\href
  {https://doi.org/10.1007/s00601-021-01649-6} {\bibfield  {journal} {\bibinfo
  {journal} {Few Body Syst.}\ }\textbf {\bibinfo {volume} {62}},\ \bibinfo
  {pages} {61} (\bibinfo {year} {2021}{\natexlab{b}})}\BibitemShut {NoStop}%
\bibitem [{\citenamefont {Dong}\ \emph
  {et~al.}(2021{\natexlab{c}})\citenamefont {Dong}, \citenamefont {Guo},\ and\
  \citenamefont {Zou}}]{Dong:2021juy}%
  \BibitemOpen
  \bibfield  {author} {\bibinfo {author} {\bibfnamefont {X.-K.}\ \bibnamefont
  {Dong}}, \bibinfo {author} {\bibfnamefont {F.-K.}\ \bibnamefont {Guo}},\ and\
  \bibinfo {author} {\bibfnamefont {B.-S.}\ \bibnamefont {Zou}},\ }\href
  {https://doi.org/10.13725/j.cnki.pip.2021.02.001} {\bibfield  {journal}
  {\bibinfo  {journal} {Progr. Phys.}\ }\textbf {\bibinfo {volume} {41}},\
  \bibinfo {pages} {65} (\bibinfo {year} {2021}{\natexlab{c}})},\ \Eprint
  {https://arxiv.org/abs/2101.01021} {arXiv:2101.01021 [hep-ph]} \BibitemShut
  {NoStop}%
\bibitem [{\citenamefont {Albaladejo}\ \emph {et~al.}(2022)\citenamefont
  {Albaladejo} \emph {et~al.}}]{JPAC:2021rxu}%
  \BibitemOpen
  \bibfield  {author} {\bibinfo {author} {\bibfnamefont {M.}~\bibnamefont
  {Albaladejo}} \emph {et~al.} (\bibinfo {collaboration} {JPAC}),\ }\href
  {https://doi.org/10.1016/j.ppnp.2022.103981} {\bibfield  {journal} {\bibinfo
  {journal} {Prog. Part. Nucl. Phys.}\ }\textbf {\bibinfo {volume} {127}},\
  \bibinfo {pages} {103981} (\bibinfo {year} {2022})},\ \Eprint
  {https://arxiv.org/abs/2112.13436} {arXiv:2112.13436 [hep-ph]} \BibitemShut
  {NoStop}%
\bibitem [{\citenamefont {Feijoo}\ \emph {et~al.}(2021)\citenamefont {Feijoo},
  \citenamefont {Liang},\ and\ \citenamefont {Oset}}]{Feijoo:2021ppq}%
  \BibitemOpen
  \bibfield  {author} {\bibinfo {author} {\bibfnamefont {A.}~\bibnamefont
  {Feijoo}}, \bibinfo {author} {\bibfnamefont {W.~H.}\ \bibnamefont {Liang}},\
  and\ \bibinfo {author} {\bibfnamefont {E.}~\bibnamefont {Oset}},\ }\href
  {https://doi.org/10.1103/PhysRevD.104.114015} {\bibfield  {journal} {\bibinfo
   {journal} {Phys. Rev. D}\ }\textbf {\bibinfo {volume} {104}},\ \bibinfo
  {pages} {114015} (\bibinfo {year} {2021})},\ \Eprint
  {https://arxiv.org/abs/2108.02730} {arXiv:2108.02730 [hep-ph]} \BibitemShut
  {NoStop}%
\bibitem [{\citenamefont {Ling}\ \emph {et~al.}(2022)\citenamefont {Ling},
  \citenamefont {Liu}, \citenamefont {Geng}, \citenamefont {Wang},\ and\
  \citenamefont {Xie}}]{Ling:2021bir}%
  \BibitemOpen
  \bibfield  {author} {\bibinfo {author} {\bibfnamefont {X.-Z.}\ \bibnamefont
  {Ling}}, \bibinfo {author} {\bibfnamefont {M.-Z.}\ \bibnamefont {Liu}},
  \bibinfo {author} {\bibfnamefont {L.-S.}\ \bibnamefont {Geng}}, \bibinfo
  {author} {\bibfnamefont {E.}~\bibnamefont {Wang}},\ and\ \bibinfo {author}
  {\bibfnamefont {J.-J.}\ \bibnamefont {Xie}},\ }\href
  {https://doi.org/10.1016/j.physletb.2022.136897} {\bibfield  {journal}
  {\bibinfo  {journal} {Phys. Lett. B}\ }\textbf {\bibinfo {volume} {826}},\
  \bibinfo {pages} {136897} (\bibinfo {year} {2022})},\ \Eprint
  {https://arxiv.org/abs/2108.00947} {arXiv:2108.00947 [hep-ph]} \BibitemShut
  {NoStop}%
\bibitem [{\citenamefont {Fleming}\ \emph {et~al.}(2021)\citenamefont
  {Fleming}, \citenamefont {Hodges},\ and\ \citenamefont
  {Mehen}}]{Fleming:2021wmk}%
  \BibitemOpen
  \bibfield  {author} {\bibinfo {author} {\bibfnamefont {S.}~\bibnamefont
  {Fleming}}, \bibinfo {author} {\bibfnamefont {R.}~\bibnamefont {Hodges}},\
  and\ \bibinfo {author} {\bibfnamefont {T.}~\bibnamefont {Mehen}},\ }\href
  {https://doi.org/10.1103/PhysRevD.104.116010} {\bibfield  {journal} {\bibinfo
   {journal} {Phys. Rev. D}\ }\textbf {\bibinfo {volume} {104}},\ \bibinfo
  {pages} {116010} (\bibinfo {year} {2021})},\ \Eprint
  {https://arxiv.org/abs/2109.02188} {arXiv:2109.02188 [hep-ph]} \BibitemShut
  {NoStop}%
\bibitem [{\citenamefont {Ren}\ \emph {et~al.}(2022)\citenamefont {Ren},
  \citenamefont {Wu},\ and\ \citenamefont {Zhu}}]{Ren:2021dsi}%
  \BibitemOpen
  \bibfield  {author} {\bibinfo {author} {\bibfnamefont {H.}~\bibnamefont
  {Ren}}, \bibinfo {author} {\bibfnamefont {F.}~\bibnamefont {Wu}},\ and\
  \bibinfo {author} {\bibfnamefont {R.}~\bibnamefont {Zhu}},\ }\href
  {https://doi.org/10.1155/2022/9103031} {\bibfield  {journal} {\bibinfo
  {journal} {Adv. High Energy Phys.}\ }\textbf {\bibinfo {volume} {2022}},\
  \bibinfo {pages} {9103031} (\bibinfo {year} {2022})},\ \Eprint
  {https://arxiv.org/abs/2109.02531} {arXiv:2109.02531 [hep-ph]} \BibitemShut
  {NoStop}%
\bibitem [{\citenamefont {Chen}\ \emph
  {et~al.}(2022{\natexlab{a}})\citenamefont {Chen}, \citenamefont {Chen},
  \citenamefont {Meng}, \citenamefont {Wang},\ and\ \citenamefont
  {Zhu}}]{Chen:2021cfl}%
  \BibitemOpen
  \bibfield  {author} {\bibinfo {author} {\bibfnamefont {K.}~\bibnamefont
  {Chen}}, \bibinfo {author} {\bibfnamefont {R.}~\bibnamefont {Chen}}, \bibinfo
  {author} {\bibfnamefont {L.}~\bibnamefont {Meng}}, \bibinfo {author}
  {\bibfnamefont {B.}~\bibnamefont {Wang}},\ and\ \bibinfo {author}
  {\bibfnamefont {S.-L.}\ \bibnamefont {Zhu}},\ }\href
  {https://doi.org/10.1140/epjc/s10052-022-10540-5} {\bibfield  {journal}
  {\bibinfo  {journal} {Eur. Phys. J. C}\ }\textbf {\bibinfo {volume} {82}},\
  \bibinfo {pages} {581} (\bibinfo {year} {2022}{\natexlab{a}})},\ \Eprint
  {https://arxiv.org/abs/2109.13057} {arXiv:2109.13057 [hep-ph]} \BibitemShut
  {NoStop}%
\bibitem [{\citenamefont {Albaladejo}(2022)}]{Albaladejo:2021vln}%
  \BibitemOpen
  \bibfield  {author} {\bibinfo {author} {\bibfnamefont {M.}~\bibnamefont
  {Albaladejo}},\ }\href {https://doi.org/10.1016/j.physletb.2022.137052}
  {\bibfield  {journal} {\bibinfo  {journal} {Phys. Lett. B}\ }\textbf
  {\bibinfo {volume} {829}},\ \bibinfo {pages} {137052} (\bibinfo {year}
  {2022})},\ \Eprint {https://arxiv.org/abs/2110.02944} {arXiv:2110.02944
  [hep-ph]} \BibitemShut {NoStop}%
\bibitem [{\citenamefont {Du}\ \emph {et~al.}(2022)\citenamefont {Du},
  \citenamefont {Baru}, \citenamefont {Dong}, \citenamefont {Filin},
  \citenamefont {Guo}, \citenamefont {Hanhart}, \citenamefont {Nefediev},
  \citenamefont {Nieves},\ and\ \citenamefont {Wang}}]{Du:2021zzh}%
  \BibitemOpen
  \bibfield  {author} {\bibinfo {author} {\bibfnamefont {M.-L.}\ \bibnamefont
  {Du}}, \bibinfo {author} {\bibfnamefont {V.}~\bibnamefont {Baru}}, \bibinfo
  {author} {\bibfnamefont {X.-K.}\ \bibnamefont {Dong}}, \bibinfo {author}
  {\bibfnamefont {A.}~\bibnamefont {Filin}}, \bibinfo {author} {\bibfnamefont
  {F.-K.}\ \bibnamefont {Guo}}, \bibinfo {author} {\bibfnamefont
  {C.}~\bibnamefont {Hanhart}}, \bibinfo {author} {\bibfnamefont
  {A.}~\bibnamefont {Nefediev}}, \bibinfo {author} {\bibfnamefont
  {J.}~\bibnamefont {Nieves}},\ and\ \bibinfo {author} {\bibfnamefont
  {Q.}~\bibnamefont {Wang}},\ }\href
  {https://doi.org/10.1103/PhysRevD.105.014024} {\bibfield  {journal} {\bibinfo
   {journal} {Phys. Rev. D}\ }\textbf {\bibinfo {volume} {105}},\ \bibinfo
  {pages} {014024} (\bibinfo {year} {2022})},\ \Eprint
  {https://arxiv.org/abs/2110.13765} {arXiv:2110.13765 [hep-ph]} \BibitemShut
  {NoStop}%
\bibitem [{\citenamefont {Baru}\ \emph {et~al.}(2022)\citenamefont {Baru},
  \citenamefont {Dong}, \citenamefont {Du}, \citenamefont {Filin},
  \citenamefont {Guo}, \citenamefont {Hanhart}, \citenamefont {Nefediev},
  \citenamefont {Nieves},\ and\ \citenamefont {Wang}}]{Baru:2021ldu}%
  \BibitemOpen
  \bibfield  {author} {\bibinfo {author} {\bibfnamefont {V.}~\bibnamefont
  {Baru}}, \bibinfo {author} {\bibfnamefont {X.-K.}\ \bibnamefont {Dong}},
  \bibinfo {author} {\bibfnamefont {M.-L.}\ \bibnamefont {Du}}, \bibinfo
  {author} {\bibfnamefont {A.}~\bibnamefont {Filin}}, \bibinfo {author}
  {\bibfnamefont {F.-K.}\ \bibnamefont {Guo}}, \bibinfo {author} {\bibfnamefont
  {C.}~\bibnamefont {Hanhart}}, \bibinfo {author} {\bibfnamefont
  {A.}~\bibnamefont {Nefediev}}, \bibinfo {author} {\bibfnamefont
  {J.}~\bibnamefont {Nieves}},\ and\ \bibinfo {author} {\bibfnamefont
  {Q.}~\bibnamefont {Wang}},\ }\href
  {https://doi.org/10.1016/j.physletb.2022.137290} {\bibfield  {journal}
  {\bibinfo  {journal} {Phys. Lett. B}\ }\textbf {\bibinfo {volume} {833}},\
  \bibinfo {pages} {137290} (\bibinfo {year} {2022})},\ \Eprint
  {https://arxiv.org/abs/2110.07484} {arXiv:2110.07484 [hep-ph]} \BibitemShut
  {NoStop}%
\bibitem [{\citenamefont {Santowsky}\ and\ \citenamefont
  {Fischer}(2022)}]{Santowsky:2021bhy}%
  \BibitemOpen
  \bibfield  {author} {\bibinfo {author} {\bibfnamefont {N.}~\bibnamefont
  {Santowsky}}\ and\ \bibinfo {author} {\bibfnamefont {C.~S.}\ \bibnamefont
  {Fischer}},\ }\href {https://doi.org/10.1140/epjc/s10052-022-10272-6}
  {\bibfield  {journal} {\bibinfo  {journal} {Eur. Phys. J. C}\ }\textbf
  {\bibinfo {volume} {82}},\ \bibinfo {pages} {313} (\bibinfo {year} {2022})},\
  \Eprint {https://arxiv.org/abs/2111.15310} {arXiv:2111.15310 [hep-ph]}
  \BibitemShut {NoStop}%
\bibitem [{\citenamefont {Deng}\ and\ \citenamefont
  {Zhu}(2022)}]{Deng:2021gnb}%
  \BibitemOpen
  \bibfield  {author} {\bibinfo {author} {\bibfnamefont {C.}~\bibnamefont
  {Deng}}\ and\ \bibinfo {author} {\bibfnamefont {S.-L.}\ \bibnamefont {Zhu}},\
  }\href {https://doi.org/10.1103/PhysRevD.105.054015} {\bibfield  {journal}
  {\bibinfo  {journal} {Phys. Rev. D}\ }\textbf {\bibinfo {volume} {105}},\
  \bibinfo {pages} {054015} (\bibinfo {year} {2022})},\ \Eprint
  {https://arxiv.org/abs/2112.12472} {arXiv:2112.12472 [hep-ph]} \BibitemShut
  {NoStop}%
\bibitem [{\citenamefont {Ke}\ \emph {et~al.}(2022)\citenamefont {Ke},
  \citenamefont {Liu},\ and\ \citenamefont {Li}}]{Ke:2021rxd}%
  \BibitemOpen
  \bibfield  {author} {\bibinfo {author} {\bibfnamefont {H.-W.}\ \bibnamefont
  {Ke}}, \bibinfo {author} {\bibfnamefont {X.-H.}\ \bibnamefont {Liu}},\ and\
  \bibinfo {author} {\bibfnamefont {X.-Q.}\ \bibnamefont {Li}},\ }\href
  {https://doi.org/10.1140/epjc/s10052-022-10092-8} {\bibfield  {journal}
  {\bibinfo  {journal} {Eur. Phys. J. C}\ }\textbf {\bibinfo {volume} {82}},\
  \bibinfo {pages} {144} (\bibinfo {year} {2022})},\ \Eprint
  {https://arxiv.org/abs/2112.14142} {arXiv:2112.14142 [hep-ph]} \BibitemShut
  {NoStop}%
\bibitem [{\citenamefont {Agaev}\ \emph {et~al.}(2022)\citenamefont {Agaev},
  \citenamefont {Azizi},\ and\ \citenamefont {Sundu}}]{Agaev:2022ast}%
  \BibitemOpen
  \bibfield  {author} {\bibinfo {author} {\bibfnamefont {S.~S.}\ \bibnamefont
  {Agaev}}, \bibinfo {author} {\bibfnamefont {K.}~\bibnamefont {Azizi}},\ and\
  \bibinfo {author} {\bibfnamefont {H.}~\bibnamefont {Sundu}},\ }\href
  {https://doi.org/10.1007/JHEP06(2022)057} {\bibfield  {journal} {\bibinfo
  {journal} {JHEP}\ }\textbf {\bibinfo {volume} {06}},\ \bibinfo {pages}
  {057}},\ \Eprint {https://arxiv.org/abs/2201.02788} {arXiv:2201.02788
  [hep-ph]} \BibitemShut {NoStop}%
\bibitem [{\citenamefont {Meng}\ \emph {et~al.}(2023)\citenamefont {Meng},
  \citenamefont {Wang}, \citenamefont {Wang},\ and\ \citenamefont
  {Zhu}}]{Meng:2022ozq}%
  \BibitemOpen
  \bibfield  {author} {\bibinfo {author} {\bibfnamefont {L.}~\bibnamefont
  {Meng}}, \bibinfo {author} {\bibfnamefont {B.}~\bibnamefont {Wang}}, \bibinfo
  {author} {\bibfnamefont {G.-J.}\ \bibnamefont {Wang}},\ and\ \bibinfo
  {author} {\bibfnamefont {S.-L.}\ \bibnamefont {Zhu}},\ }\href
  {https://doi.org/10.1016/j.physrep.2023.04.003} {\bibfield  {journal}
  {\bibinfo  {journal} {Phys. Rept.}\ }\textbf {\bibinfo {volume} {1019}},\
  \bibinfo {pages} {1} (\bibinfo {year} {2023})},\ \Eprint
  {https://arxiv.org/abs/2204.08716} {arXiv:2204.08716 [hep-ph]} \BibitemShut
  {NoStop}%
\bibitem [{\citenamefont {Abreu}(2022)}]{Abreu:2022sra}%
  \BibitemOpen
  \bibfield  {author} {\bibinfo {author} {\bibfnamefont {L.~M.}\ \bibnamefont
  {Abreu}},\ }\href {https://doi.org/10.1016/j.nuclphysb.2022.115994}
  {\bibfield  {journal} {\bibinfo  {journal} {Nucl. Phys. B}\ }\textbf
  {\bibinfo {volume} {985}},\ \bibinfo {pages} {115994} (\bibinfo {year}
  {2022})},\ \Eprint {https://arxiv.org/abs/2206.01166} {arXiv:2206.01166
  [hep-ph]} \BibitemShut {NoStop}%
\bibitem [{\citenamefont {Chen}\ \emph
  {et~al.}(2022{\natexlab{b}})\citenamefont {Chen}, \citenamefont {Shi},
  \citenamefont {Chen}, \citenamefont {Gong}, \citenamefont {Liu},
  \citenamefont {Sun},\ and\ \citenamefont {Zhang}}]{Chen:2022vpo}%
  \BibitemOpen
  \bibfield  {author} {\bibinfo {author} {\bibfnamefont {S.}~\bibnamefont
  {Chen}}, \bibinfo {author} {\bibfnamefont {C.}~\bibnamefont {Shi}}, \bibinfo
  {author} {\bibfnamefont {Y.}~\bibnamefont {Chen}}, \bibinfo {author}
  {\bibfnamefont {M.}~\bibnamefont {Gong}}, \bibinfo {author} {\bibfnamefont
  {Z.}~\bibnamefont {Liu}}, \bibinfo {author} {\bibfnamefont {W.}~\bibnamefont
  {Sun}},\ and\ \bibinfo {author} {\bibfnamefont {R.}~\bibnamefont {Zhang}},\
  }\href {https://doi.org/10.1016/j.physletb.2022.137391} {\bibfield  {journal}
  {\bibinfo  {journal} {Phys. Lett. B}\ }\textbf {\bibinfo {volume} {833}},\
  \bibinfo {pages} {137391} (\bibinfo {year} {2022}{\natexlab{b}})},\ \Eprint
  {https://arxiv.org/abs/2206.06185} {arXiv:2206.06185 [hep-lat]} \BibitemShut
  {NoStop}%
\bibitem [{\citenamefont {Albaladejo}\ and\ \citenamefont
  {Nieves}(2022)}]{Albaladejo:2022sux}%
  \BibitemOpen
  \bibfield  {author} {\bibinfo {author} {\bibfnamefont {M.}~\bibnamefont
  {Albaladejo}}\ and\ \bibinfo {author} {\bibfnamefont {J.}~\bibnamefont
  {Nieves}},\ }\href {https://doi.org/10.1140/epjc/s10052-022-10695-1}
  {\bibfield  {journal} {\bibinfo  {journal} {Eur. Phys. J. C}\ }\textbf
  {\bibinfo {volume} {82}},\ \bibinfo {pages} {724} (\bibinfo {year} {2022})},\
  \Eprint {https://arxiv.org/abs/2203.04864} {arXiv:2203.04864 [hep-ph]}
  \BibitemShut {NoStop}%
\bibitem [{\citenamefont {Dai}\ \emph {et~al.}(2023{\natexlab{a}})\citenamefont
  {Dai}, \citenamefont {Abreu}, \citenamefont {Feijoo},\ and\ \citenamefont
  {Oset}}]{Dai:2023cyo}%
  \BibitemOpen
  \bibfield  {author} {\bibinfo {author} {\bibfnamefont {L.~R.}\ \bibnamefont
  {Dai}}, \bibinfo {author} {\bibfnamefont {L.~M.}\ \bibnamefont {Abreu}},
  \bibinfo {author} {\bibfnamefont {A.}~\bibnamefont {Feijoo}},\ and\ \bibinfo
  {author} {\bibfnamefont {E.}~\bibnamefont {Oset}},\ }\href
  {https://doi.org/10.1140/epjc/s10052-023-12159-6} {\bibfield  {journal}
  {\bibinfo  {journal} {Eur. Phys. J. C}\ }\textbf {\bibinfo {volume} {83}},\
  \bibinfo {pages} {983} (\bibinfo {year} {2023}{\natexlab{a}})},\ \Eprint
  {https://arxiv.org/abs/2304.01870} {arXiv:2304.01870 [hep-ph]} \BibitemShut
  {NoStop}%
\bibitem [{\citenamefont {Wang}\ \emph {et~al.}(2024)\citenamefont {Wang},
  \citenamefont {Yang}, \citenamefont {Wu}, \citenamefont {Oka},\ and\
  \citenamefont {Zhu}}]{Wang:2023ovj}%
  \BibitemOpen
  \bibfield  {author} {\bibinfo {author} {\bibfnamefont {G.-J.}\ \bibnamefont
  {Wang}}, \bibinfo {author} {\bibfnamefont {Z.}~\bibnamefont {Yang}}, \bibinfo
  {author} {\bibfnamefont {J.-J.}\ \bibnamefont {Wu}}, \bibinfo {author}
  {\bibfnamefont {M.}~\bibnamefont {Oka}},\ and\ \bibinfo {author}
  {\bibfnamefont {S.-L.}\ \bibnamefont {Zhu}},\ }\href
  {https://doi.org/10.1016/j.scib.2024.07.012} {\bibfield  {journal} {\bibinfo
  {journal} {Sci. Bull.}\ }\textbf {\bibinfo {volume} {69}},\ \bibinfo {pages}
  {3036} (\bibinfo {year} {2024})},\ \Eprint {https://arxiv.org/abs/2306.12406}
  {arXiv:2306.12406 [hep-ph]} \BibitemShut {NoStop}%
\bibitem [{\citenamefont {Acharya}\ \emph
  {et~al.}(2019{\natexlab{a}})\citenamefont {Acharya} \emph
  {et~al.}}]{ALICE:2018ysd}%
  \BibitemOpen
  \bibfield  {author} {\bibinfo {author} {\bibfnamefont {S.}~\bibnamefont
  {Acharya}} \emph {et~al.} (\bibinfo {collaboration} {ALICE}),\ }\href
  {https://doi.org/10.1103/PhysRevC.99.024001} {\bibfield  {journal} {\bibinfo
  {journal} {Phys. Rev. C}\ }\textbf {\bibinfo {volume} {99}},\ \bibinfo
  {pages} {024001} (\bibinfo {year} {2019}{\natexlab{a}})},\ \Eprint
  {https://arxiv.org/abs/1805.12455} {arXiv:1805.12455 [nucl-ex]} \BibitemShut
  {NoStop}%
\bibitem [{\citenamefont {Acharya}\ \emph
  {et~al.}(2019{\natexlab{b}})\citenamefont {Acharya} \emph
  {et~al.}}]{ALICE:2018nnl}%
  \BibitemOpen
  \bibfield  {author} {\bibinfo {author} {\bibfnamefont {S.}~\bibnamefont
  {Acharya}} \emph {et~al.} (\bibinfo {collaboration} {ALICE}),\ }\href
  {https://doi.org/10.1016/j.physletb.2018.12.033} {\bibfield  {journal}
  {\bibinfo  {journal} {Phys. Lett. B}\ }\textbf {\bibinfo {volume} {790}},\
  \bibinfo {pages} {22} (\bibinfo {year} {2019}{\natexlab{b}})},\ \Eprint
  {https://arxiv.org/abs/1809.07899} {arXiv:1809.07899 [nucl-ex]} \BibitemShut
  {NoStop}%
\bibitem [{\citenamefont {Acharya}\ \emph
  {et~al.}(2020{\natexlab{a}})\citenamefont {Acharya} \emph
  {et~al.}}]{ALICE:2019gcn}%
  \BibitemOpen
  \bibfield  {author} {\bibinfo {author} {\bibfnamefont {S.}~\bibnamefont
  {Acharya}} \emph {et~al.} (\bibinfo {collaboration} {ALICE}),\ }\href
  {https://doi.org/10.1103/PhysRevLett.124.092301} {\bibfield  {journal}
  {\bibinfo  {journal} {Phys. Rev. Lett.}\ }\textbf {\bibinfo {volume} {124}},\
  \bibinfo {pages} {092301} (\bibinfo {year} {2020}{\natexlab{a}})},\ \Eprint
  {https://arxiv.org/abs/1905.13470} {arXiv:1905.13470 [nucl-ex]} \BibitemShut
  {NoStop}%
\bibitem [{\citenamefont {Acharya}\ \emph
  {et~al.}(2021{\natexlab{a}})\citenamefont {Acharya} \emph
  {et~al.}}]{ALICE:2020wvi}%
  \BibitemOpen
  \bibfield  {author} {\bibinfo {author} {\bibfnamefont {S.}~\bibnamefont
  {Acharya}} \emph {et~al.} (\bibinfo {collaboration} {ALICE}),\ }\href
  {https://doi.org/10.1103/PhysRevC.103.055201} {\bibfield  {journal} {\bibinfo
   {journal} {Phys. Rev. C}\ }\textbf {\bibinfo {volume} {103}},\ \bibinfo
  {pages} {055201} (\bibinfo {year} {2021}{\natexlab{a}})},\ \Eprint
  {https://arxiv.org/abs/2005.11124} {arXiv:2005.11124 [nucl-ex]} \BibitemShut
  {NoStop}%
\bibitem [{\citenamefont {Acharya}\ \emph
  {et~al.}(2022{\natexlab{a}})\citenamefont {Acharya} \emph
  {et~al.}}]{ALICE:2021njx}%
  \BibitemOpen
  \bibfield  {author} {\bibinfo {author} {\bibfnamefont {S.}~\bibnamefont
  {Acharya}} \emph {et~al.} (\bibinfo {collaboration} {ALICE}),\ }\href
  {https://doi.org/10.1016/j.physletb.2022.137272} {\bibfield  {journal}
  {\bibinfo  {journal} {Phys. Lett. B}\ }\textbf {\bibinfo {volume} {833}},\
  \bibinfo {pages} {137272} (\bibinfo {year} {2022}{\natexlab{a}})},\ \Eprint
  {https://arxiv.org/abs/2104.04427} {arXiv:2104.04427 [nucl-ex]} \BibitemShut
  {NoStop}%
\bibitem [{\citenamefont {Acharya}\ \emph
  {et~al.}(2022{\natexlab{b}})\citenamefont {Acharya} \emph
  {et~al.}}]{ALICE:2021ovd}%
  \BibitemOpen
  \bibfield  {author} {\bibinfo {author} {\bibfnamefont {S.}~\bibnamefont
  {Acharya}} \emph {et~al.} (\bibinfo {collaboration} {ALICE}),\ }\href
  {https://doi.org/10.1016/j.physletb.2022.137335} {\bibfield  {journal}
  {\bibinfo  {journal} {Phys. Lett. B}\ }\textbf {\bibinfo {volume} {833}},\
  \bibinfo {pages} {137335} (\bibinfo {year} {2022}{\natexlab{b}})},\ \Eprint
  {https://arxiv.org/abs/2111.06611} {arXiv:2111.06611 [nucl-ex]} \BibitemShut
  {NoStop}%
\bibitem [{\citenamefont {Acharya}\ \emph
  {et~al.}(2020{\natexlab{b}})\citenamefont {Acharya} \emph
  {et~al.}}]{ALICE:2019buq}%
  \BibitemOpen
  \bibfield  {author} {\bibinfo {author} {\bibfnamefont {S.}~\bibnamefont
  {Acharya}} \emph {et~al.} (\bibinfo {collaboration} {ALICE}),\ }\href
  {https://doi.org/10.1016/j.physletb.2020.135419} {\bibfield  {journal}
  {\bibinfo  {journal} {Phys. Lett. B}\ }\textbf {\bibinfo {volume} {805}},\
  \bibinfo {pages} {135419} (\bibinfo {year} {2020}{\natexlab{b}})},\ \Eprint
  {https://arxiv.org/abs/1910.14407} {arXiv:1910.14407 [nucl-ex]} \BibitemShut
  {NoStop}%
\bibitem [{\citenamefont {Acharya}\ \emph
  {et~al.}(2019{\natexlab{c}})\citenamefont {Acharya} \emph
  {et~al.}}]{ALICE:2019eol}%
  \BibitemOpen
  \bibfield  {author} {\bibinfo {author} {\bibfnamefont {S.}~\bibnamefont
  {Acharya}} \emph {et~al.} (\bibinfo {collaboration} {ALICE}),\ }\href
  {https://doi.org/10.1016/j.physletb.2019.134822} {\bibfield  {journal}
  {\bibinfo  {journal} {Phys. Lett. B}\ }\textbf {\bibinfo {volume} {797}},\
  \bibinfo {pages} {134822} (\bibinfo {year} {2019}{\natexlab{c}})},\ \Eprint
  {https://arxiv.org/abs/1905.07209} {arXiv:1905.07209 [nucl-ex]} \BibitemShut
  {NoStop}%
\bibitem [{\citenamefont {Acharya}\ \emph
  {et~al.}(2019{\natexlab{d}})\citenamefont {Acharya} \emph
  {et~al.}}]{ALICE:2019hdt}%
  \BibitemOpen
  \bibfield  {author} {\bibinfo {author} {\bibfnamefont {S.}~\bibnamefont
  {Acharya}} \emph {et~al.} (\bibinfo {collaboration} {ALICE}),\ }\href
  {https://doi.org/10.1103/PhysRevLett.123.112002} {\bibfield  {journal}
  {\bibinfo  {journal} {Phys. Rev. Lett.}\ }\textbf {\bibinfo {volume} {123}},\
  \bibinfo {pages} {112002} (\bibinfo {year} {2019}{\natexlab{d}})},\ \Eprint
  {https://arxiv.org/abs/1904.12198} {arXiv:1904.12198 [nucl-ex]} \BibitemShut
  {NoStop}%
\bibitem [{\citenamefont {Acharya}\ \emph
  {et~al.}(2021{\natexlab{b}})\citenamefont {Acharya} \emph
  {et~al.}}]{ALICE:2021cpv}%
  \BibitemOpen
  \bibfield  {author} {\bibinfo {author} {\bibfnamefont {S.}~\bibnamefont
  {Acharya}} \emph {et~al.} (\bibinfo {collaboration} {ALICE}),\ }\href
  {https://doi.org/10.1103/PhysRevLett.127.172301} {\bibfield  {journal}
  {\bibinfo  {journal} {Phys. Rev. Lett.}\ }\textbf {\bibinfo {volume} {127}},\
  \bibinfo {pages} {172301} (\bibinfo {year} {2021}{\natexlab{b}})},\ \Eprint
  {https://arxiv.org/abs/2105.05578} {arXiv:2105.05578 [nucl-ex]} \BibitemShut
  {NoStop}%
\bibitem [{\citenamefont {Acharya}\ \emph
  {et~al.}(2023{\natexlab{a}})\citenamefont {Acharya} \emph
  {et~al.}}]{ALICE:2022yyh}%
  \BibitemOpen
  \bibfield  {author} {\bibinfo {author} {\bibfnamefont {S.}~\bibnamefont
  {Acharya}} \emph {et~al.} (\bibinfo {collaboration} {ALICE}),\ }\href
  {https://doi.org/10.1140/epjc/s10052-023-11476-0} {\bibfield  {journal}
  {\bibinfo  {journal} {Eur. Phys. J. C}\ }\textbf {\bibinfo {volume} {83}},\
  \bibinfo {pages} {340} (\bibinfo {year} {2023}{\natexlab{a}})},\ \Eprint
  {https://arxiv.org/abs/2205.15176} {arXiv:2205.15176 [nucl-ex]} \BibitemShut
  {NoStop}%
\bibitem [{\citenamefont {Acharya}\ \emph
  {et~al.}(2022{\natexlab{c}})\citenamefont {Acharya} \emph
  {et~al.}}]{ALICE:2021cyj}%
  \BibitemOpen
  \bibfield  {author} {\bibinfo {author} {\bibfnamefont {S.}~\bibnamefont
  {Acharya}} \emph {et~al.} (\bibinfo {collaboration} {ALICE}),\ }\href
  {https://doi.org/10.1016/j.physletb.2022.137060} {\bibfield  {journal}
  {\bibinfo  {journal} {Phys. Lett. B}\ }\textbf {\bibinfo {volume} {829}},\
  \bibinfo {pages} {137060} (\bibinfo {year} {2022}{\natexlab{c}})},\ \Eprint
  {https://arxiv.org/abs/2105.05190} {arXiv:2105.05190 [nucl-ex]} \BibitemShut
  {NoStop}%
\bibitem [{\citenamefont {Acharya}\ \emph
  {et~al.}(2021{\natexlab{c}})\citenamefont {Acharya} \emph
  {et~al.}}]{ALICE:2021szj}%
  \BibitemOpen
  \bibfield  {author} {\bibinfo {author} {\bibfnamefont {S.}~\bibnamefont
  {Acharya}} \emph {et~al.} (\bibinfo {collaboration} {ALICE}),\ }\href
  {https://doi.org/10.1016/j.physletb.2021.136708} {\bibfield  {journal}
  {\bibinfo  {journal} {Phys. Lett. B}\ }\textbf {\bibinfo {volume} {822}},\
  \bibinfo {pages} {136708} (\bibinfo {year} {2021}{\natexlab{c}})},\ \Eprint
  {https://arxiv.org/abs/2105.05683} {arXiv:2105.05683 [nucl-ex]} \BibitemShut
  {NoStop}%
\bibitem [{\citenamefont {Acharya}\ \emph
  {et~al.}(2023{\natexlab{b}})\citenamefont {Acharya} \emph
  {et~al.}}]{ALICE:2023wjz}%
  \BibitemOpen
  \bibfield  {author} {\bibinfo {author} {\bibfnamefont {S.}~\bibnamefont
  {Acharya}} \emph {et~al.} (\bibinfo {collaboration} {ALICE}),\ }\href
  {https://doi.org/10.1016/j.physletb.2023.138145} {\bibfield  {journal}
  {\bibinfo  {journal} {Phys. Lett. B}\ }\textbf {\bibinfo {volume} {845}},\
  \bibinfo {pages} {138145} (\bibinfo {year} {2023}{\natexlab{b}})},\ \Eprint
  {https://arxiv.org/abs/2305.19093} {arXiv:2305.19093 [nucl-ex]} \BibitemShut
  {NoStop}%
\bibitem [{\citenamefont {Acharya}\ \emph
  {et~al.}(2024{\natexlab{a}})\citenamefont {Acharya} \emph
  {et~al.}}]{ALICE:2023eyl}%
  \BibitemOpen
  \bibfield  {author} {\bibinfo {author} {\bibfnamefont {S.}~\bibnamefont
  {Acharya}} \emph {et~al.} (\bibinfo {collaboration} {ALICE}),\ }\href
  {https://doi.org/10.1016/j.physletb.2024.138915} {\bibfield  {journal}
  {\bibinfo  {journal} {Phys. Lett. B}\ }\textbf {\bibinfo {volume} {856}},\
  \bibinfo {pages} {138915} (\bibinfo {year} {2024}{\natexlab{a}})},\ \Eprint
  {https://arxiv.org/abs/2312.12830} {arXiv:2312.12830 [hep-ex]} \BibitemShut
  {NoStop}%
\bibitem [{\citenamefont {Acharya}\ \emph
  {et~al.}(2022{\natexlab{d}})\citenamefont {Acharya} \emph
  {et~al.}}]{ALICE:2022enj}%
  \BibitemOpen
  \bibfield  {author} {\bibinfo {author} {\bibfnamefont {S.}~\bibnamefont
  {Acharya}} \emph {et~al.} (\bibinfo {collaboration} {ALICE}),\ }\href
  {https://doi.org/10.1103/PhysRevD.106.052010} {\bibfield  {journal} {\bibinfo
   {journal} {Phys. Rev. D}\ }\textbf {\bibinfo {volume} {106}},\ \bibinfo
  {pages} {052010} (\bibinfo {year} {2022}{\natexlab{d}})},\ \Eprint
  {https://arxiv.org/abs/2201.05352} {arXiv:2201.05352 [nucl-ex]} \BibitemShut
  {NoStop}%
\bibitem [{\citenamefont {Acharya}\ \emph
  {et~al.}(2024{\natexlab{b}})\citenamefont {Acharya} \emph
  {et~al.}}]{ALICE:2024bhk}%
  \BibitemOpen
  \bibfield  {author} {\bibinfo {author} {\bibfnamefont {S.}~\bibnamefont
  {Acharya}} \emph {et~al.} (\bibinfo {collaboration} {ALICE}),\ }\href
  {https://doi.org/10.1103/PhysRevD.110.032004} {\bibfield  {journal} {\bibinfo
   {journal} {Phys. Rev. D}\ }\textbf {\bibinfo {volume} {110}},\ \bibinfo
  {pages} {032004} (\bibinfo {year} {2024}{\natexlab{b}})},\ \Eprint
  {https://arxiv.org/abs/2401.13541} {arXiv:2401.13541 [nucl-ex]} \BibitemShut
  {NoStop}%
\bibitem [{\citenamefont {Lisa}\ and\ \citenamefont
  {Pratt}(2010)}]{Lisa:2008gf}%
  \BibitemOpen
  \bibfield  {author} {\bibinfo {author} {\bibfnamefont {M.~A.}\ \bibnamefont
  {Lisa}}\ and\ \bibinfo {author} {\bibfnamefont {S.}~\bibnamefont {Pratt}},\
  }\bibinfo {title} {{Femtoscopically Probing the Freeze-out Configuration in
  Heavy Ion Collisions}},\ in\ \href
  {https://doi.org/10.1007/978-3-642-01539-7_21} {\emph {\bibinfo {booktitle}
  {{Relativistic Heavy Ion Physics}}}},\ \bibinfo {editor} {edited by\ \bibinfo
  {editor} {\bibfnamefont {R.}~\bibnamefont {Stock}}}\ (\bibinfo {year}
  {2010})\ \Eprint {https://arxiv.org/abs/0811.1352} {arXiv:0811.1352
  [nucl-ex]} \BibitemShut {NoStop}%
\bibitem [{\citenamefont {Verde}\ \emph {et~al.}(2003)\citenamefont {Verde},
  \citenamefont {Danielewicz}, \citenamefont {Lynch}, \citenamefont {Brown},
  \citenamefont {Gelbke},\ and\ \citenamefont {Tsang}}]{Verde:2003cx}%
  \BibitemOpen
  \bibfield  {author} {\bibinfo {author} {\bibfnamefont {G.}~\bibnamefont
  {Verde}}, \bibinfo {author} {\bibfnamefont {P.}~\bibnamefont {Danielewicz}},
  \bibinfo {author} {\bibfnamefont {W.~G.}\ \bibnamefont {Lynch}}, \bibinfo
  {author} {\bibfnamefont {D.~A.}\ \bibnamefont {Brown}}, \bibinfo {author}
  {\bibfnamefont {C.~K.}\ \bibnamefont {Gelbke}},\ and\ \bibinfo {author}
  {\bibfnamefont {M.~B.}\ \bibnamefont {Tsang}},\ }\href
  {https://doi.org/10.1103/PhysRevC.67.034606} {\bibfield  {journal} {\bibinfo
  {journal} {Phys. Rev. C}\ }\textbf {\bibinfo {volume} {67}},\ \bibinfo
  {pages} {034606} (\bibinfo {year} {2003})},\ \Eprint
  {https://arxiv.org/abs/nucl-ex/0301013} {arXiv:nucl-ex/0301013} \BibitemShut
  {NoStop}%
\bibitem [{\citenamefont {Acharya}\ \emph
  {et~al.}(2020{\natexlab{c}})\citenamefont {Acharya} \emph
  {et~al.}}]{ALICE:2020ibs}%
  \BibitemOpen
  \bibfield  {author} {\bibinfo {author} {\bibfnamefont {S.}~\bibnamefont
  {Acharya}} \emph {et~al.} (\bibinfo {collaboration} {ALICE}),\ }\href
  {https://doi.org/10.1016/j.physletb.2020.135849} {\bibfield  {journal}
  {\bibinfo  {journal} {Phys. Lett. B}\ }\textbf {\bibinfo {volume} {811}},\
  \bibinfo {pages} {135849} (\bibinfo {year} {2020}{\natexlab{c}})},\ \Eprint
  {https://arxiv.org/abs/2004.08018} {arXiv:2004.08018 [nucl-ex]} \BibitemShut
  {NoStop}%
\bibitem [{\citenamefont {Acharya}\ \emph
  {et~al.}(2023{\natexlab{c}})\citenamefont {Acharya} \emph
  {et~al.}}]{ALICE:2023sjd}%
  \BibitemOpen
  \bibfield  {author} {\bibinfo {author} {\bibfnamefont {S.}~\bibnamefont
  {Acharya}} \emph {et~al.} (\bibinfo {collaboration} {ALICE}),\ }\href@noop {}
  {\  (\bibinfo {year} {2023}{\natexlab{c}})},\ \Eprint
  {https://arxiv.org/abs/2311.14527} {arXiv:2311.14527 [hep-ph]} \BibitemShut
  {NoStop}%
\bibitem [{\citenamefont {Taylor}(2006)}]{Taylor:2006}%
  \BibitemOpen
  \bibfield  {author} {\bibinfo {author} {\bibfnamefont {J.}~\bibnamefont
  {Taylor}},\ }\href@noop {} {\emph {\bibinfo {title} {{Scattering Theory: The
  Quantum Theory of Nonrelativistic Collisions}}}}\ (\bibinfo  {publisher}
  {Dover Publications},\ \bibinfo {year} {2006})\BibitemShut {NoStop}%
\bibitem [{\citenamefont {Galindo}\ and\ \citenamefont
  {Pascual}(2012)}]{Pascual:2012}%
  \BibitemOpen
  \bibfield  {author} {\bibinfo {author} {\bibfnamefont {A.}~\bibnamefont
  {Galindo}}\ and\ \bibinfo {author} {\bibfnamefont {P.}~\bibnamefont
  {Pascual}},\ }\href@noop {} {\emph {\bibinfo {title} {{Quantum Mechanics
  II}}}}\ (\bibinfo  {publisher} {Springer-Verlag},\ \bibinfo {year}
  {2012})\BibitemShut {NoStop}%
\bibitem [{\citenamefont {K\"all\'en}(1964)}]{Kallen:1964lxa}%
  \BibitemOpen
  \bibfield  {author} {\bibinfo {author} {\bibfnamefont {G.}~\bibnamefont
  {K\"all\'en}},\ }\href@noop {} {\emph {\bibinfo {title} {{Elementary particle
  physics}}}}\ (\bibinfo  {publisher} {Addison-Wesley},\ \bibinfo {address}
  {Reading, MA},\ \bibinfo {year} {1964})\BibitemShut {NoStop}%
\bibitem [{\citenamefont {Nieves}\ and\ \citenamefont
  {Ruiz~Arriola}(2000)}]{Nieves:1999bx}%
  \BibitemOpen
  \bibfield  {author} {\bibinfo {author} {\bibfnamefont {J.}~\bibnamefont
  {Nieves}}\ and\ \bibinfo {author} {\bibfnamefont {E.}~\bibnamefont
  {Ruiz~Arriola}},\ }\href {https://doi.org/10.1016/S0375-9474(00)00321-3}
  {\bibfield  {journal} {\bibinfo  {journal} {Nucl. Phys. A}\ }\textbf
  {\bibinfo {volume} {679}},\ \bibinfo {pages} {57} (\bibinfo {year} {2000})},\
  \Eprint {https://arxiv.org/abs/hep-ph/9907469} {arXiv:hep-ph/9907469}
  \BibitemShut {NoStop}%
\bibitem [{\citenamefont {Ballot}\ and\ \citenamefont
  {Richard}(1983)}]{Ballot:1983iv}%
  \BibitemOpen
  \bibfield  {author} {\bibinfo {author} {\bibfnamefont {J.~l.}\ \bibnamefont
  {Ballot}}\ and\ \bibinfo {author} {\bibfnamefont {J.~M.}\ \bibnamefont
  {Richard}},\ }\href {https://doi.org/10.1016/0370-2693(83)90991-7} {\bibfield
   {journal} {\bibinfo  {journal} {Phys. Lett. B}\ }\textbf {\bibinfo {volume}
  {123}},\ \bibinfo {pages} {449} (\bibinfo {year} {1983})}\BibitemShut
  {NoStop}%
\bibitem [{\citenamefont {Zouzou}\ \emph {et~al.}(1986)\citenamefont {Zouzou},
  \citenamefont {Silvestre-Brac}, \citenamefont {Gignoux},\ and\ \citenamefont
  {Richard}}]{Zouzou:1986qh}%
  \BibitemOpen
  \bibfield  {author} {\bibinfo {author} {\bibfnamefont {S.}~\bibnamefont
  {Zouzou}}, \bibinfo {author} {\bibfnamefont {B.}~\bibnamefont
  {Silvestre-Brac}}, \bibinfo {author} {\bibfnamefont {C.}~\bibnamefont
  {Gignoux}},\ and\ \bibinfo {author} {\bibfnamefont {J.~M.}\ \bibnamefont
  {Richard}},\ }\href {https://doi.org/10.1007/BF01557611} {\bibfield
  {journal} {\bibinfo  {journal} {Z. Phys. C}\ }\textbf {\bibinfo {volume}
  {30}},\ \bibinfo {pages} {457} (\bibinfo {year} {1986})}\BibitemShut
  {NoStop}%
\bibitem [{\citenamefont {Dai}\ \emph {et~al.}(2023{\natexlab{b}})\citenamefont
  {Dai}, \citenamefont {Song},\ and\ \citenamefont {Oset}}]{Dai:2023kwv}%
  \BibitemOpen
  \bibfield  {author} {\bibinfo {author} {\bibfnamefont {L.~R.}\ \bibnamefont
  {Dai}}, \bibinfo {author} {\bibfnamefont {J.}~\bibnamefont {Song}},\ and\
  \bibinfo {author} {\bibfnamefont {E.}~\bibnamefont {Oset}},\ }\href
  {https://doi.org/10.1016/j.physletb.2023.138200} {\bibfield  {journal}
  {\bibinfo  {journal} {Phys. Lett. B}\ }\textbf {\bibinfo {volume} {846}},\
  \bibinfo {pages} {138200} (\bibinfo {year} {2023}{\natexlab{b}})},\ \Eprint
  {https://arxiv.org/abs/2306.01607} {arXiv:2306.01607 [hep-ph]} \BibitemShut
  {NoStop}%
\bibitem [{\citenamefont {Sun}\ \emph {et~al.}(2023)\citenamefont {Sun},
  \citenamefont {Fan},\ and\ \citenamefont {Cao}}]{Sun:2022cxf}%
  \BibitemOpen
  \bibfield  {author} {\bibinfo {author} {\bibfnamefont {B.-X.}\ \bibnamefont
  {Sun}}, \bibinfo {author} {\bibfnamefont {Y.-Y.}\ \bibnamefont {Fan}},\ and\
  \bibinfo {author} {\bibfnamefont {Q.-Q.}\ \bibnamefont {Cao}},\ }\href
  {https://doi.org/10.1088/1572-9494/acc31d} {\bibfield  {journal} {\bibinfo
  {journal} {Commun. Theor. Phys.}\ }\textbf {\bibinfo {volume} {75}},\
  \bibinfo {pages} {055301} (\bibinfo {year} {2023})},\ \Eprint
  {https://arxiv.org/abs/2206.02961} {arXiv:2206.02961 [hep-ph]} \BibitemShut
  {NoStop}%
\bibitem [{\citenamefont {Chizzali}\ \emph {et~al.}(2024)\citenamefont
  {Chizzali}, \citenamefont {Kamiya}, \citenamefont {Del~Grande}, \citenamefont
  {Doi}, \citenamefont {Fabbietti}, \citenamefont {Hatsuda},\ and\
  \citenamefont {Lyu}}]{Chizzali:2022pjd}%
  \BibitemOpen
  \bibfield  {author} {\bibinfo {author} {\bibfnamefont {E.}~\bibnamefont
  {Chizzali}}, \bibinfo {author} {\bibfnamefont {Y.}~\bibnamefont {Kamiya}},
  \bibinfo {author} {\bibfnamefont {R.}~\bibnamefont {Del~Grande}}, \bibinfo
  {author} {\bibfnamefont {T.}~\bibnamefont {Doi}}, \bibinfo {author}
  {\bibfnamefont {L.}~\bibnamefont {Fabbietti}}, \bibinfo {author}
  {\bibfnamefont {T.}~\bibnamefont {Hatsuda}},\ and\ \bibinfo {author}
  {\bibfnamefont {Y.}~\bibnamefont {Lyu}},\ }\href
  {https://doi.org/10.1016/j.physletb.2023.138358} {\bibfield  {journal}
  {\bibinfo  {journal} {Phys. Lett. B}\ }\textbf {\bibinfo {volume} {848}},\
  \bibinfo {pages} {138358} (\bibinfo {year} {2024})},\ \Eprint
  {https://arxiv.org/abs/2212.12690} {arXiv:2212.12690 [nucl-ex]} \BibitemShut
  {NoStop}%
\bibitem [{\citenamefont {Feijoo}\ \emph
  {et~al.}(2024{\natexlab{b}})\citenamefont {Feijoo}, \citenamefont {Sarti},
  \citenamefont {Nieves}, \citenamefont {Ramos},\ and\ \citenamefont
  {Vida\~na}}]{Feijoo:2024qqg}%
  \BibitemOpen
  \bibfield  {author} {\bibinfo {author} {\bibfnamefont {A.}~\bibnamefont
  {Feijoo}}, \bibinfo {author} {\bibfnamefont {V.~M.}\ \bibnamefont {Sarti}},
  \bibinfo {author} {\bibfnamefont {J.}~\bibnamefont {Nieves}}, \bibinfo
  {author} {\bibfnamefont {A.}~\bibnamefont {Ramos}},\ and\ \bibinfo {author}
  {\bibfnamefont {I.}~\bibnamefont {Vida\~na}},\ }\href@noop {} {\  (\bibinfo
  {year} {2024}{\natexlab{b}})},\ \Eprint {https://arxiv.org/abs/2411.10245}
  {arXiv:2411.10245 [hep-ph]} \BibitemShut {NoStop}%
\bibitem [{\citenamefont {Preston}\ and\ \citenamefont
  {Bhaduri}(1993)}]{Preston:1993}%
  \BibitemOpen
  \bibfield  {author} {\bibinfo {author} {\bibfnamefont {M.}~\bibnamefont
  {Preston}}\ and\ \bibinfo {author} {\bibfnamefont {R.~K.}\ \bibnamefont
  {Bhaduri}},\ }\href@noop {} {\emph {\bibinfo {title} {{Structure of the
  Nucleus}}}}\ (\bibinfo  {publisher} {Westview Press},\ \bibinfo {year}
  {1993})\BibitemShut {NoStop}%
\bibitem [{\citenamefont {Gamermann}\ \emph {et~al.}(2010)\citenamefont
  {Gamermann}, \citenamefont {Nieves}, \citenamefont {Oset},\ and\
  \citenamefont {Ruiz~Arriola}}]{Gamermann:2009uq}%
  \BibitemOpen
  \bibfield  {author} {\bibinfo {author} {\bibfnamefont {D.}~\bibnamefont
  {Gamermann}}, \bibinfo {author} {\bibfnamefont {J.}~\bibnamefont {Nieves}},
  \bibinfo {author} {\bibfnamefont {E.}~\bibnamefont {Oset}},\ and\ \bibinfo
  {author} {\bibfnamefont {E.}~\bibnamefont {Ruiz~Arriola}},\ }\href
  {https://doi.org/10.1103/PhysRevD.81.014029} {\bibfield  {journal} {\bibinfo
  {journal} {Phys. Rev. D}\ }\textbf {\bibinfo {volume} {81}},\ \bibinfo
  {pages} {014029} (\bibinfo {year} {2010})},\ \Eprint
  {https://arxiv.org/abs/0911.4407} {arXiv:0911.4407 [hep-ph]} \BibitemShut
  {NoStop}%
\bibitem [{\citenamefont {Weinberg}(1965)}]{Weinberg:1965zz}%
  \BibitemOpen
  \bibfield  {author} {\bibinfo {author} {\bibfnamefont {S.}~\bibnamefont
  {Weinberg}},\ }\href {https://doi.org/10.1103/PhysRev.137.B672} {\bibfield
  {journal} {\bibinfo  {journal} {Phys. Rev.}\ }\textbf {\bibinfo {volume}
  {137}},\ \bibinfo {pages} {B672} (\bibinfo {year} {1965})}\BibitemShut
  {NoStop}%
\end{thebibliography}
%

\end{document}